\documentclass[printer]{aa}

\usepackage{natbib}
\bibpunct{(}{)}{;}{a}{}{,}

\usepackage{graphicx}
\usepackage{subfigure}
\usepackage{amsmath}

\newlength{\figwidth}
\usepackage{color}
\newcommand{\mybold}[1]{#1}

\begin{document}

\title{Magnetic confinement of the solar tachocline: \\ II. Coupling to a
  convection zone}
\titlerunning{Magnetic confinement of the tachocline : II. Coupling to a
  convection zone}

\author{A. Strugarek \inst{1}, A. S. Brun \inst{1} \and J.-P. Zahn \inst{2,1}}
\authorrunning{Strugarek, Brun, \& Zahn}
\offprints{A. Strugarek \\ \email{antoine.strugarek@cea.fr}}

\institute{
Laboratoire AIM Paris-Saclay, CEA/Irfu Universit\'e Paris-Diderot
CNRS/INSU, F-91191 Gif-sur-Yvette, France
\and
LUTH, Observatoire de Paris, CNRS-Universit\'e Paris Diderot, Place Jules Janssen, 92195 Meudon, France }

\abstract{The reason for the observed thinness of the solar tachocline is
  still not well understood. One of the explanations that have been
  proposed is that a primordial magnetic field renders the rotation uniform in the radiation zone.}
{We test here the validity of this magnetic scenario through 3D numerical 
MHD simulations that encompass both the radiation zone and the convection zone. }
{The numerical simulations are performed with the
anelastic spherical harmonics (ASH) code. The computational domain extends from $0.07\;R_\odot$ to $0.97\;R_\odot$.}
{In the parameter regime we explored, a dipolar fossil field aligned with the rotation axis cannot remain confined in the radiation zone. When the field lines are allowed to interact with turbulent unstationary convective motions at the base of the convection zone, 3D effects prevent the field confinement.}
{In agreement with previous work, we find that a dipolar fossil field,
 even when it is initially buried deep inside the radiation zone, will spread into the convective zone. According to
Ferraro's law of iso-rotation, it then imprints on the radiation zone the latitudinal differential rotation of the convection zone, which is not observed.}

\date{Received 14 january 2011 / Accepted 12 June 2011}


\maketitle

\section{Introduction}
\label{sec:introduction}
The discovery of the solar tachocline can be considered as one of the
great achievements of helioseismology (\textit{see}
\citealt{Brown:1989p7}). In this layer the rotation switches from
differential (i.e., varying with latitude) in the convection zone to nearly uniform in the radiation zone below.  Its extreme thinness (less than 5 \% of the solar radius) came as a surprise  (\textit{see} \citealt{Charbonneau:1999p19}) and 
has still not  been explained properly. 

\citet{Spiegel:1992p5} made the first attempt to model the 
tachocline, and they showed that it should spread deep into the radiation
zone owing to thermal diffusion. As a result, the
differential rotation should extend down to $0.3\, R_\odot$ after
$4.5$ Gyr, contrary to what is inferred from helioseismology
\citep{Schou:1998p3791,Thompson:2003p2511}. The authors suggested 
one way to confine the tachocline, namely to counter-balance thermal diffusion by an anisotropic turbulence, which would smooth the differential rotation in latitude. 
This model was successfully simulated in 2D by \citet{Elliott:1997p822}.

\citet{Gough:1998p34} (GM98 hereafter) argued that such an anisotropic
turbulent momentum transport is not able to erode a large-scale
latitudinal shear, citing examples from geophysical
studies (\textit{see} \citealt{starr1968physics} and more
  recently \citealt{Dritschel:2008p4287} for a comprehensive review). For instance, the quasi-biennal oscillation in the Earth's
stratosphere arises from the anti-diffusive behavior of anisotropic
turbulence. But the question is hardly settled because \citet{Miesch:2003p523} found in his numerical simulations that this turbulence is indeed anti-diffusive in the vertical direction, but the shear is reduced in latitude. Similar conclusions were reached through theoretical (analytical) studies by \citet{Kim:2005p623}, \citet{Leprovost:2006p549}, and \citet{Kim:2007p555}.

Gough \& McIntyre proposed an alternate
model, in which
a primordial magnetic field is buried in the radiative zone
and inhibits the spread of the tachocline. This is achieved through a 
balance between the confined magnetic field and a meridional flow pervading the base of the convection zone. 
Such a fossil field was also invoked by
\citet{Rudiger:1997p17} to explain the thinness of the tachocline and
by \citet{Barnes:1999p2302} to interpret the lithium-7 depletion at the
solar surface. Moreover, it would easily account for the quasi uniform rotation of the radiative interior,
which the hydrodynamic model of Spiegel \& Zahn does not (at least in its original formulation). 
However, one may expect that such a field would spread by ohmic diffusion, and that it would eventually connect with the convection zone, thus imposing the differential rotation of that zone on the radiation zone
below.

In order to test the magnetic confinement scenario and
  to put constraints on the existence of such a primordial field, several numerical simulations of the solar radiation zone were
performed by \citet{Garaud:2002p27} in 2D, and by \citet{Brun:2006p24}
(\textit{i.e.} paper I, hereafter BZ06) in 3D. These simulations exhibit a propagation of the differential rotation into the radiation zone
(primarily in the polar region). The primordial magnetic field
successfully connects to the base of the convection zone where the shear is imposed and, according to Ferraro's
law of iso-rotation (\citealt{Ferraro:1937p830}), the angular velocity is transmitted 
along the magnetic field lines from the convection zone into the
radiative interior, much faster than it would through thermal diffusion.
However, in the aforementioned calculations the base of the convection
zone was treated 
as an impenetrable wall, \textit{i.e.} radial motions could not penetrate in
either direction (from top down or from bottom up). No provision was
made for a meridional circulation originating in the convection zone that would penetrate into the radiation zone and could
prevent the upward spread of the magnetic field. 
This penetration was taken into account in numerical
simulations by
\citet{Sule:2005p31}, \citet{Rudiger:2007p40}, \citet{Garaud:2008p36}, in a
 model of the polar region by \citet{Wood:2007p4288}, and in
 simulations coupling the radiative zone to a convective zone by \citet{Rogers:2011p1234} to
properly represent the GM98 scenario. In most of these
  models, an axisymmetric
and stationary meridional circulation was imposed at the top of the radiation zone, strong enough  to bend the magnetic
field lines of the primordial field. As could be expected, various profiles of
meridional circulation resulted in different confinement properties, thus emphasizing the need
for a more self-consistent approach. We thus propose here to treat the problem
by letting a genuine
convective envelope dynamically generate its meridional circulation
and differential rotation and to study how these nonlinearly
  generated flows interact with a fossil field.

Our 3D simulations treat the nonlinear coupling of the convection and radiation zones,
and the model includes all physical ingredients
that play a role in the tachocline confinement (thermal and viscous
diffusion, meridional circulation, convective penetration, solar-like
stratification, magnetic fields, pumping, ...).
The paper is organized as follows. In Sect. \ref{sec:model} we describe the model
we use to address the question of the magnetic confinement of the
tachocline. We then emphasize in Sect. \ref{sec:general-evolution}
the global trend of our simulations, and examine in Sect. \ref{sec:analysis}
the dynamics of the tachocline region. Discussions and conclusions are
reported in Sect. \ref{sec:disc-concl}.

\section{The model}
\label{sec:model}

We use the well tested ASH code (anelastic spherical harmonics, see
\citealt{Clune:1999p42}, \citealt{Miesch:2000p29}, \citealt{Brun:2004p1}). Originally designed
to model solar convection, it has been adapted to include the
radiation zone as well (see BZ06). With this code, \citet{Brun:2010p1234}
performed the first 3D MHD simulations of the whole sun
from $r=0.07\, R_\odot$ to $r=0.97\, R_\odot$ with realistic
stratification, thus nonlinearly coupling the
radiation zone with the convection zone. Convective motions at the base of the convective envelope (around $r=0.72\, R_\odot$)
  penetrate into the radiative zone over a distance of about
$0.04\, R_\odot$, exciting internal waves that propagate over the
entire radiative interior. The model maintains a radiation zone  in 
solid body rotation and develops a differentially rotating convective zone
that agrees well with helioseismic inversions (\textit{see}
\citealt{Thompson:2003p2511}).
Although approximately three times thicker than observed, the simulated tachocline
possesses both a latitudinal and radial shear. This tachocline spreads
downward first because of thermal diffusion, and later because of viscous diffusion, as explained in SZ92.
We do not assume here
any anisotropic turbulent diffusivity to prevent the spread of the
tachocline.

\begin{figure}[!htbp]
  \centering
 \subfigure[]{
   \label{fig:dsdr}
  \includegraphics[width=.48\figwidth]{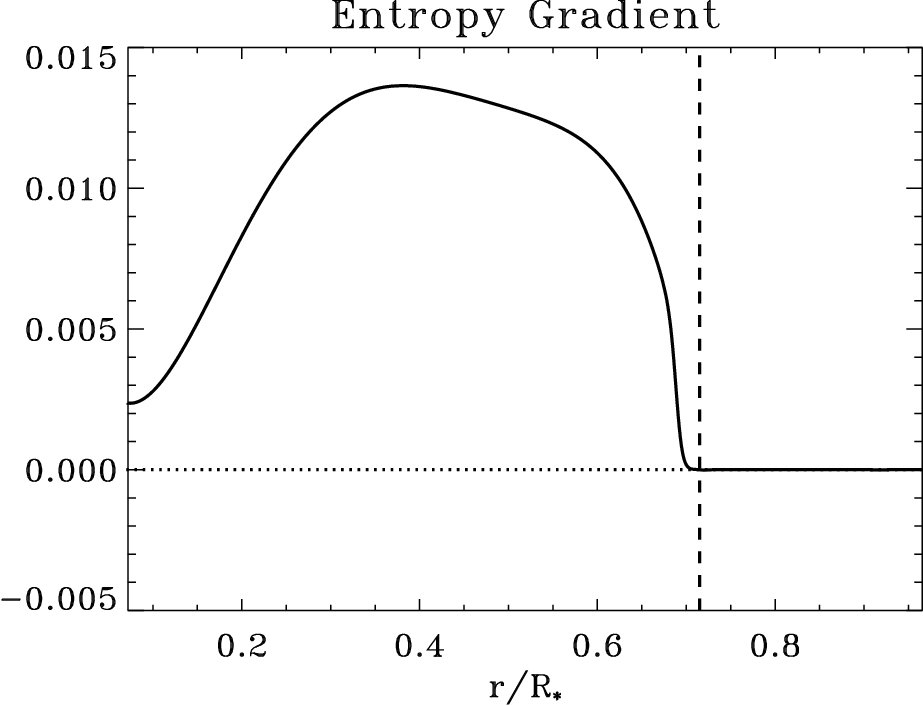}
 \includegraphics[width=.5\figwidth]{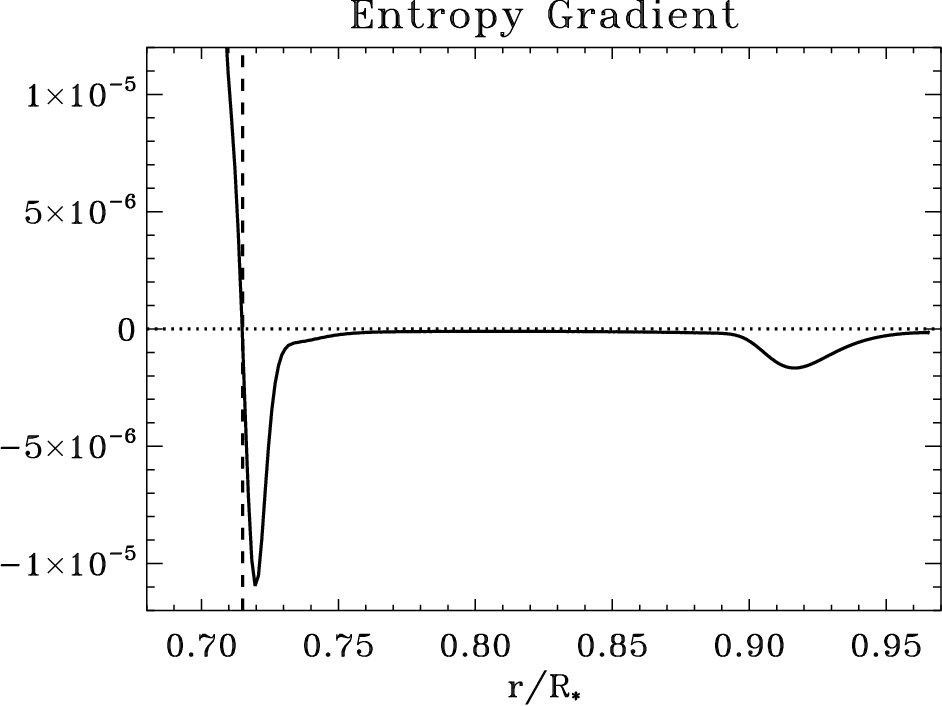} 
}
 \subfigure[]{
   \label{fig:vr_shsl}
  \includegraphics[width=.5\figwidth,angle=90]{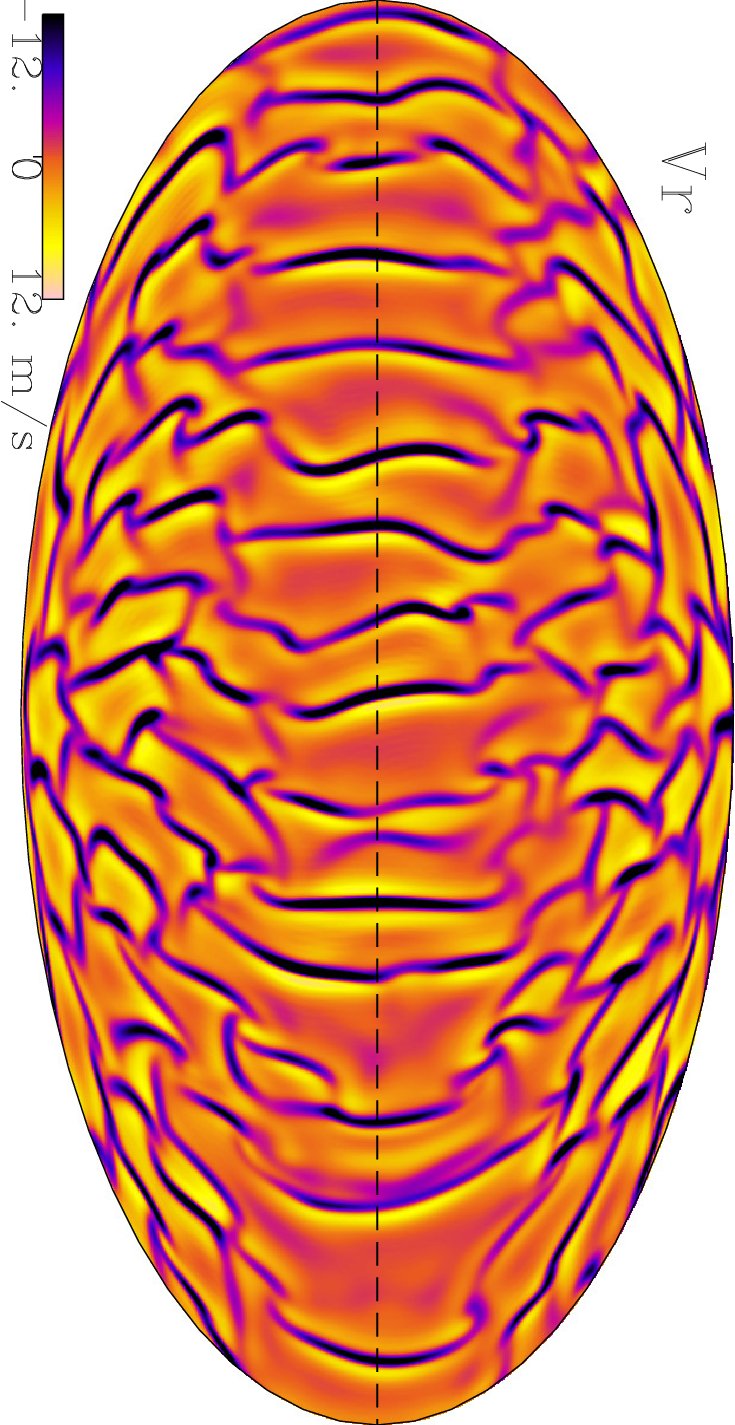}
}
 \subfigure[]{
   \label{fig:rms_vels}
  \includegraphics[width=\figwidth]{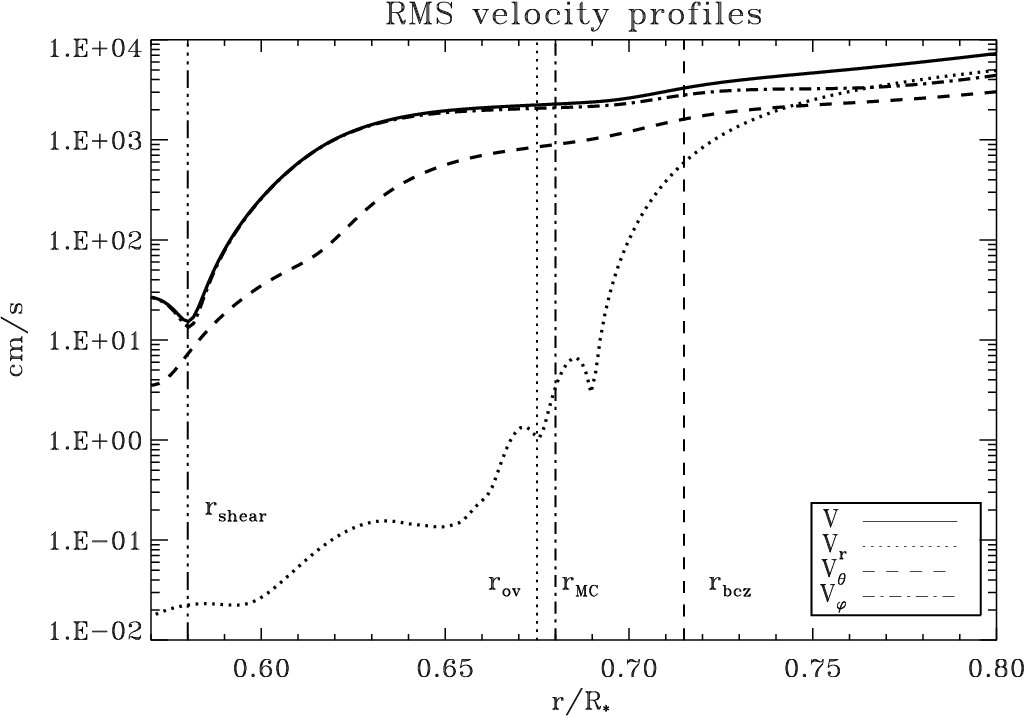}
}
  \caption{(a) Mean entropy gradient profile 
      and zoom in the region where the entropy gradient
    changes sign. Entropy gradients are plotted in
    $erg\, g^{-1}\, K^{-1}\, cm^{-1}$.
    (b) Radial velocity on a spherical shell
    near the top of the
    convection zone ($0.96\, R_\odot$). Dark colors represent the
    downflows, while bright-yellow colors denote the upflows.
    (c) rms velocities profiles at the interface between
  radiative and convective zones. The four vertical bars label
  the base of the convection zone (where the entropy
  gradient changes sign), the penetration depth of meridional circulation, the overshooting depth of convective
  motions, and the shear depth where differential rotation vanishes.}
  \label{fig:dsdr_plus_conv}
\end{figure}

In the present work, we start our simulation from a mature
  model with a convection and a radiation zone. We then
introduce an axisymmetric dipole-like magnetic field deep in the inner
radiation zone. This work is meant to explore further the mechanisms
considered in BZ06. The system is then evolved self-consistently
through the interplay of fluid and magnetic field dynamics.
As in \citet{Brun:2010p1234}, the computational domain extends from $r_{bottom}=0.07R_\odot$ to
$r_{top}=0.97R_\odot$, and the boundary between radiation and convection zone 
is defined at $r_{bcz}\sim 0.715\, R_\odot$, based on the initial entropy
profile (\textit{see} Figs. \ref{fig:dsdr} and \ref{fig:rms_vels})
that is assumed to setup the background equilibrium state. We
also define in Fig. \ref{fig:rms_vels} the
convective overshooting depth $r_{ov} \sim 0.675\, R_\odot$ (\textit{see}
Sect. \ref{sec:char-background-model}), the
penetration depth of the meridional circulation $r_{MC} \sim 0.68\, R_\odot$ (\textit{see}
Sect. \ref{sec:tach-regi-evol}), and the shear depth $r_{shear}\sim 0.58\,
R_\odot$ (\textit{see} Sect. \ref{sec:char-background-model}).

\begin{figure}[h]
  \centering
  \includegraphics[width=.75\figwidth]{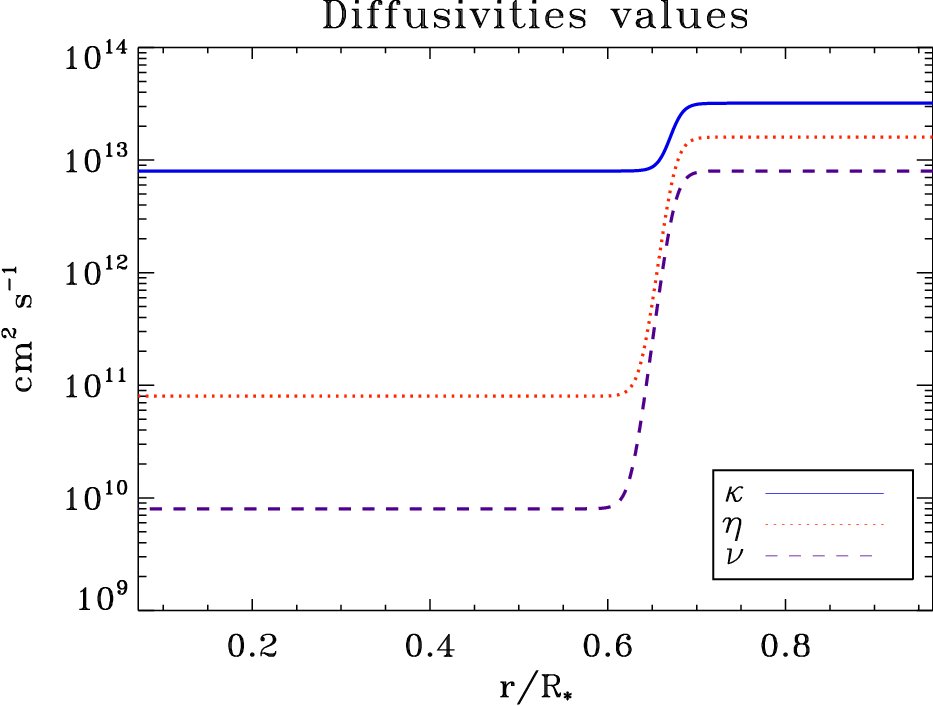}
  \caption{Radial profile of the diffusivities used in our model (the
    values are given in cgs units). The
    $\kappa$ profile corresponds to thermal diffusivity in equation (\ref{eq:entropy_conserv}), acting on the entropy fluctuations.}
  \label{fig:diffs}
\end{figure}

The diffusivities (Fig. \ref{fig:diffs}) were chosen to achieve
a realistic rotation profile
in the convective
  envelope (see Sect. \ref{sec:char-background-model}) by means of an efficient angular momentum redistribution by
  Reynolds stresses \citep{Brun:2002p996,Miesch:2006p1957}. In the
  radiation zone, they have the
same values as in BZ06 for consistency with previous studies. We define the
following $step$ function to construct our diffusivity profiles:
\begin{eqnarray}
  step(r) &=& \frac{1}{2}\left( 1 + \tanh\left(10^{-10}\frac{r-r_\nu}{\nu_s} \right) \right),\nonumber \\
 \nu &=& \nu_e\nu_t + \nu_t\left(1-\nu_e\right)\, step,
  \label{eq:nu_eq} 
\end{eqnarray}
where $r_\nu = 4.7\, 10^{10} \mbox{ cm}$ and $\nu_s = 0.1$ control the radial localization and
the thickness of the diffusivity jump. The parameter $\nu_t=8.0\,
10^{12} \mbox{ cm}^2\, \mbox{s}^{-1}$
controls the viscosity value in the convection zone and
$\nu_e = 10^{-3}$ controls the size of the viscosity jump. The thermal and
magnetic diffusivities are similarly defined with the same formulation
as for the step function, but with different $\kappa_t=3.2\, 10^{13}
\mbox{ cm}^2\, \mbox{s}^{-1},
\eta_t=1.6\, 10^{13} \mbox{ cm}^2\, \mbox{s}^{-1}$ and
$\kappa_e=0.25, \eta_e=5.0\, 10^{-3}$ parameters to generate the diffusivity profiles plotted in
Fig. \ref{fig:diffs}.
We used a $N_r\times N_\theta \times N_\varphi = 1024 \times 256 \times 512$ grid on massively parallel computers to
compute this model.
The Prandtl number $\nu/ \kappa$ is $10^{-3}$ in
the radiative zone and $0.25$ in the convective zone; and the
magnetic Prandtl number $\nu / \eta $ is $0.1$ in the radiative zone and $0.5$ in the
convective zone. The hierarchy between the various diffusivities is
thus respected even if their amplitudes are higher than in the
Sun. As will be seen in Sect. \ref{sec:general-evolution},
  they were chosen in a way that nonlinear processes act efficiently
  in the regions of interest.

\subsection{Governing equations}
\label{sec:governing-equations}
ASH solves the 3D MHD set of equations (\ref{eq:mass_consrv}-\ref{eq:inductiom_eq}). It uses the anelastic
approximation to filter out the sound waves, and the LES approach (large
eddy simulation) with parameterization to take into account subgrid
motions. The mean state is described by mean profiles of
density $\bar{\rho}$, temperature $\bar{T}$, pressure
$\bar{P}$, and entropy $\bar{S}$. The reference state is taken
from a thermally relaxed 1D solar structure model (\citealt{Brun:2002p831}) and is regularly
updated.  Fluctuations around the reference state are denoted
without bars. The governing equations are written in electromagnetic units, in the reference frame rotating at angular velocity $\mathbf{\Omega}_0=\Omega_0\mathbf{e}_z$:
\begin{align}
 \label{eq:mass_consrv}
 \boldsymbol{\nabla}\cdot\left(\bar{\rho}\mathbf{v}\right) &= 0 \\
   \label{eq:Maxw1}
   \boldsymbol{\nabla}\cdot\mathbf{B} &= 0 \\
   \bar{\rho}\left[\partial_t \mathbf{v} + \left(\mathbf{v}\cdot \mbox{\boldmath $\nabla$}
     \right)\mathbf{v} + 2\mbox{\boldmath $\Omega_0$}\times\mathbf{v} \right] &=-\mbox{\boldmath $\nabla$} P +
   \rho \mathbf{g}  \nonumber \\
   +\frac{1}{4\pi}\left(\mbox{\boldmath $\nabla$}\times\mathbf{B}\right)\times\mathbf{B} &-
   \mbox{\boldmath $\nabla$}\cdot\mathbf{\mathcal{D}} - \left[\mbox{\boldmath $\nabla$}\bar{P}-\bar{\rho}\mathbf{g} \right]
          \label{eq:mom_conserv} \\
          \bar{\rho}\bar{T}\left[\partial_t S +
         \mathbf{v}\cdot\mbox{\boldmath
           $\nabla$}\left(\bar{S}+S\right) \right] &=
              \boldsymbol{\nabla}\cdot\left[
                \kappa_r\bar{\rho}c_p\mbox{\boldmath $\nabla$}\left(\bar{T}+T\right)\right. \nonumber
                  \\ 
                  + \left. \kappa_{0}\bar{\rho}\bar{T}\mbox{\boldmath $\nabla$}\bar{S} +
                    \kappa\bar{\rho}\bar{T}\mbox{\boldmath $\nabla$} S \right] &+
                  \frac{4\pi\eta}{c^2}\mathbf{J}^2 \nonumber \\ 
                  + 2\bar{\rho}\nu [ e_{ij}e_{ij}
                  -\frac{1}{3}\left(\boldsymbol{\nabla}\cdot\mathbf{v}\right)^2
                  ] &+ \bar{\rho}\epsilon
   \label{eq:entropy_conserv} \\
   \partial_t \mathbf{B} =
   \mbox{\boldmath $\nabla$}\times\left(\mathbf{v}\times\mathbf{B}\right) &- \mbox{\boldmath $\nabla$}\times\left(\eta\mbox{\boldmath $\nabla$}\times\mathbf{B}\right),
   \label{eq:inductiom_eq} 
\end{align}
where $\mathbf{v}$ is the local velocity, $\mathbf{B}$ the magnetic field, $\kappa$, $\nu$ and
$\eta$ are respectively the effective thermal diffusivity, eddy viscosity,
and magnetic diffusivity. 
The radiative diffusivity $\kappa_r$ is deduced from a 1D model and
  adjusted in the radiative zone and in the overshooting layer
  to achieve an equilibrated radial flux balance (\textit{see} Fig. \ref{fig:radial_flux_bal}).
The thermal diffusion coefficient $\kappa_0$
plays a role at the top of the convective zone (where convective motions vanish) to ensure the heat transport through the surface. This term proportional to $d\bar{S}/dr$ is
part of our subgrid scale treatment in the convection zone. The
unresolved flux only acts in the upper part of the convection zone
(\textit{see} Fig. \ref{fig:radial_flux_bal}),
and $\kappa_0$ is chosen to be small enough that this flux does not
play any role in the radiative interior (besides eventual numerical stabilization).
$\bar{\rho}\epsilon$ is the rate of nuclear energy generation chosen
to have the correct integrated luminosity (\textit{see} \citealt{Brun:2010p1234}). 
$\mathbf{J}=(c/4\pi)\mbox{\boldmath $\nabla$}\times\mathbf{B}$
is the current density, and the viscous stress tensor $\mathcal{D}$
is defined by
\begin{equation}
  \label{eq:visc_stress_tensor}
 \mathcal{D}_{ij} = -2\bar{\rho}\nu\left[e_{ij}-\frac{1}{3}\left(\mbox{\boldmath $\nabla$}\cdot\mathbf{v}\right)\delta_{ij}\right] \, .
\end{equation}
The system is closed by using the linearized ideal gas law:
\begin{equation}
  \label{eq:gas_law}
  \frac{\rho}{\bar{\rho}} = \frac{P}{\bar{P}} - \frac{T}{\bar{T}} =
  \frac{P}{\gamma \bar{P}} - \frac{S}{c_p},
\end{equation}
with $c_p$ the specific heat at constant pressure and $\gamma$ the
adiabatic exponent.

We chose rigid and stress-free conditions at the boundary shells
for the velocity. We also imposed constant mean entropy gradient at the
boundaries, matched the magnetic field to an external potential
magnetic field at the top, and treated the bottom boundary as a
perfect conductor.

\subsection{Characteristics of the background model and choice of the initial magnetic field}
\label{sec:char-background-model}

Integrated solar models built with the ASH code that couple a deep
radiative interior to a convective zone have been previously
described in \citet{Brun:2010p1234}; here we need only to stress some
properties that are of particular interest for our study of the tachocline
dynamics. 
\begin{figure}[!htbp]
  \centering
 \includegraphics[width=.65\figwidth,angle=90]{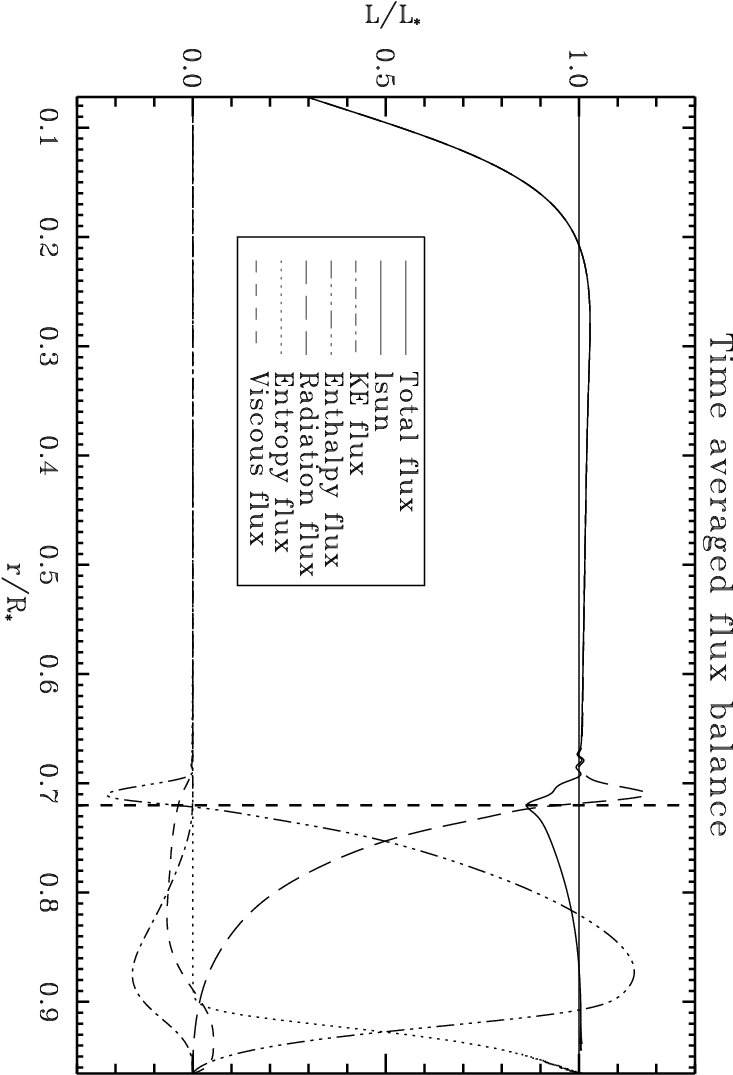}
 \caption{Radial flux balance. In the convection zone, energy is
    mainly carried by the enthalpy flux; the entropy flux represents
   the flux carried by the unresolved motions. Note the penetration of
   convective motions below the convection zone (represented by the dashed line).}
  \label{fig:radial_flux_bal}
\end{figure}

\begin{figure*}[!htb]
  \centering
 \subfigure[]{
  \label{fig:Omega_hydro}
  \includegraphics[width=.45\figwidth]{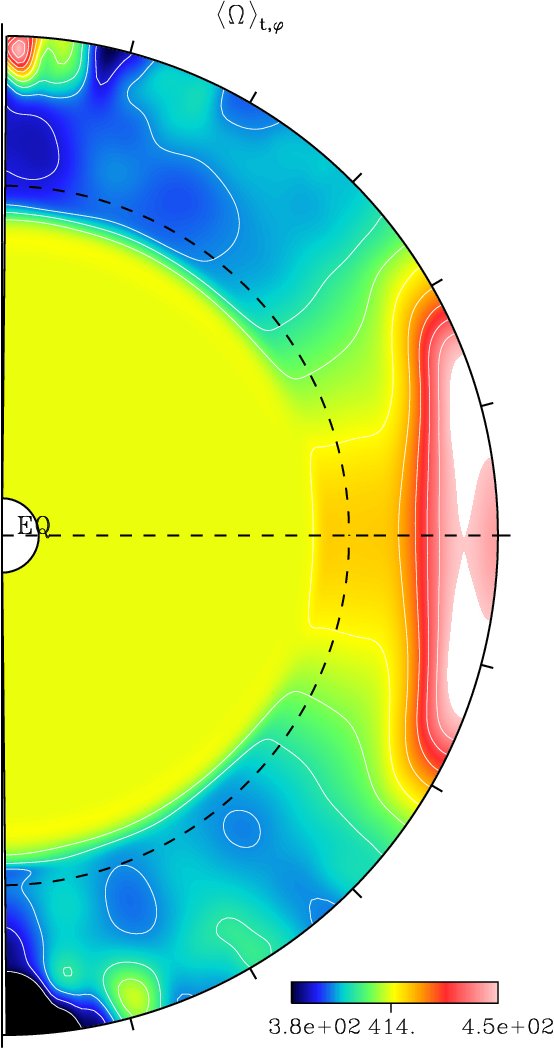}
}
  \subfigure[]{
  \label{fig:Omega_hydro_several_lats}
 \includegraphics[width=.4\figwidth]{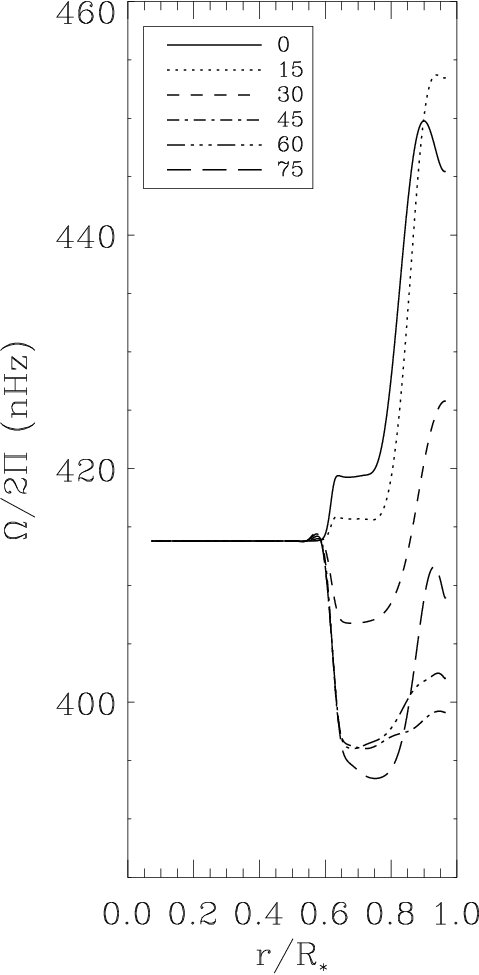}  
}
 \subfigure[]{
  \label{fig:MC_hydro}
  \includegraphics[width=.85\figwidth,angle=90]{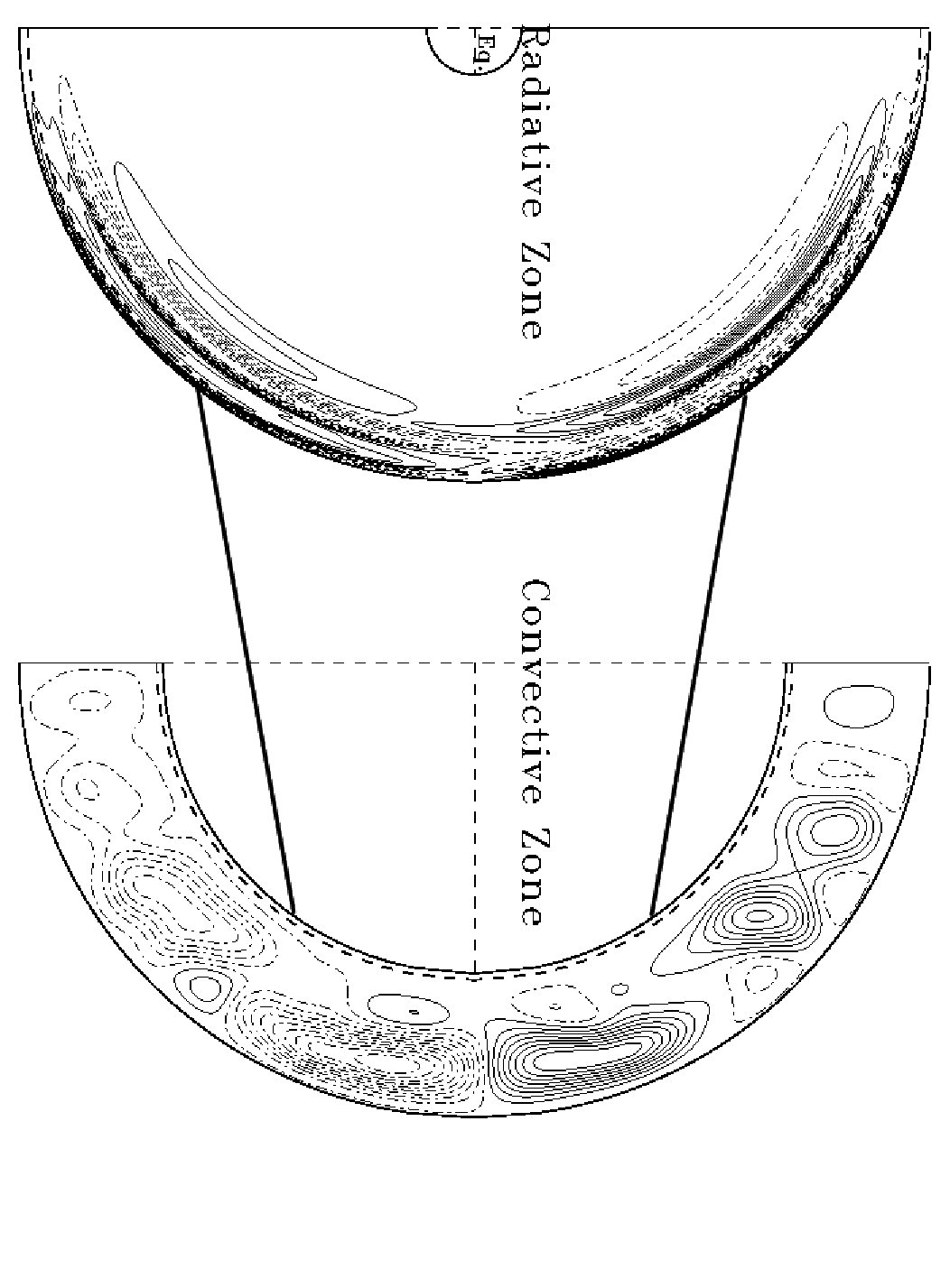}
}
  \caption{(a) Azimuthally and temporally averaged rotation
   frequency. Yellow denotes the rotation rate of the interior, red is
    the higher rotation rate and blue the lower rotation rate. (b) Radial profile of rotation frequency at different
    latitudes. (c) Meridional circulation in the radiation zone (left) and in the convection zone (right),
    evaluated from the poloidal mass flux averaged over time and
    longitude. Solid contours denote counterclockwise circulation, dashed
    contours clockwise circulation.}
  \label{fig:rot_prof}
\end{figure*} 

The starting point of our model is the choice of thermodynamic
quantity profiles that will allow us to treat the
convective and radiative zones together. Using a standard solar model
calibrated to seismic data (\citealt{Brun:2002p831}) computed by the
CESAM code (\citealt{Morel:1997p1788}), we chose to use
the real solar stratification in the radiation zone and take a constant negative initial entropy gradient in
the convection zone.
These profiles are closely in agreement 
the guess profile after a Newton-Raphson solve. After setting a stable/unstable
stratification, we perturbed the background state assuming a
Rayleigh number well above the critical Rayleigh number for the onset
of convection. A sample of the convective motions realized in the
model is shown in Fig. \ref{fig:vr_shsl}, where one can recognize
the characteristic banana pattern near the equator and a patchy
behavior at high latitudes (\textit{see} \citealt{Brun:2002p996}). We
then let the model evolve toward a mature state in two steps. First we
let the overshooting layer and the differential rotation develop. Then
we sped up the thermal relaxation of the system by adding a bump
on the radiative flux in the overshooting region (\textit{see} \citealt{Brun:2010p1234} for
more details). Thanks to this procedure, we quickly achieved good radial
flux balance (Fig. \ref{fig:radial_flux_bal}). The energy provided by
nuclear reactions is
transported through radiation in the radiation zone, and
is essentially transported by the enthalpy flux in the convection
zone. The overshoot region, defined here as the region of negative
enthalpy flux, extends from $r_{bcz}=0.715\, R_\odot$ to $r_{ov}=0.675\, R_\odot$,
leading to $d_{ov} \approx 0.04\, R_\odot \approx 0.4\, H_p$ (where
$H_p$ is the local pressure scale height).
In the convection zone, the
radiative flux quickly goes to zero, while viscous and kinetic energy
fluxes tend to oppose the outward transport of heat of the
enthalpy flux associated to the convective motions.
Near the top of the domain, convective motions vanish due to our
choice of impenetrable wall; thus the enthalpy flux
decreases, and the energy is then transported by the entropy flux.
 
With our choice of diffusion coefficients (Fig. \ref{fig:diffs}), the system reached 
a state of fully developed convection with a rotation profile that
agreed well with the results of helioseismology
(\textit{e.g.} \citealt{Thompson:2003p2511}): the convection zone rotates faster
at the equator than at  the poles, and the rotation profiles
are conical at mid-latitudes (see Fig. \ref{fig:Omega_hydro}).  We
display the radial profile of $\Omega$ at indicated latitudes in Figure
\ref{fig:Omega_hydro_several_lats}. We clearly see the presence of a tachocline.
Future work will aim at reducing the thickness of the jump of the
diffusivities (Fig. \ref{fig:diffs}) with a more
flexible radial discretization (currently under development) 
to model a thinner hydrodynamic tachocline.
Finally, the meridional circulation in the convection zone is composed by a
major cell in each hemisphere (Fig. \ref{fig:MC_hydro}), and by smaller counter-rotating cells at
the poles.

When the hydrodynamic model reached an equilibrated state (as
reported in \citealt{Brun:2010p1234}), we introduced a dipolar axisymmetric magnetic field buried in the
radiative zone. It is set to vanish at the base of the tachocline, below the level
where differential rotation starts, and we imposed the functional form 
$\mathbf{B} = B_0\left(B_r\mathbf{e}_r+
  B_\theta\mathbf{e}_\theta\right)$, with 
\begin{equation}
  \label{eq:mag_field_init}
  B_r = \frac{1}{r^2\sin\theta}\partial_\theta\Psi \, , \hspace{1cm}
  B_\theta = -\frac{1}{r\sin\theta}\partial_r\Psi \, .
\end{equation}
$\Psi(r,\theta)$ is constant on field lines. As in BZ06, we chose
\begin{eqnarray}
  \label{eq:Psi_Choice}
  \Psi &= \left({r}/{R}\right)^2(r-R_b)^2\sin^2\theta & \mbox{for
  } r\le R_b \nonumber\\
  &= 0 & \mbox{for } r\ge R_b
\end{eqnarray}
where $R_b = 0.57\, R_\odot$ is the bounding radius of the confined
field. According to \citet{Gough:1998p34}, the amplitude of
  the magnetic field controls the scalings of the
  tachocline and the magnetopause (a thin layer of intense magnetic field).
  With our choice of parameters, these scalings would require an
  initial seed magnetic field of the order of
$10^4\, G$ (\textit{i.e.}, $\delta/R \sim
  \Delta/R \sim 10^{-2}$, in GM98's notations). Here we
set $B_0$ to $4.2\,10^4\,G$ to be at equipartition of
energy in the radiative zone between magnetic energy and
  total kinetic energy in the rotating reference frame. 

We stress here that such a magnetic field is subject to the high $m$ Tayler
instability (\textit{see} \citealt{Tayler:1973p1727};
\citealt{Brun:2006p24,Brun:2007p2626}) because it has no azimuthal component
to start with.
One could prefer to choose a mixed
poloidal/toroidal magnetic field because it is the only possible stable
configuration of magnetic field in radiative stellar interiors
\citep{Tayler:1973p1727,Braithwaite:2004p1944,Duez:2010p1947}.
However, this has no consequences for the problem at hand, because
a toroidal field quickly develops and stabilizes the magnetic configuration.

\section{General evolution}
\label{sec:general-evolution}

The first evolutionary phase of our simulation is similar to that in
BZ06, with the magnetic field diffusing through the tachocline. But here we
follow its evolution as it enters into the convection zone, and we witness
the back-reaction of convective motions on the field. The outcome is
the same in the end, as we shall see in
Sect. \ref{sec:glob-evol-magn}.
More details on tachocline dynamics are
given in Sect. \ref{sec:tach-regi-evol}.

\begin{figure*}[!htbp]
  \centering
  \subfigure[]{
  \label{fig:vphi_plus_Bstrl1}
  \includegraphics[width=.5\figwidth]{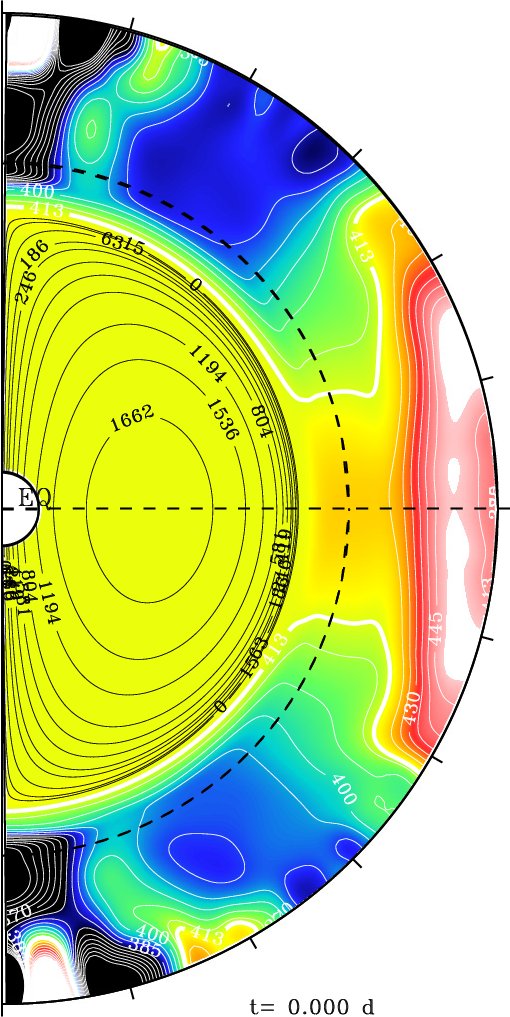}
}
  \subfigure[]{
  \label{fig:vphi_plus_Bstrl2}
  \includegraphics[width=.5\figwidth]{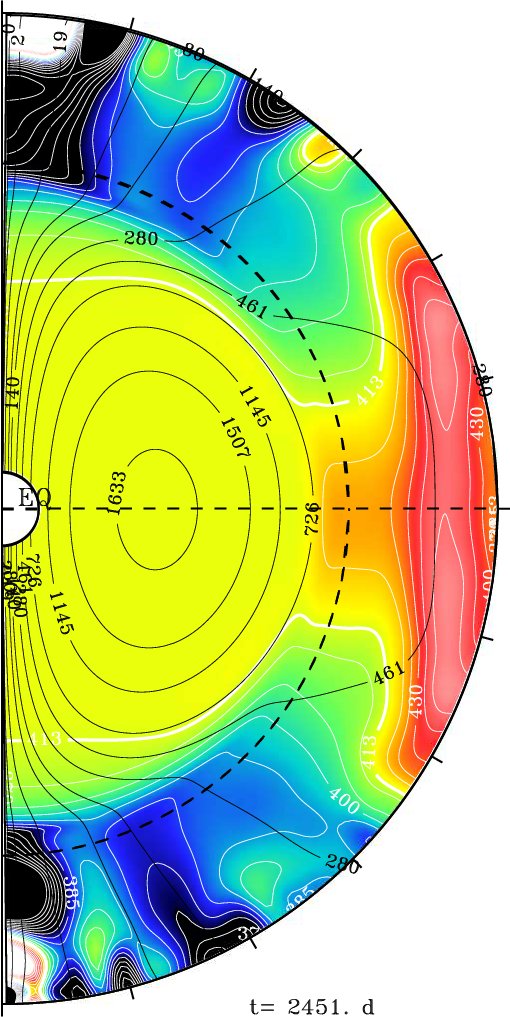}
}
  \subfigure[]{
  \label{fig:vphi_plus_Bstrl3}
  \includegraphics[width=.5\figwidth]{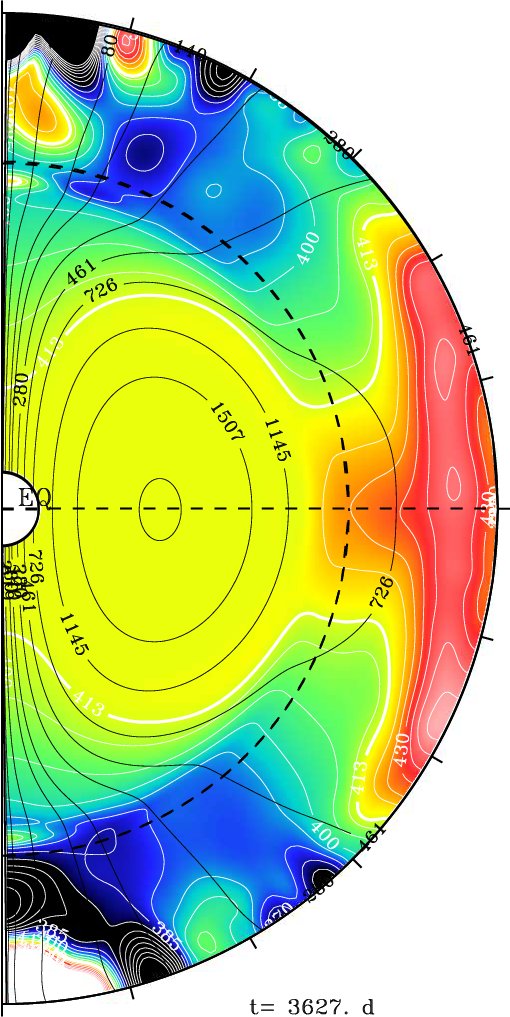}
}
  \subfigure[]{
  \label{fig:vphi_plus_Bstrl4}
  \includegraphics[width=.5\figwidth]{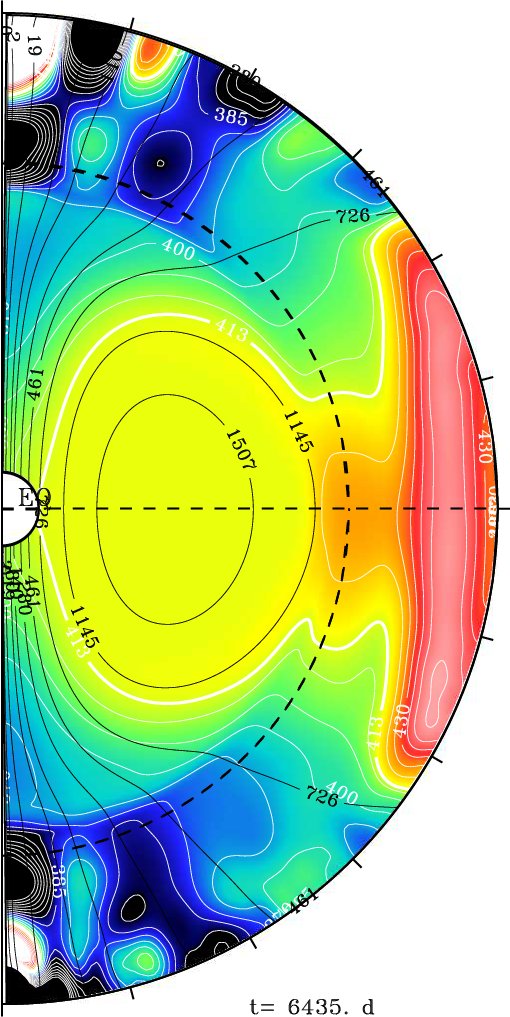}
}
  \caption{Poloidal snapshots of azimuthal averages of the rotation profile
    (colored background) and magnetic field lines (black). Color table
  is the same as in Fig. \ref{fig:Omega_hydro}.\vspace{1cm}}
  \label{fig:vphi_plus_Bstrl}
\end{figure*}
\begin{figure*}[!htbp]
  \centering
  \subfigure[]{
  \label{fig:Bphi_p_MC1}
  \includegraphics[width=.5\figwidth]{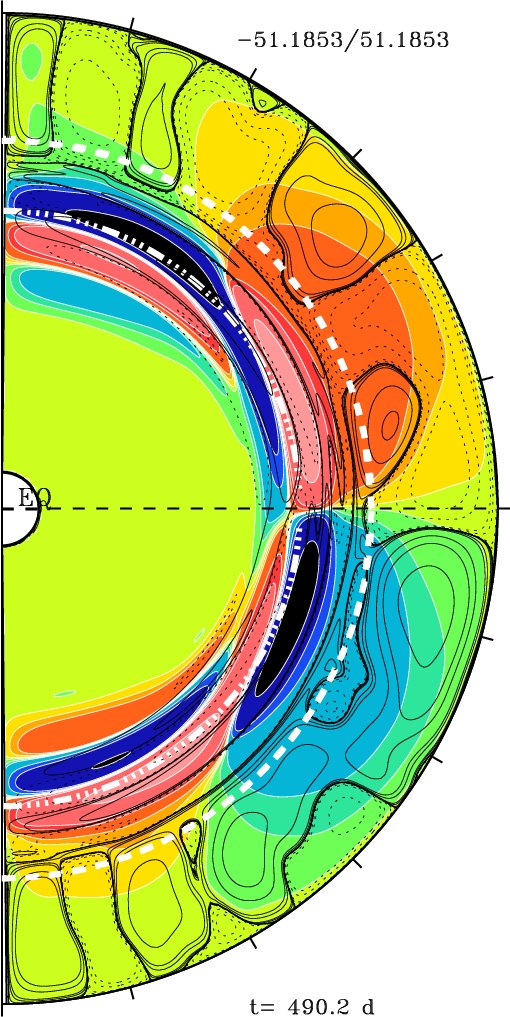}
}
  \subfigure[]{
  \label{fig:Bphi_p_MC2}
  \includegraphics[width=.5\figwidth]{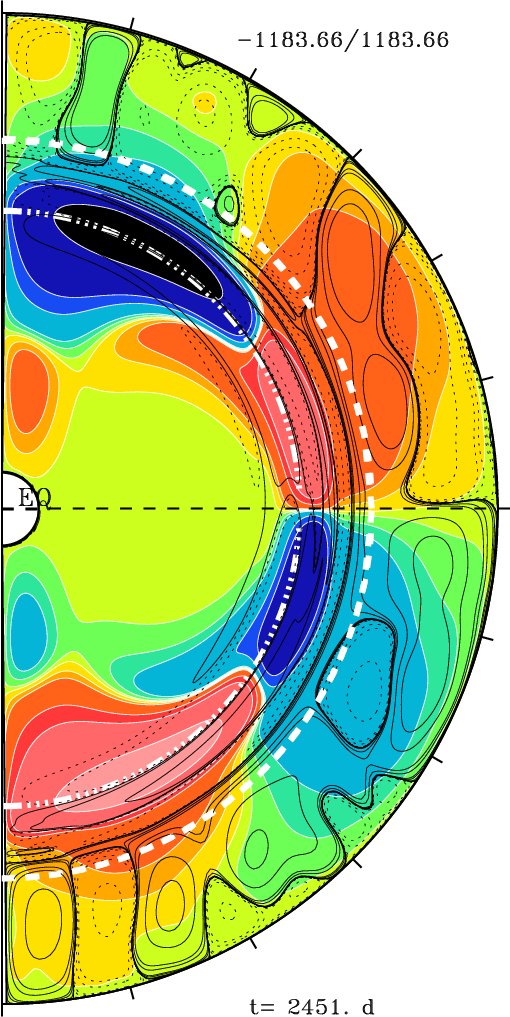}
}
  \subfigure[]{
  \label{fig:Bphi_p_MC3}
  \includegraphics[width=.5\figwidth]{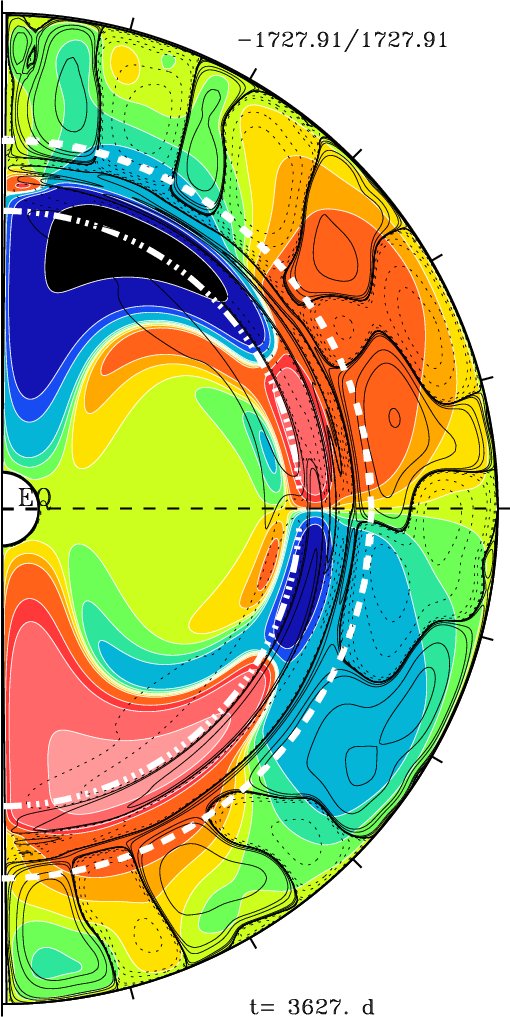}
}
  \subfigure[]{
  \label{fig:Bphi_p_MC4}
  \includegraphics[width=.5\figwidth]{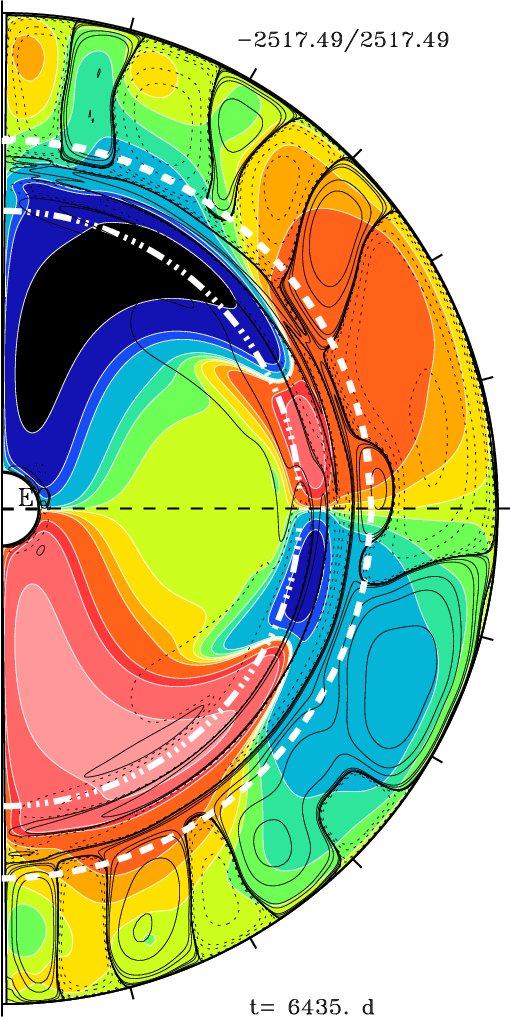}
}
  \caption{Poloidal snapshots of $\langle B_\varphi \rangle$
    (colored background) and instant meridional circulation (black
    lines). Initially, the magnetic field is purely poloidal. The first
    figure on the left is taken just after the beginning iteration
    (not the same time as in Fig. \ref{fig:vphi_plus_Bstrl1}) to demonstrate
    how the longitudinal magnetic field is primarily created. Red is
    the positive magnetic field, and blue the negative magnetic
    field. In order to display the azimuthal magnetic field 
      in the radiation zone and the convection zone (where it is
      much weaker), a logarithmic scale was used for the color
      table. The two white lines indicate the base of the convection zone
      $r_{bcz}$ and the shear depth $r_{shear}$ (\textit{cf} Fig. \ref{fig:rms_vels}).}
  \label{fig:Bphi_p_MC}
\end{figure*}

\subsection{Global evolution of the magnetic field}
\label{sec:glob-evol-magn}

We started the simulation by burying the primordial field
  deep within the radiative interior and below the tachocline. 
As the simulation proceeded, the field expanded through the tachocline into the convection zone, where it
experienced a complex evolution under the combined action of
advection by the convective motions, shearing through the differential
rotation and ohmic diffusion.
It finally reached the domain boundary (Fig.
\ref{fig:vphi_plus_Bstrl2}) where it connected to an
external potential field.  

The distortion of the field lines produces magnetic torques
that force the angular velocity  $\Omega$ to tend toward constant
along the field lines of the mean poloidal field, thus \mybold{leading
to}
Ferraro's law of iso-rotation. 
This is illustrated in Fig. \ref{fig:vphi_plus_Bstrl}, where the iso-contour
$\Omega = 414 \mbox{ nHz}$ (\textit{i.e.}, the initial rotation frequency of the whole radiation zone)
is \mybold{thicker and more intense}.
As a result, \mybold{the radiation zone slows down under the influence
  of the magnetic torques (Fig. \ref{fig:vphi_plus_Bstrl3}). Our
  choice of
  torque-free boundary conditions implies that the total
angular momentum is conserved. It is extracted from the radiative zone
and redistributed within the convective envelope.}
\begin{figure*}[!htbp]
  \centering
  \subfigure[]{
  \label{fig:3D_1}
  \includegraphics[width=.68\figwidth]{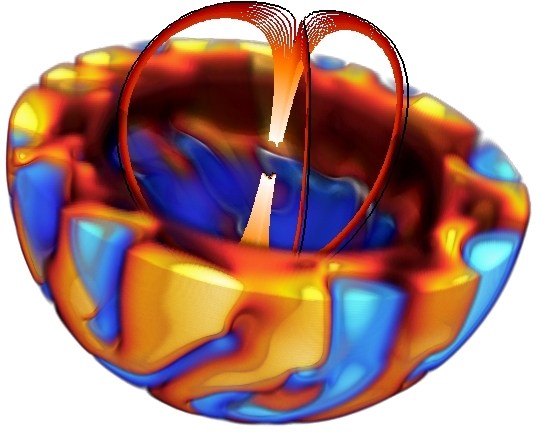}  
}
  \subfigure[]{
  \label{fig:3D_2}
  \includegraphics[width=.68\figwidth]{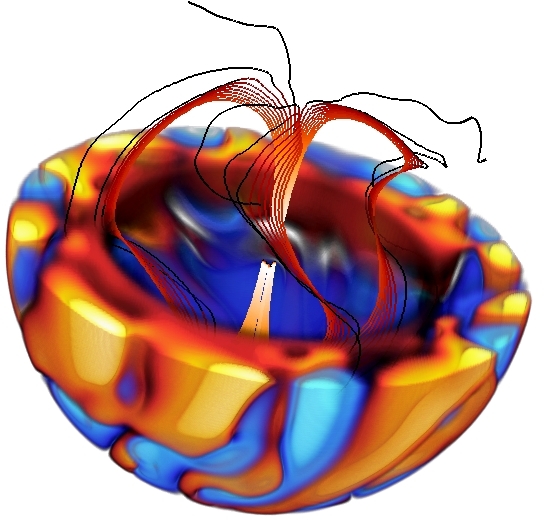}  
}
  \subfigure[]{
  \label{fig:3D_3}
  \includegraphics[width=.68\figwidth]{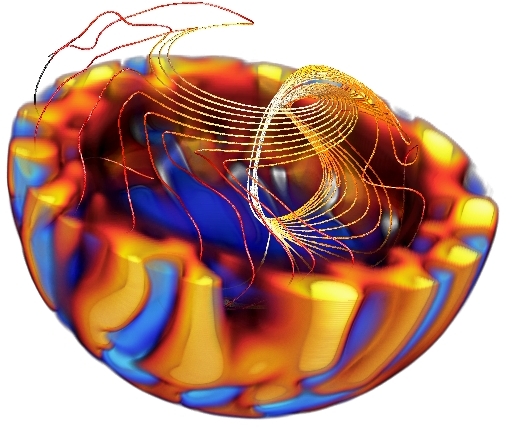}  
}
  \caption{Three-dimensional views of the longitudinal velocity in the
    rotating frame. Red are the positive velocities, and blue the negative
    velocities. Picture (a) is the initial time where we added the
    magnetic field, and picture (b) is taken after two solar rotation periods.
    Picture (c) is taken much later in the simulation, when the
    magnetic layer became stable.
    Colored lines are the magnetic field lines we
    constructed from only a few seed points to be able to
    distinguish them. Dark colors denote lower magnetic field
    intensity. Note that for the third image, we only put in seed
    points at one longitude value to distinguish the field
    lines that explore almost the entire $\varphi$ domain in the
    radiation zone.}
  \label{fig:3D}
\end{figure*}
\begin{figure*}[!htbp]
  \centering
 \includegraphics[width=2.2\figwidth]{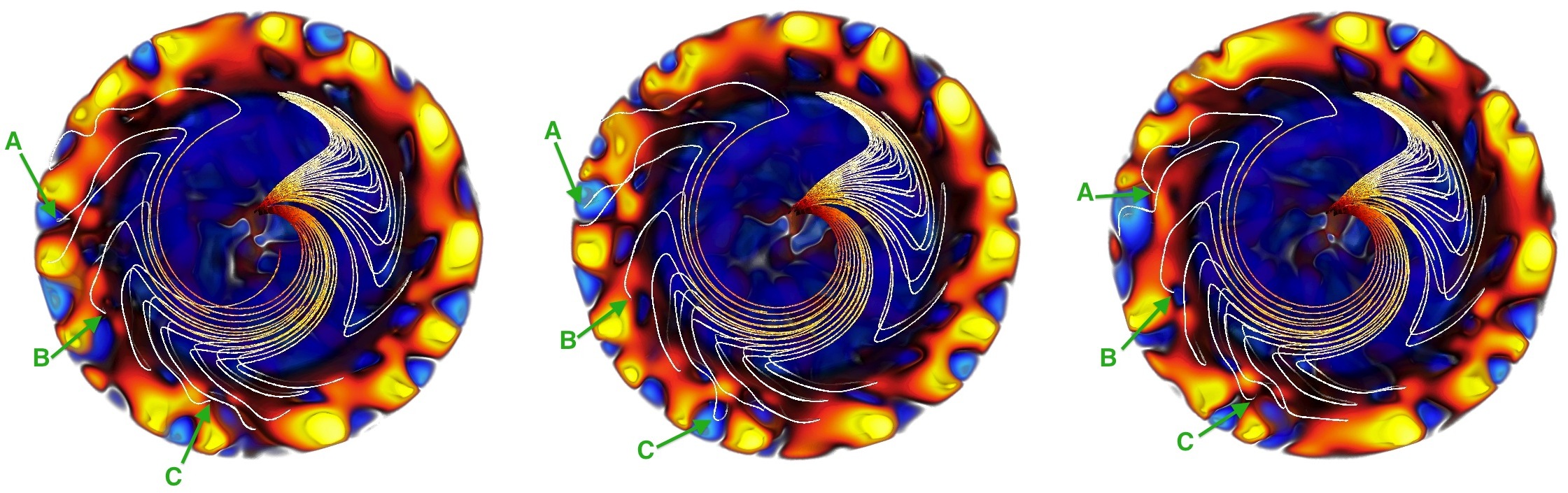}  
 \caption{Three-dimensional views from the north pole. The left
    picture corresponds to Fig. \ref{fig:3D_3}, and the two following
    snapshots are taken $2$ and $4$ days later. We
    reversed the color table of the field lines to emphasize their
    movement. The green arrows indicate the movement of three field
    lines during this period.
  }
  \label{fig:3D_up}
\end{figure*}

As the initial magnetic field meets the angular velocity shear, toroidal
magnetic field is created in the tachocline region \textit{via} the
$\Omega$-effect (\textit{see} \citet{Moffatt:1978p1953}), consistent with
the magnetic layer introduced in GM98. Near the initial time
(Fig. \ref{fig:Bphi_p_MC1}) the the magnetic layer
exhibits a mixed $l=1,l=3$ configuration. The axisymmetric longitudinal
magnetic field then presents a $l=3$ structure
(Fig. \ref{fig:Bphi_p_MC2}), consistent with the action of
differential rotation on the dipolar magnetic field through the
induction equation. This layer of axisymmetric azimuthal
magnetic field spreads downward, accompanying the spread of
differential rotation
(Figs. \ref{fig:Bphi_p_MC3}-\ref{fig:Bphi_p_MC4}). We also observe
that the upper radial localization of maximum azimuthal magnetic field remains remarkably localized at the
base of our initial tachocline. Some azimuthal field does penetrate
into the convective envelope, partly because of some local $\Omega$-effect
that is at work in this region.

Meridional circulations exist in the radiative and convective
zones and are self-consistently generated by the convective
motions and local dynamics (Fig. \ref{fig:Bphi_p_MC}). Observe that the evolution of the
magnetic layer greatly perturbs the pattern of the meridional
circulations in the radiative zone.
Comparing Fig. \ref{fig:Bphi_p_MC} to the hydrodynamic
model (Fig. \ref{fig:MC_hydro}), we observe that even if temporal
averages of the meridional circulation produce large hemispherical
cells in the convection zone, the instantaneous pattern of this flow
is multicellular both in longitude and latitude. Those 3D  motions are unable
to prevent the magnetic field from spreading into the
convective zone, in contrast to what GM98 and
\citet{Garaud:2008p36} proposed in their $2$D scenario. \mybold{Note that
\citet{Rogers:2011p1234} confirm the
  absence of magnetic field confinement in recent 2D simulations,
  where the author
  coupled a radiative interior to a convective envelope.}

Azimuthally averaged views can mislead the reader here bacause from those
figures one cannot deduce the real impact of the turbulent convective motions
on the magnetic field. Figure \ref{fig:3D} displays
3D visualizations that emphasize at the same time the
advective and diffusive processes that act on the magnetic field in
the convective zone. Magnetic field lines are twisted and sheared by
convective motions, but still the magnetic field connects the
radiative and convective zones. One clearly observes the
\mybold{mostly horizontal} magnetic layer in the tachocline on Fig. \ref{fig:3D_3}.
In that panel also the radiation zone starts to show the `footprint'
of the differential rotation
of the convection zone. In order to follow one particular field
line, we display in Fig. \ref{fig:3D_up} a \mybold{3D rendering time
  evolution using a high sampling cadence,} viewed from the north pole. We observe that particular
field lines (labeled A, B, and C) are strongly influenced by
convective motions and exhibit a
complex behavior, far from just simple diffusion. \mybold{The primary
  importance of 3D motions will also be discussed in
  Sect. \ref{sec:tach-regi-evol}.}

We stress here that the magnetic field lines first penetrate into the
convective zone through the equator. One may wonder how this can be achieved
in spite of the magnetic pumping (\textit{see}
\citealt{Tobias:2001p472,Weiss:2004p464}) that should be acting
in this region. Let us mention several reasons why the magnetic
pumping may not be efficient enough in our simulation, and by
extension perhaps in the Sun:
\begin{itemize}
\renewcommand{\labelitemi}{$\bullet$}
\item Magnetic pumping has been proven to be much less
efficient when strong rotation is present \mybold{and enhances
horizontal mixing. This corresponds to a low Rossby number 
regime, which is the case for the Sun and our simulation.}
Indeed,
our Rossby number $\displaystyle R_o =
{\omega_{rms}}/{2\Omega}$ \mybold{varies from} $0.1$ \mybold{to} $1$
in the convection zone.
\item It has also been proven to be much less efficient as the
underlying region (here the tachocline) is more stably stratified. We
use here a realistic (strong) solar-like stratification.
\item The meridional circulation tends to advect the
magnetic field upward at the equator, opposing the effect
of the magnetic pumping. 
However, our simulation is perhaps not turbulent
enough to counterbalance this effect, even though $R_m>1$.
\item \mybold{Our relatively high diffusivities may facilitate the connection between the two
  zones by favoring the field lines' expansion. Nevertheless, it is also known that diffusion helps the
  'slipping' of the field lines (\textit{e.g.},
  \citealt{Zanni:2009p2526} from ideal flux frozen scenario). This
  implies that even though the field lines have pervaded into the
  convection zone, they may not be efficiently enough anchored to
  transport angular momentum.
 More details will be given in Sect. \ref{sec:tach-regi-evol}}.
\item \mybold{The nature of penetration at the base of the convection
    zone is influenced by the Peclet number. In our case, the Peclet
    number is on the order of $1$ at the base of the convection zone,
    implying a regime of overshooting rather than of penetrative
    convection \citep{Zahn:1991p2459}. This translates into a more
    extended region of mixing than what is occurring in the Sun. Hence,
    the motions in our simulation are likely to be more vigorous at $r_{ov}$.}
\end{itemize}
For all these reasons, the magnetic
pumping near the equator in a fossil buried magnetic field scenario
\mybold{\citep{Wood:2011p1259}}
 is quite complex and cannot be taken for granted as
a mechanism to impermeably confine an inner magnetic field near the
equator (detailed magnetic field evolution may be found in
Sect. \ref{sec:tach-regi-evol}).

\begin{figure*}[!htbp]
  \centering
  \subfigure[]{
  \label{fig:vr0}
  \includegraphics[width=.25\figwidth,angle=90]{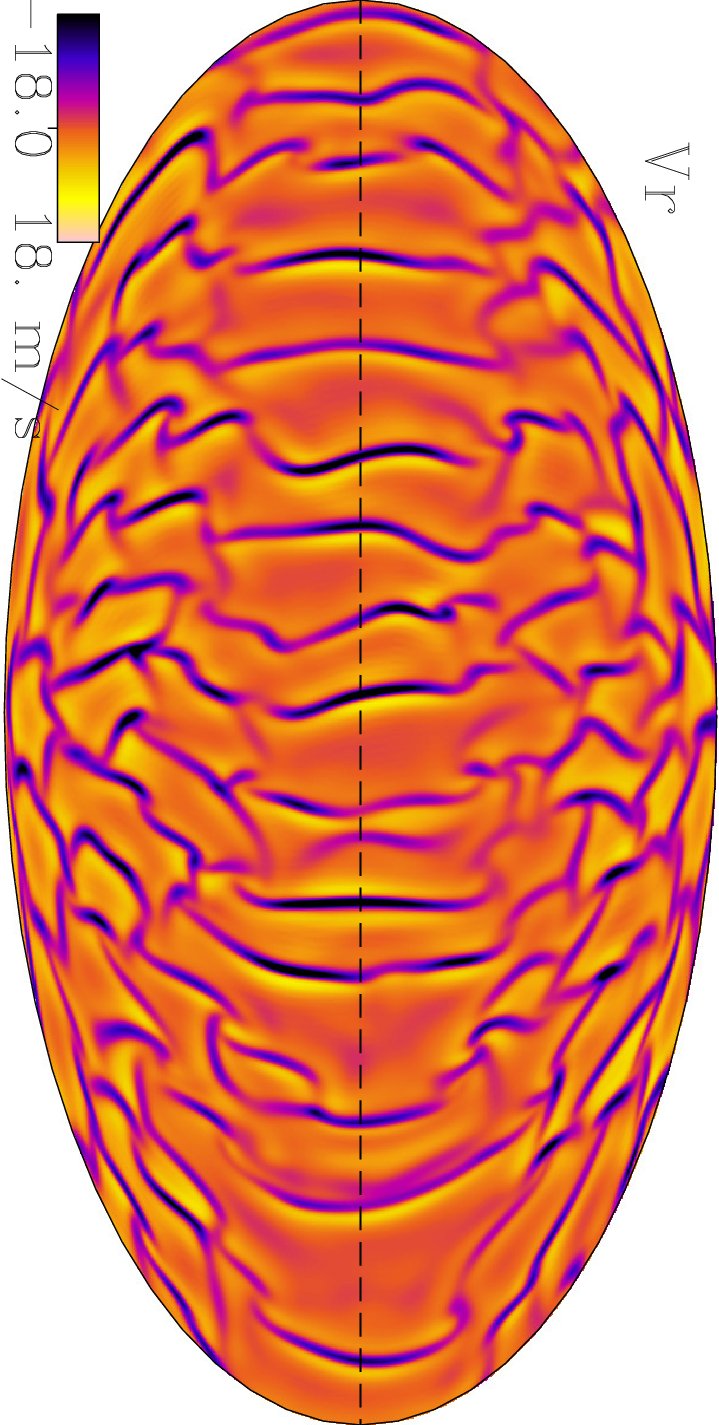}
}
  \subfigure[]{
  \label{fig:vr1950}
  \includegraphics[width=.25\figwidth,angle=90]{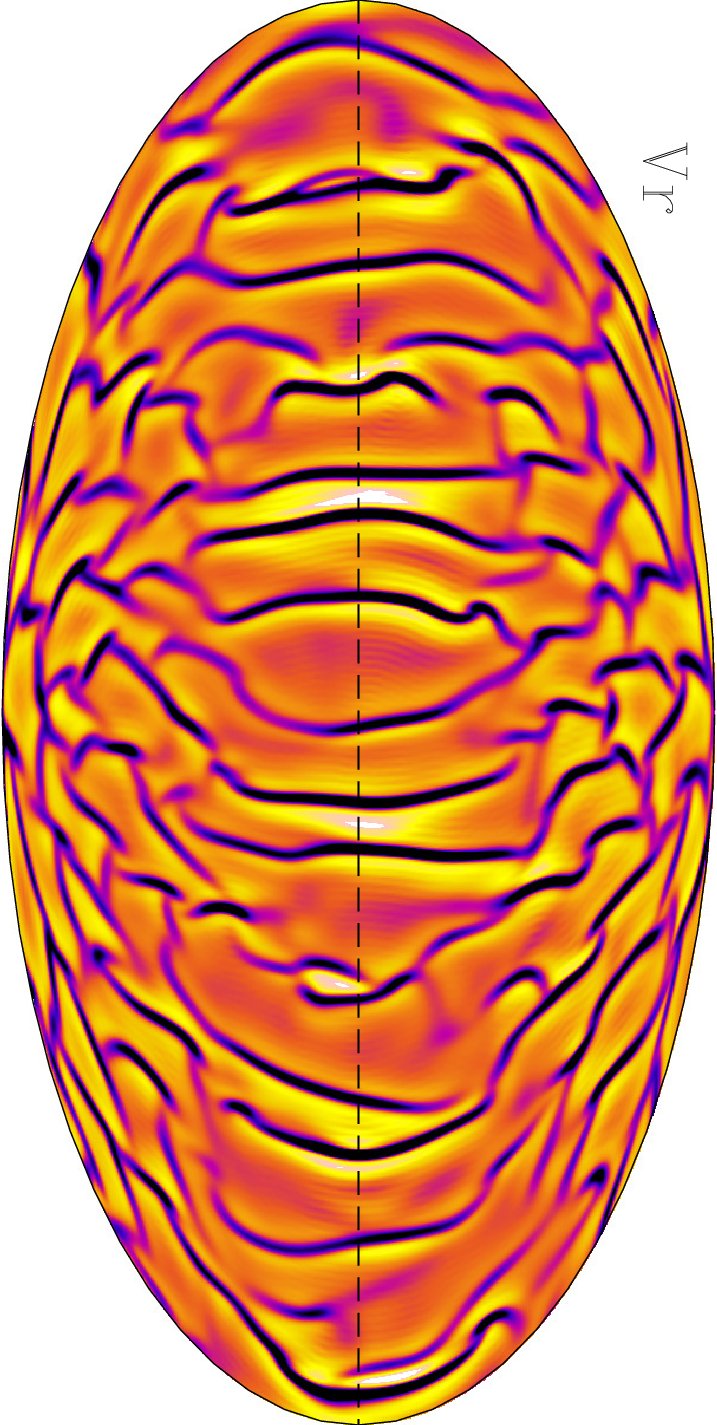}
}
  \subfigure[]{
  \label{fig:vr2550}
  \includegraphics[width=.25\figwidth,angle=90]{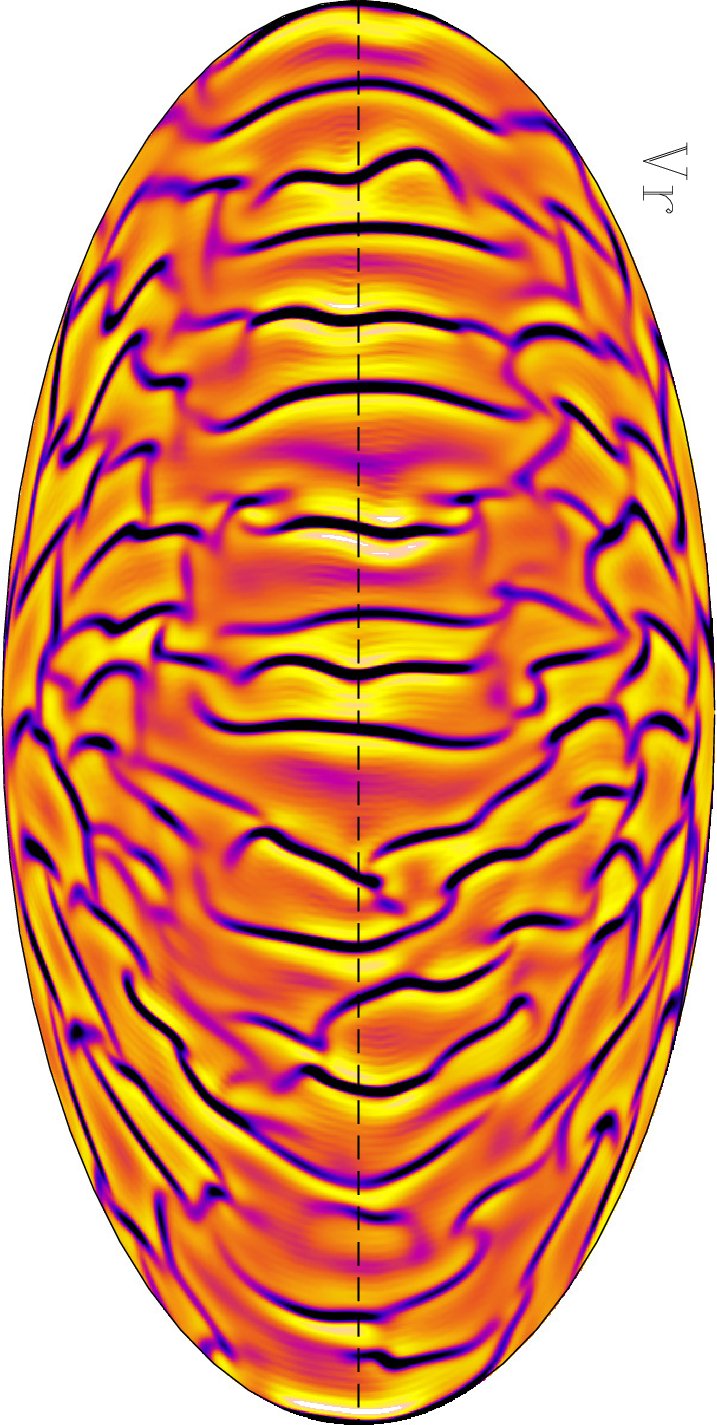}
}
  \subfigure[]{
  \label{fig:vr4100}
  \includegraphics[width=.25\figwidth,angle=90]{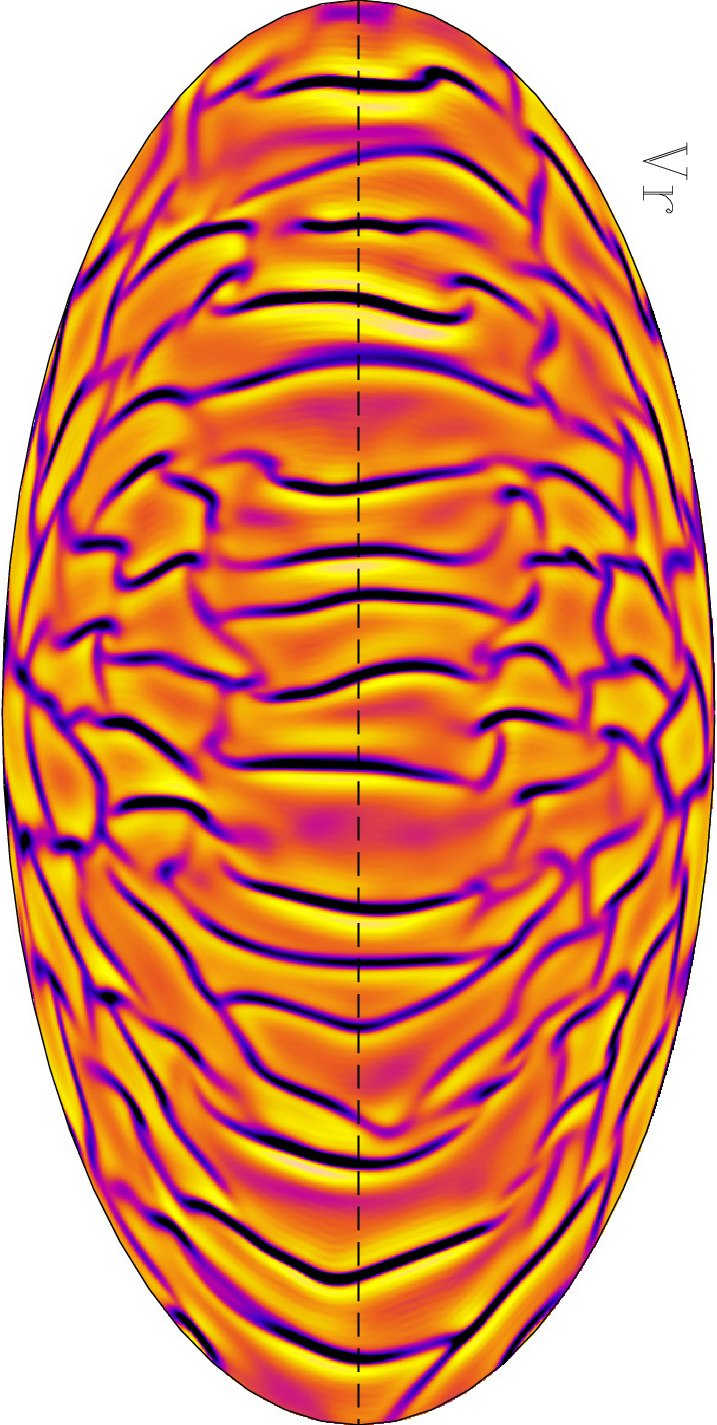}
}
 \subfigure[]{
  \label{fig:Br0}
  \includegraphics[width=.25\figwidth,angle=90]{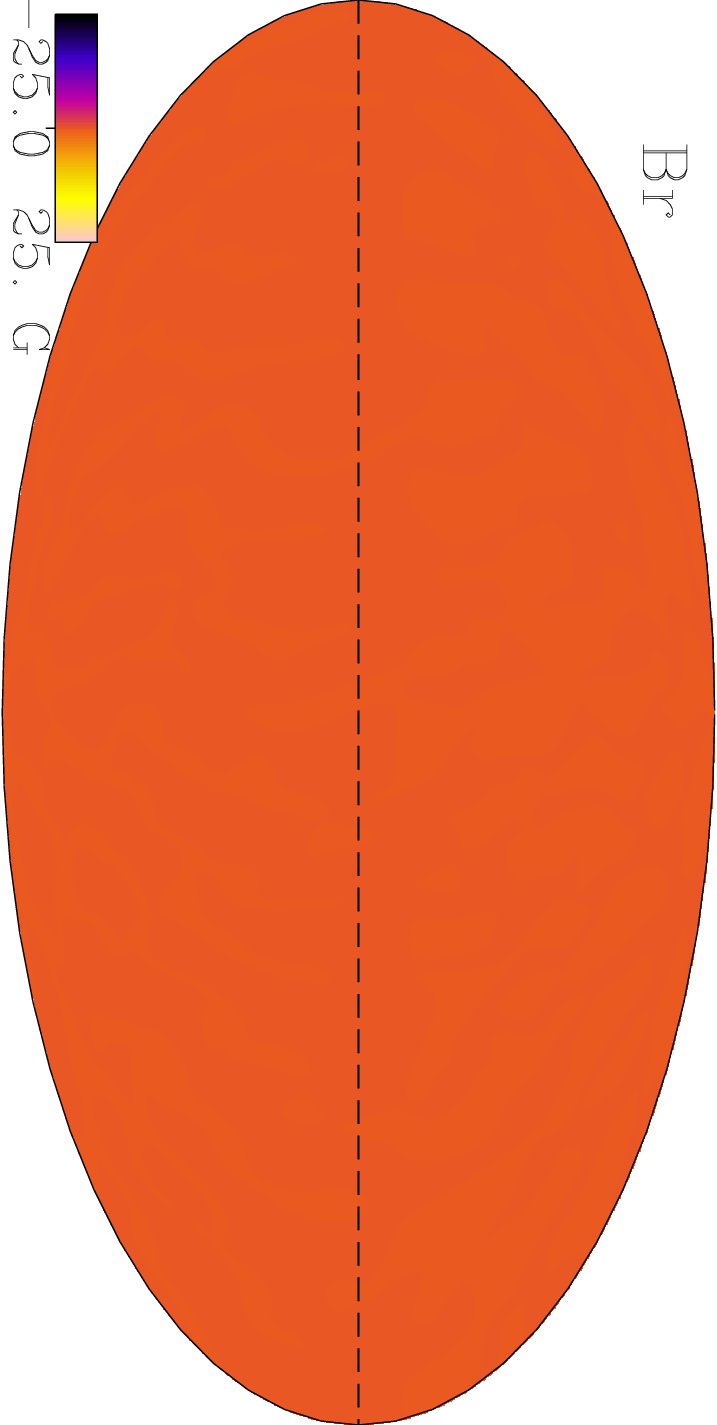}
}
  \subfigure[]{
  \label{fig:br1950}
  \includegraphics[width=.25\figwidth,angle=90]{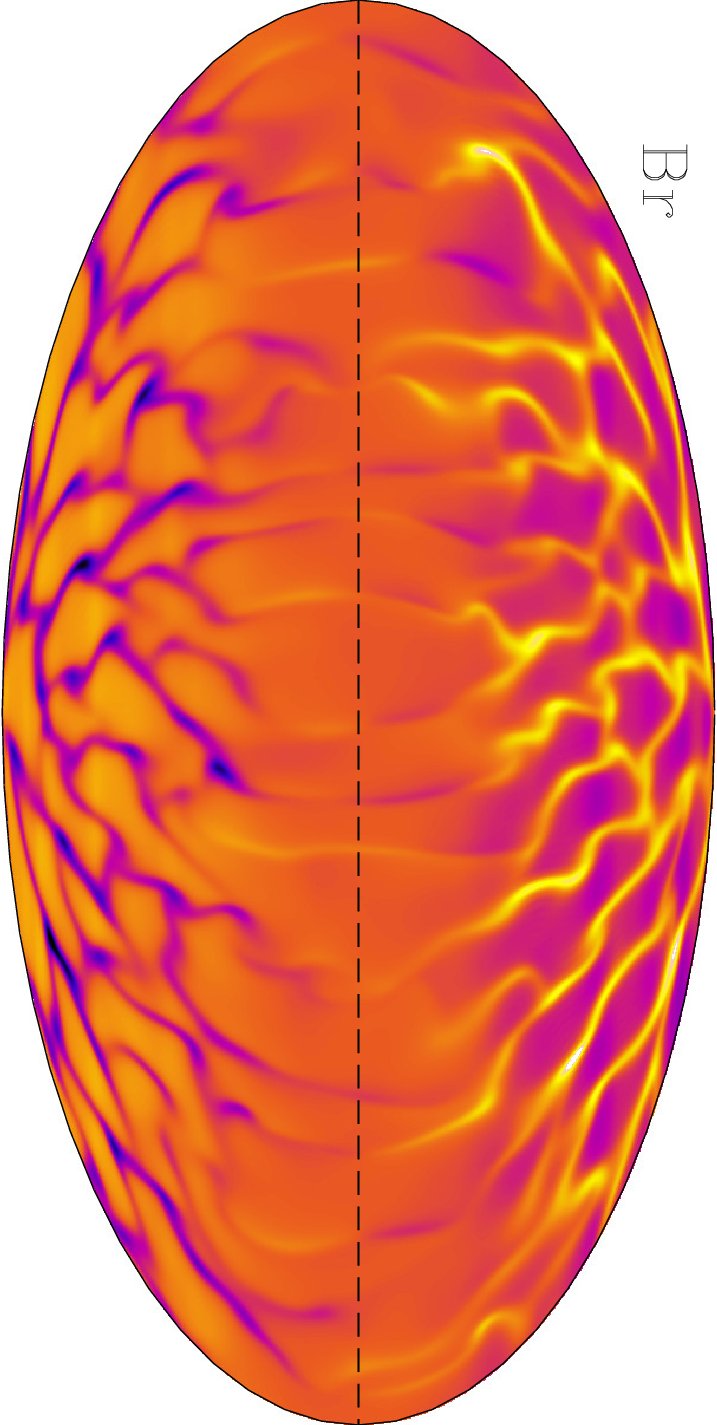}
}
  \subfigure[]{
  \label{fig:br2550}
  \includegraphics[width=.25\figwidth,angle=90]{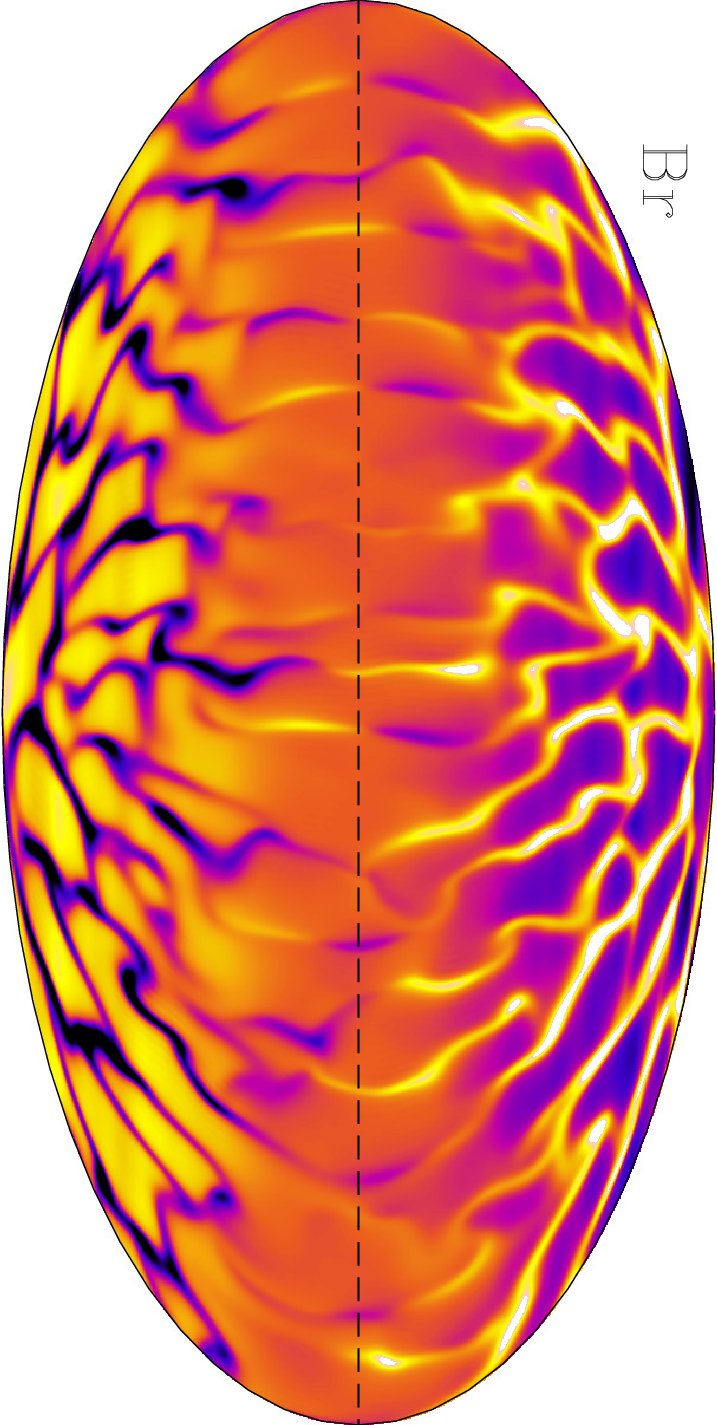}
}
  \subfigure[]{
  \label{fig:br4100}
  \includegraphics[width=.25\figwidth,angle=90]{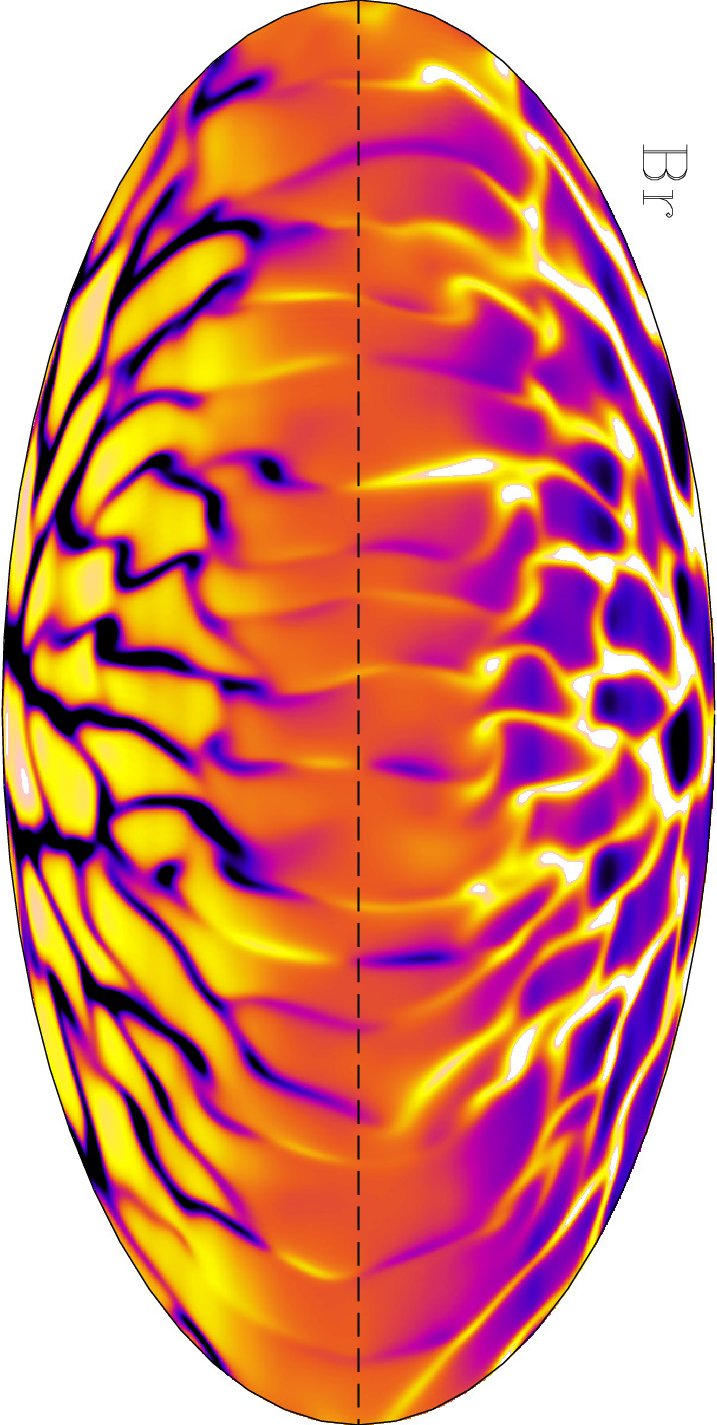}
}
  \caption{Time evolution of $v_r$ and $B_r$ for the shell
    $r=0.96\, R_\odot$. The different times are exactly the same as in
  Fig. \ref{fig:vphi_plus_Bstrl}. We subtracted the $m=0$
  component for $B_r$. Dark colors denote negative velocities
  and radial magnetic field, and bright colors positive
  velocities and radial magnetic field.}
  \label{fig:Shsls_evol}
\end{figure*}

\begin{figure}[!htbp]
  \centering
 \includegraphics[width=\figwidth]{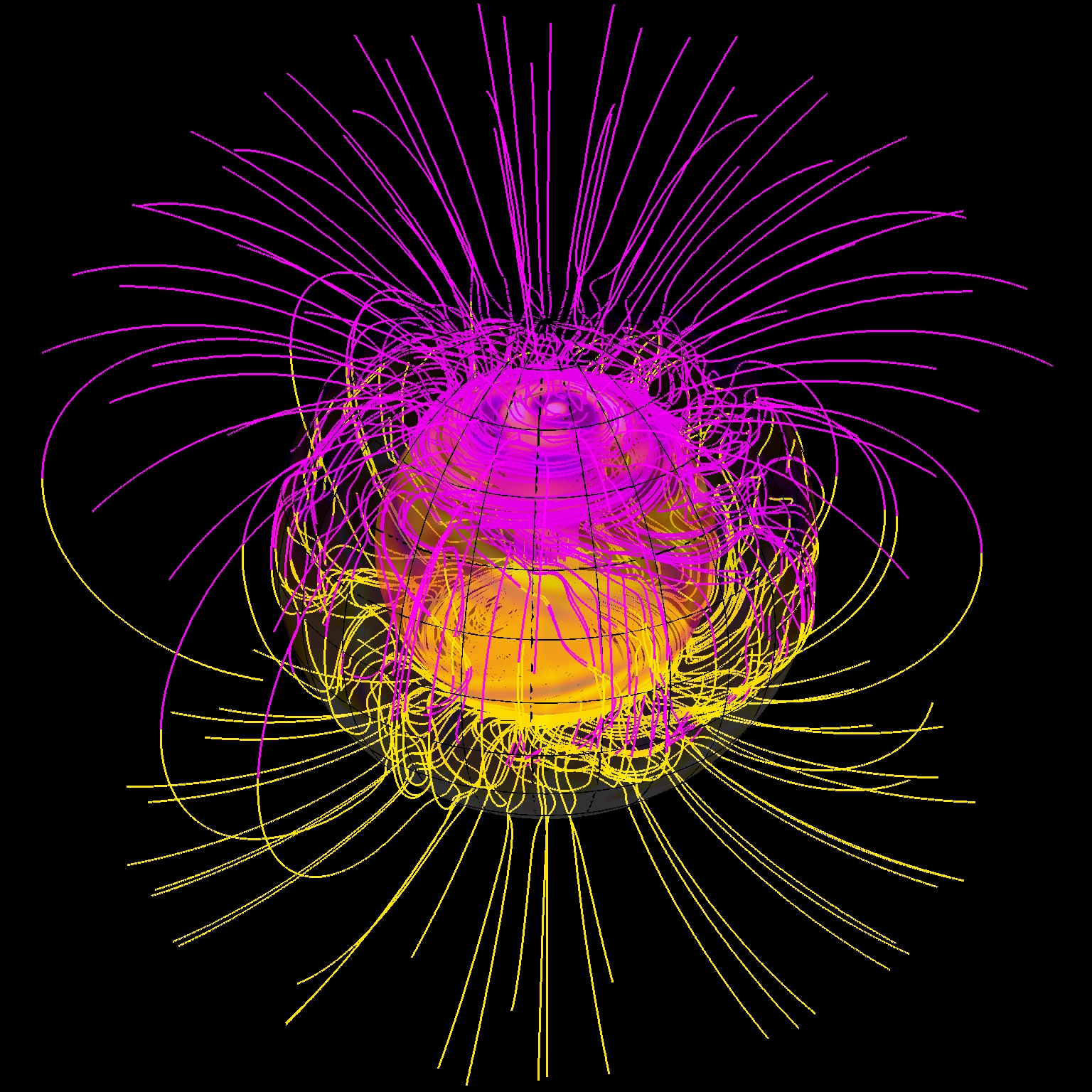}
 \caption{Magnetic field lines from the center
   to the corona. For $r>0.96\; R_\odot$, a potential extrapolation is
   assumed. This
   snapshot is taken at the same time as Fig. \ref{fig:br2550}. The two very
 transparent spherical shells are set at $r=0.62R_\odot$ and
 $r=R_\odot$, we display in them the contours of the non-axisymmetric $B_\varphi$ and total $B_r$.
 Magenta denotes the positive radial
 magnetic field, and yellow denotes the negative radial magnetic field.}
  \label{fig:3D_reconstruct}
\end{figure}

In order to watch how the primordial magnetic field manifests its presence
in the convective zone, we tracked the
evolution of $B_r$ near the surface (at $r=0.96\, R_\odot$). 
Our choice of parameters ($P_m=0.5$) for this simulation prevents the development
of a dynamo-generated magnetic field in the convection zone because the resulting magnetic Reynolds number
is less than one hundred there. The threshold for
dynamo action is likely to be above that value, as shown by
\citet{Brun:2004p1}. As a consequence, the presence of a magnetic
field at the surface of the model (as seen in
Fig. \ref{fig:Shsls_evol}) is solely caused by the spreading and
reshuffling of
the inner magnetic field through the convective envelope. Since no dynamo action is operating in the model,
 the magnetic field decays on an ohmic time scale and will eventually vanish.
However, we did not run the simulation long enough to
 access this decaying phase across the whole Sun. 
We also stress that the ratio
$ME/KE$ is only few $10^{-3}$ throughout the convection zone during the late evolution of our model;
it is still far too low to observe any influence of the magnetic field
on the convective patterns (\textit{see} \citealt{Cattaneo:2003p2833}).
Note that the non-axisymmetric
pattern of $B_r$ on Fig. \ref{fig:Shsls_evol} is strongly correlated with $v_r$, the vertical velocity 
pattern, with  opposite sign of $B_r$ in the
northern and southern hemispheres.
Because the latitudinal derivative of $v_r$
changes sign at the equator, we deduce that it is the shear 
term $\left(\mathbf{B}\cdot\boldsymbol{\nabla}\right)\mathbf{v}$
of the induction equation that dictates the evolution of $B_r$ in the
convection zone. An interesting diagnostic is provided by reconstructing
(with a potential field extrapolation) the magnetic field outside our
simulated star, up to $2\, R_\odot$
(Fig. \ref{fig:3D_reconstruct}). One easily recognizes the magnetic
layer in the tachocline and the rapidly evolving magnetic field in
the convection zone. The reconstructed outer magnetic field looks mainly
dipolar and does not give any hints for its complicated inner
structure. 

\begin{figure}[!htbp]
  \centering
  \subfigure[]{
  \label{fig:Bth}
  \includegraphics[width=.24\figwidth,angle=90]{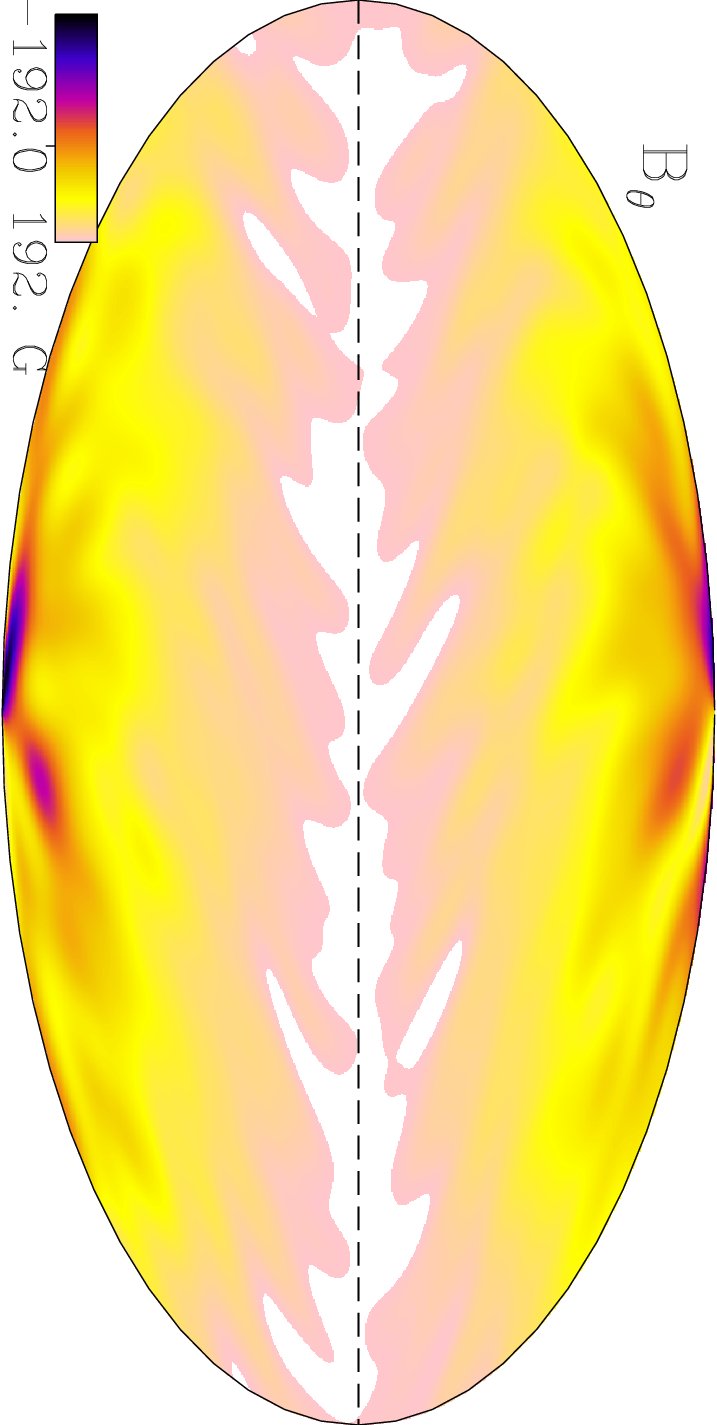}
}
  \subfigure[]{
  \label{fig:Bth_mm}
  \includegraphics[width=.24\figwidth,angle=90]{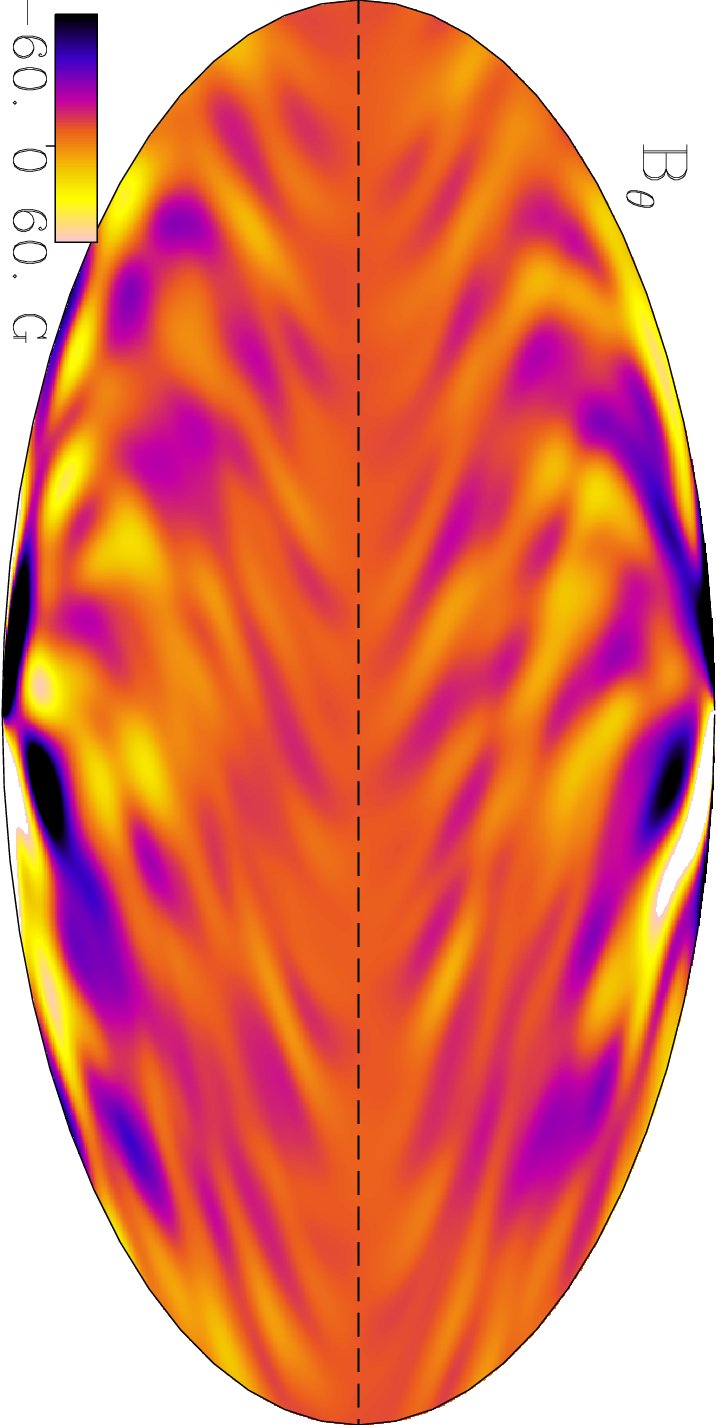}
}
  \subfigure[]{
  \label{fig:Bphi}
  \includegraphics[width=.24\figwidth,angle=90]{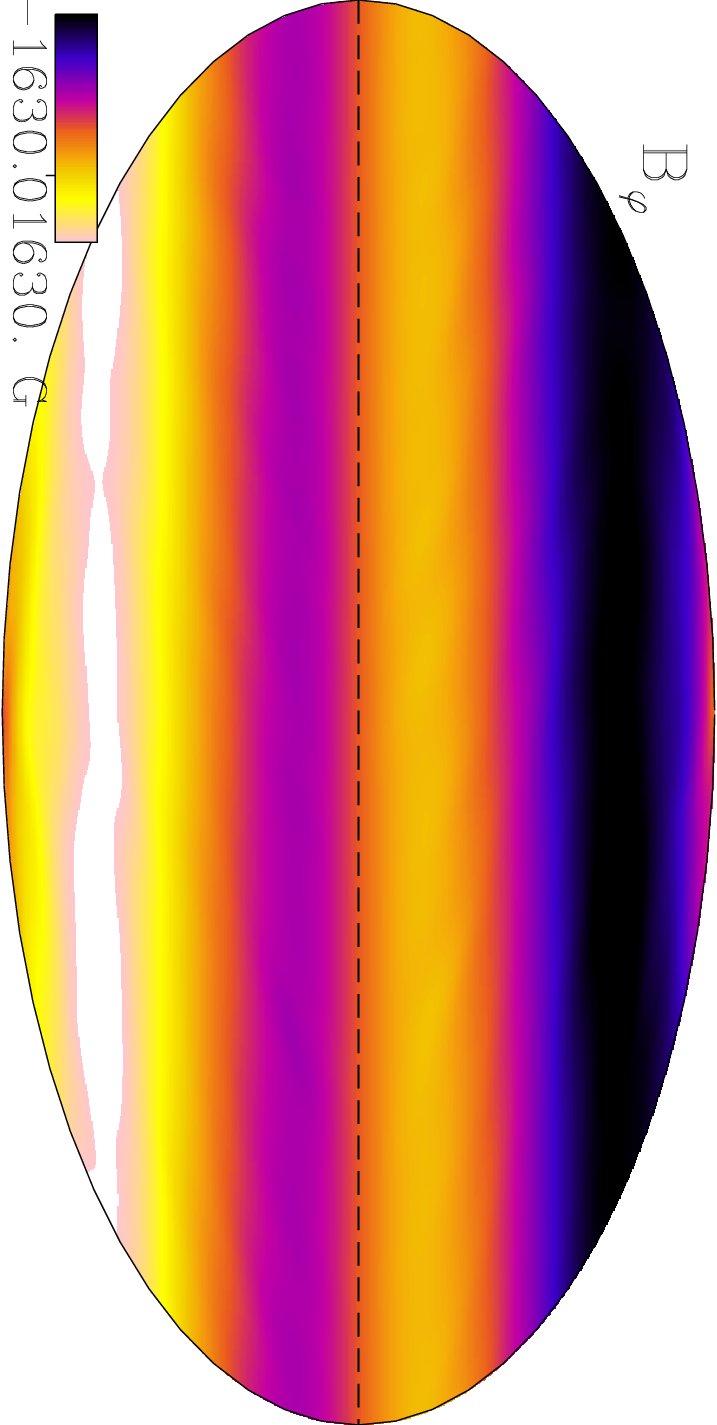}
}
  \subfigure[]{
  \label{fig:Bphi_mm}
  \includegraphics[width=.24\figwidth,angle=90]{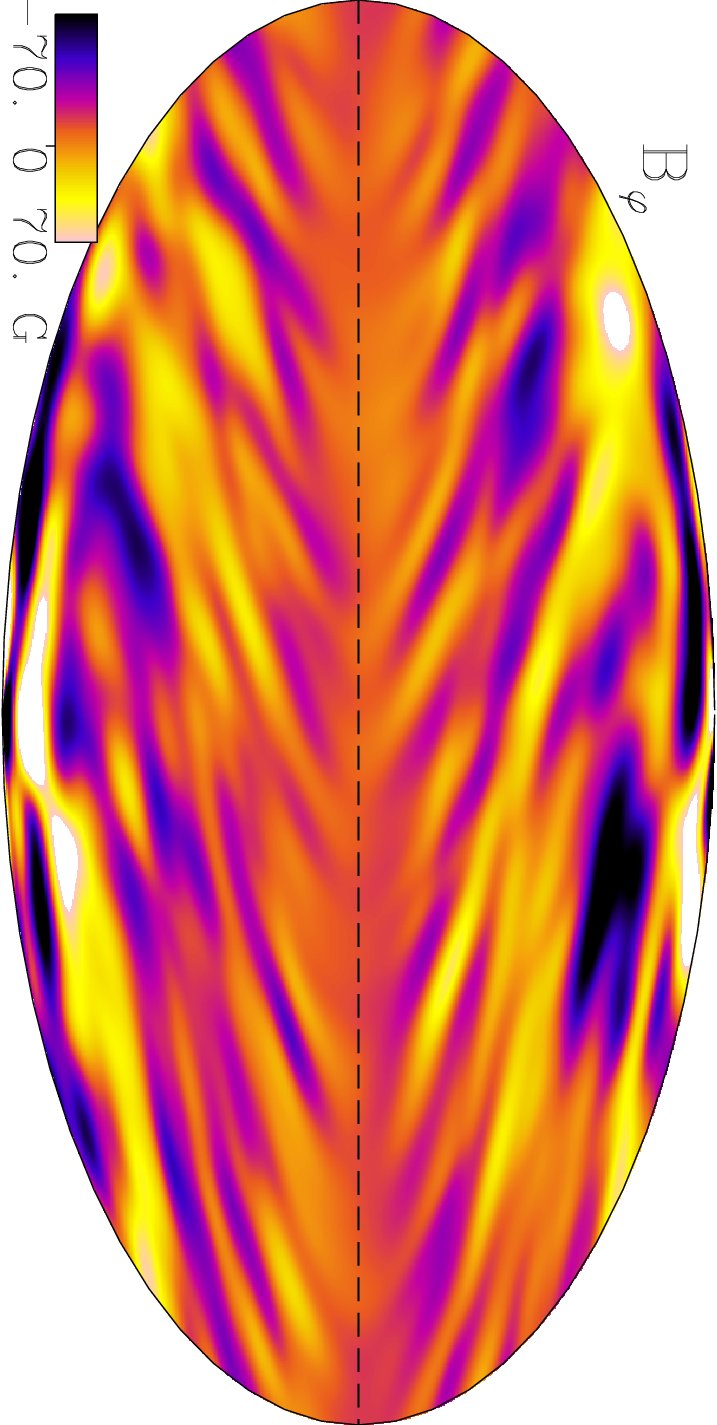}
}
  \caption{Shell slices of the horizontal components of the
    magnetic field at $r=0.60\, R_\odot$ in the late evolution of our
    model. Left figures represent
    the full fields, while the azimuthal mean was substracted in
    the right figures. Dark colors denote negative magnetic fields, and
    bright colors positive magnetic fields.}
  \label{fig:Shsls}
\end{figure}

As mentioned above, our initial magnetic field is subject to
high $m$ instabilities because it starts as a purely poloidal field. Quickly the
development of a toroidal component stabilizes the poloidal field,
resulting in a complex poloidal-toroidal topology. In the magnetic layer, the
azimuthal component of the toroidal field can locally be much larger
than the poloidal field. Such a toroidal field is subject to low $m$
instabilities (\textit{e.g.} \citealt{Tayler:1973p1727,Zahn:2007p1749,Brun:2007p2626}). We observe the
recurring of the $m=1$ pattern  at high latitudes on Fig. \ref{fig:Bphi_mm}. The profiles
of the horizontal components of the magnetic field are shown in
Figs. \ref{fig:Bth} and \ref{fig:Bphi}. In the two right panels of
Fig. \ref{fig:Shsls}, we subtracted the
axisymmetric part of the field to render the instability more
obvious. Note also that \mybold{at the time chosen to generate
the figure}, the longitudinal component of the magnetic field is ten times
stronger than its latitudinal component. 
  
\subsection{Tachocline and local magnetic field evolution}
\label{sec:tach-regi-evol}

\begin{figure}[!htbp]
  \centering
 \includegraphics[width=\figwidth]{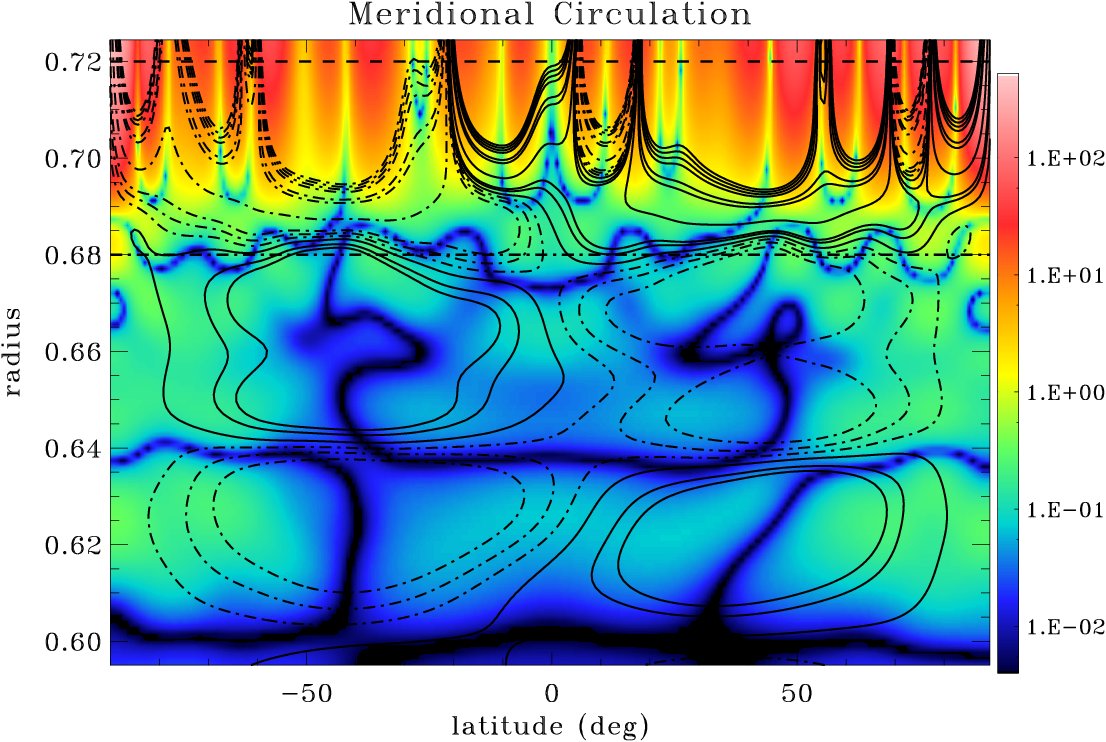}    
  \caption{
    Zoom of the \mybold{time-averaged} meridional circulation pattern \mybold{before the
      introduction of the magnetic field} in the
    tachocline area. The \mybold{stream} function is plotted in
    iso-contours, solid lines denoting clockwise circulation profiles,
    and dotted lines anti-clockwise. The background color map corresponds
    to the \textit{rms} strength of the meridional circulation flow in
    $m\, s^{-1}$ with a logarithmic scale. 
    Red colors denote the strongest downflows.
  }
  \label{fig:MC_tacho}
\end{figure}

Previous studies of the GM98 scenario have shed light on the
importance of the penetration of the meridional circulation into the
tachocline \citep{Sule:2005p31,Garaud:2008p36}. Numerical simulations
in 2D have shown that this large-scale flow was able to deflect the magnetic field,
and could prevent it from entering into the convection
zone. \mybold{We plot in Fig. \ref{fig:MC_tacho} the meridional
  circulation patterns realized in the bulk of the tachocline. We observe both
  the meridional circulation penetration (coming from above
  $r_{bcz}\sim 0.72\, R_\odot$), and the meridional circulation cells
  generated in the tachocline and at the top of the radiation
  zone. In our model, the meridional circulation penetrates approximately $4\%$ of the solar
radius below the base of the convection zone.
The \textit{rms} strength of the meridional velocity (shown
with a logarithmic colormap) strongly decreases with depth in the tachocline.
For instance, the meridional velocity loses three orders of magnitude
over $0.03$ solar radius by dropping from $18\,
m\, s^{-1}$ to $0.06\, m\, s^{-1}$ near the poles at the base of the
convective zone (\textit{see} also Fig. \ref{fig:rms_vels}). Again,
because our Peclet number is on the order of 1, this penetration is likely overestimated.}

\mybold{To evaluate the impact of the flow on the
magnetic field evolution in the penetrative layer, we
computed the magnetic Reynolds number $R_m={V^{rms}\Delta}/{\eta}$,
where $V^{rms}$ is the \mybold{total} \textit{rms} velocity,
$\Delta$ the convective penetration depth under the
convection zone, and $\eta$ the magnetic
diffusivity. We take $\Delta = r_{bcz}-r_{ov} \approx 0.04 \, R_\odot$ (\textit{see}
Figs. \ref{fig:rms_vels} and \ref{fig:MC_tacho}).
We obtained a magnetic Reynolds number on the order of $1$ to a few
in this penetration zone.  This means that we are in
a regime where shear and advection are on the order of, or slightly
greater than diffusion in the magnetic field evolution equation.
Different definitions of the magnetic Reynolds number may lead to
different values. Indeed, if one retains only 
the radial component of the velocity field (\textit{i.e.}, in the sense
of \citealt{Garaud:2008p36}) or only the meridional component of the
flow, we obtain a magnetic Reynolds number slightly lower than one. Conversely,
if one uses the total radius of the
star (\textit{i.e.}, in the sense of \citealt{Sule:2005p31}), we obtain much higher Reynolds number values. In the
end, it is the relative amplitude and the spatial structure of the various
terms in the induction equation at any given location that actually
determine the evolution of the system. The quantitative analysis of
those terms demonstrates that we are in an
advective regime, as will be seen in
Figs. \ref{fig:btheta_evol_init}-\ref{fig:Bphi_creation_begin}.}

{In addition, we also
  simulated the evolution of the same magnetic field in a purely
  diffusive case (\textit{i.e.}, without any velocity field). The
  evolution of the magnetic field \mybold{deep} in the radiation zone,
  \mybold{\textit{i.e.}, below $r_{ov}$}, is equivalent
  in the two simulations and is dominated by diffusion, its amplitude
  decreasing with the square root of time. However, the penetration of the magnetic
  field into the convective zone is faster in the full MHD
  case, \mybold{as will be seen in Fig. \ref{fig:B_evol_0p86_t}}.

\mybold{In order to demonstrate the role played by time-dependent 3D
  convective motions and nonlinearities in our simulation, we now turn to a detailed analysis of
the evolution of the mean poloidal and toroidal fields.}
Following \citet{Brown:2010p371}, we \mybold{first evaluate} the different
\mybold{terms that contribute} to the creation and maintenance of the axisymmetric
\mybold{latitudinal} magnetic field $\langle B_\theta \rangle$. It is
essentially \mybold{amplified by}
shear, transported by advection, and opposed by diffusion. Starting from the induction equation,
one may write
\begin{equation}
  \label{eq:dtbphi}
  \partial_t \langle B_\theta\rangle = P_{FS} + P_{MS} + P_{FA} +
  P_{MA} + P_{C} + P_{MD},
\end{equation}
with $P_{MS}$ and $P_{MA}$ representing the production by the mean
shearing and advecting flows, $P_{FS}$ and $P_{FA}$ by fluctuating
shear and advection, $P_{C}$ by compression, and $P_{MD}$ by mean ohmic
diffusion. \mybold{The symbol $\langle \rangle$ stands for
  azimuthal average.} Those six terms are
\begin{eqnarray}
 P_{FS} &=& \left. \left\langle\left( \bf{B}'
    \cdot\boldsymbol{\nabla} \right) \bf{v}' \right\rangle
\right|_\theta \nonumber \\ 
  P_{MS} &=& \left. \left( \left\langle \bf{B} \right\rangle
    \cdot\boldsymbol{\nabla} \right) \left\langle \bf{v} \right\rangle
\right|_\theta \nonumber\\
 P_{FA} &=& -\left. \left\langle\left( \bf{v}'
    \cdot\boldsymbol{\nabla} \right) \bf{B}' \right\rangle
\right|_\theta  \label{eq:prod_of_magfield}\\
 P_{MA} &=& -\left. \left( \left\langle \bf{v} \right\rangle
    \cdot\boldsymbol{\nabla} \right) \left\langle \bf{B} \right\rangle
\right|_\theta \nonumber\\
  P_{C} &=& \left(\left\langle v_r \right\rangle \left\langle
      B_\theta\right \rangle + \left\langle v_r'B_\theta'\right\rangle
    \right)
\frac{d}{dr}\ln{\bar{\rho}} \nonumber\\
  P_{MD} &=& -\left. \boldsymbol{\nabla}\times\left(\eta\boldsymbol{\nabla}
    \times  \left\langle\bf{B} \right\rangle \right)
\right|_\theta \, . \nonumber
\end{eqnarray}

These definitions are then easily transposed to the radial and
azimuthal components of the magnetic field.

\mybold{We plot in
Figs. \ref{fig:btheta_evol_init}-\ref{fig:btheta_evol_end} 
the mean latitudinal field $\langle B_\theta
\rangle$ and the various terms of the right hand side of
Eq. \eqref{eq:prod_of_magfield} at two instants in the simulation. We first
show in Fig. \ref{fig:btheta_evol_init} the
early phase of evolution of $\langle B_\theta
\rangle$, and a later phase of evolution in
Fig. \ref{fig:btheta_evol_end}.
At radius $r=0.6\, R_\odot$ (\textit{i.e.}, deeper than $r_{ov}$), we
initially observe two trends. First, the mean advective and shear contributions
effectively cancel one another, possibly because of our choice of
impenetrable and perfect conductor lower boundary condition ($\left\langle V_r \right\rangle = 0$ and $\left\langle
  B_\theta \right\rangle = 0$ at $r=r_{bot}$).} 
The same effect is observed for the $\langle
B_r\rangle$ component (not shown here). 
Second, because those two terms cancel
each other, it is the magnetic diffusion that dictates the evolution
of $\langle B_\theta \rangle$ in this deep region. This is not a surprise, considering our 
diffusivity values \mybold{and the lack of strong motions in this deep
  (inner) part of the model}.
\begin{figure}[!htbp]
  \centering
 \includegraphics[width=\figwidth,angle=90]{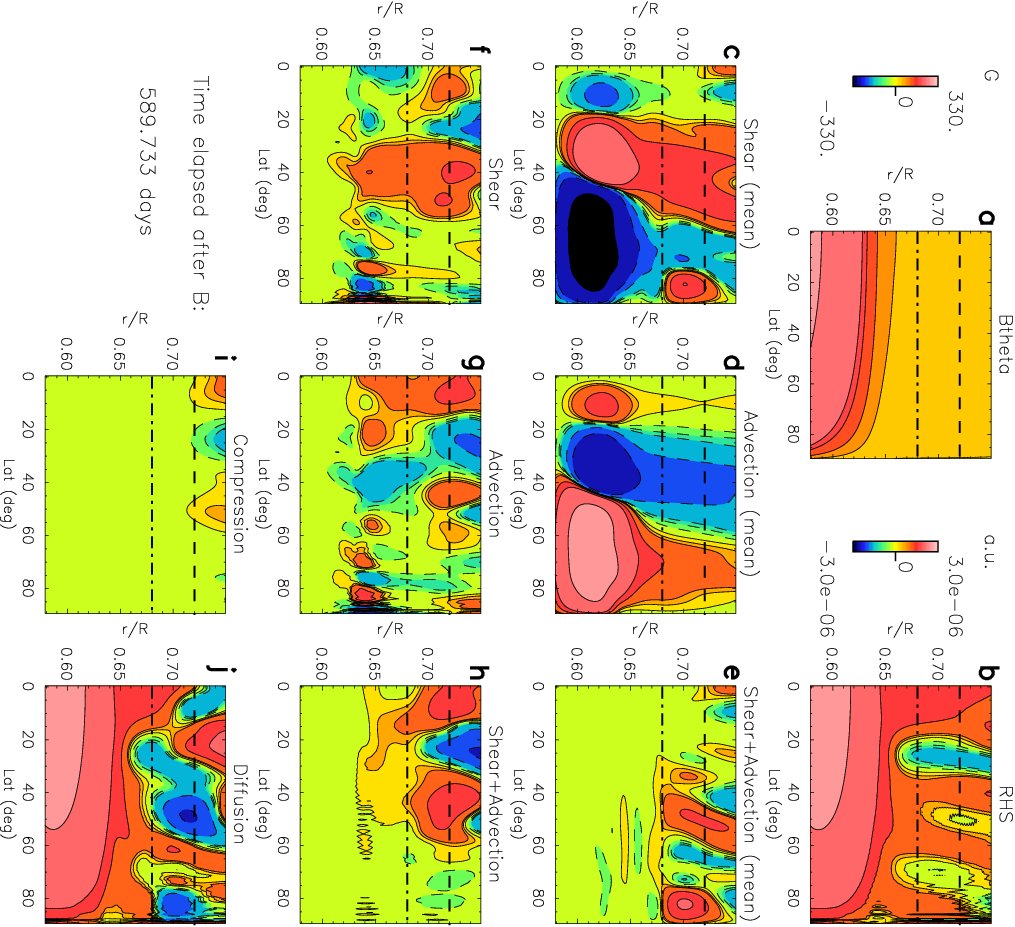}    
\caption{
Major terms of production of $\langle B_\theta \rangle$. The abscissa spans
the northern hemisphere, the ordinate is a zoom in the
tachocline. The left color bar corresponds to instantaneous $\langle
B_\theta \rangle$, while the right
color bar stands for the \mybold{nine} other panels. The RHS panel is the
sum of
the \mybold{panels e, h, j, and i}, and represents the net evolution of the mean latitudinal
magnetic field. Panels \mybold{c, d, f, g, i, and j} correspond to
$P_{MS}$, $P_{MA}$, $P_{FS}$, $P_{FA}$, $P_{C}$, and $P_{MD}$, respectively (\textit{see}
Eq. \eqref{eq:prod_of_magfield}). \mybold{The panels e and h
correspond to the sum of panels c and d, and f and g, respectively.} Color levels are not linear in
order to
see at the same time the contributions of the terms in the different
regions. Red colors denote the positive contribution to the magnetic
field and blue colors the negative contributions. \mybold{The
  two horizontal lines in each panel represent $r_{bcz}$ (dash) and
  $r_{MC}$ (dot-dash), as in Fig. \ref{fig:rms_vels}.}}
\label{fig:btheta_evol_init}
\end{figure}

\begin{figure}[!htbp]
  \centering
 \includegraphics[width=\figwidth,angle=90]{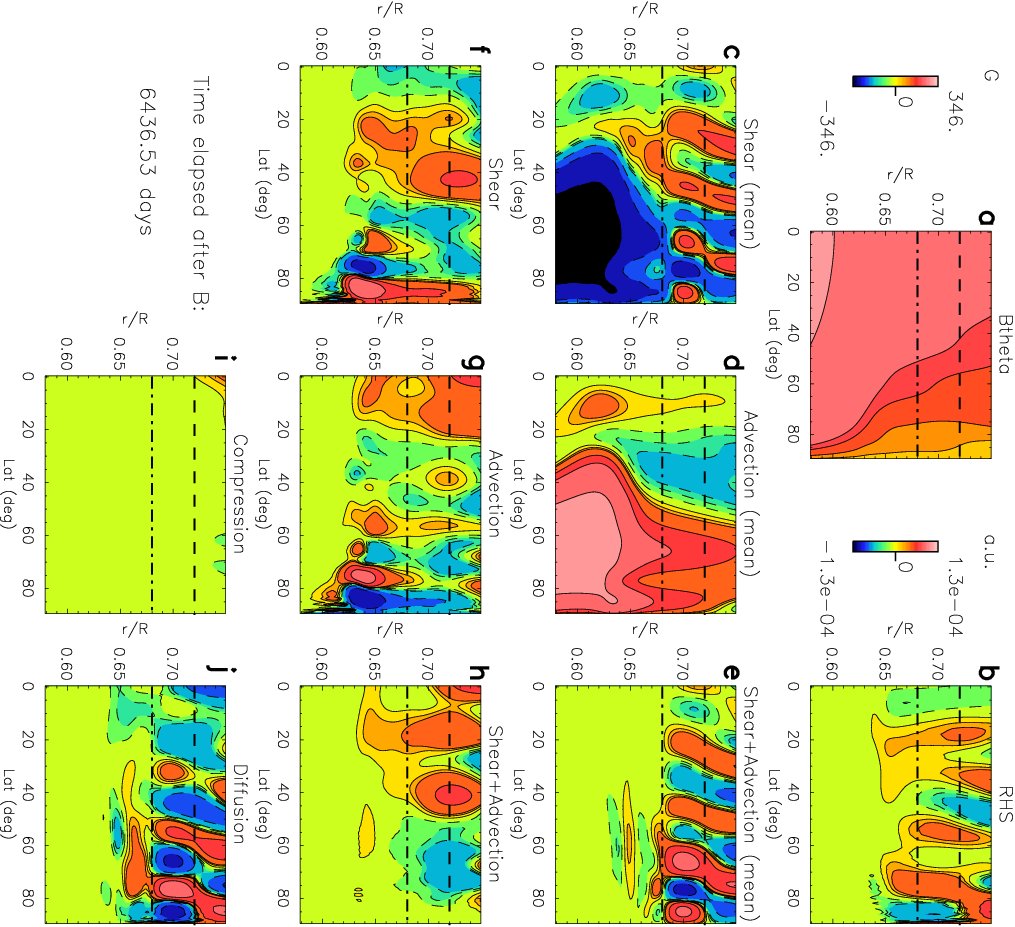}      
\caption{
Major terms of production of $\langle B_\theta \rangle$. \mybold{The panels
represent the same as in Fig. \ref{fig:btheta_evol_init}, but
at a later time.}}
\label{fig:btheta_evol_end}
\end{figure}

 \mybold{On the other hand, the patterns
   are much more complicated in the penetration layer (between $r_{ov}=0.675\, R_\odot$ and
   $r_{bcz}\sim 0.72\, R_\odot$)}. 
\mybold{Mean advection and shear are important, their sum (panel e)
  mainly acts at latitudes higher than $40^\circ$. Diffusion (panel j) acts
  everywhere in the penetration layer, but its action does not
  dominate the other contributions (panel b). The non-axisymmetric
  contributions (panels f, g, and h) are very important at latitudes lower than $60$ in the penetration
  layer. Therefore, 3D motions are obviously
responsible in our model for the
failure of the magnetic field confinement below the convection
zone at low latitude. This result contrasts with previous 2D studies, where
these motions were not taken into account.}

\mybold{During the late evolution of our model
  (Fig. \ref{fig:btheta_evol_end}), the complex balance between
  axisymmetric motions, non-axisymmetric motions and diffusion is
  still operating in the penetration layer. We observe (panel a)
  that the magnetic field has completely pervaded the tachocline and
  extends into the convection zone. Again, the relative importance of
  the three panels e, h, and j
  indicates that diffusion is not the dominant process transporting
  the field. Indeed, we find again that the non-axisymmetric
  components significantly act at low latitude in this layer. We also
  note here that the levels of the color table were modified between 
figures \ref{fig:btheta_evol_init} and \ref{fig:btheta_evol_end}, making
the diffusive term (which is
still operating around $r=0.6\, R_\odot$) harder to be seen in panel
j. Finally, we note that magnetic pumping is proven inefficient to
confine the magnetic field at
both times.}
\begin{figure}[!htbp]
  \centering
  \includegraphics[width=\figwidth]{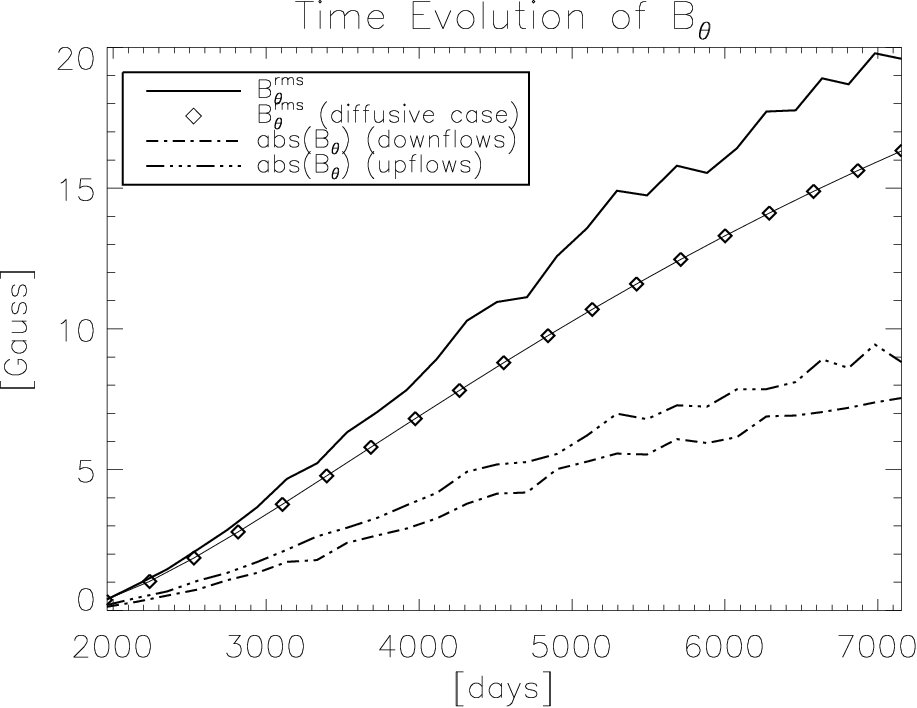}    
  \caption{
\mybold{Rms and absolute value of the latitudinal magnetic field amplitude evolution at $r=0.86\,
    R_\odot$. The dotted and dashed lines correspond to
    the absolute value of the latitudinal component of the magnetic field in the downflows and in the
    upflows. We selected the downflows part by setting the magnetic
    field to zero at the points
    corresponding to a positive $v_r$ on the $r=0.86\; R_\odot$
    spherical shell. We averaged the remaining values over the
  shell to obtain $B_\theta$ in downflows, and did the opposite for the
  upflows selection. The diamond line represents the evolution of
  the rms latitudinal magnetic field in the purely diffusive (test) case.}}
  \label{fig:B_evol_0p86_t}
\end{figure}

\mybold{In order to estimate the pumping, we plot in
Fig. \ref{fig:B_evol_0p86_t} the evolution of the amplitude of the
latitudinal magnetic field at
$r=0.86\, R_\odot$ (in the convection zone) after its introduction. As previously
seen, the amplitude rises. We plot on the same figure the evolution
of the absolute value of the latitudinal component of the magnetic field both in the downflows (dot-dash) and in
the upflows (triple dot-dash). We observe that the amplitude
rises slower in the downflows than in the upflows.
In addition, the temporal lag between the up- and
downflow regions increases with time. This agrees with the action of magnetic
pumping by downflows. However, it is a minor
effect in the global evolution of the magnetic field.}

\mybold{We also plot in Fig. \ref{fig:B_evol_0p86_t}
  the evolution of the rms latitudinal component of the magnetic field in the
  convection zone in both the nonlinear simulation discussed in the
  paper (solid line) and in another purely diffusive test case (diamonds).
 Once more, this stresses that the field evolution is
  different in our simulation than in the purely diffusive test
  case. The larger amplitude of $B_\theta$ in our convection
  simulation clearly demonstrates the role of advection in pulling the
  field inside the convection zone.
The trends in Fig. \ref{fig:B_evol_0p86_t} are similar between $r=r_{ov}$ and $r_{top}$,
\textit{i.e.}, in the whole convection zone and even in the penetration layer.}

\begin{figure}[!htbp]
  \centering
  \includegraphics[width=\figwidth,angle=90]{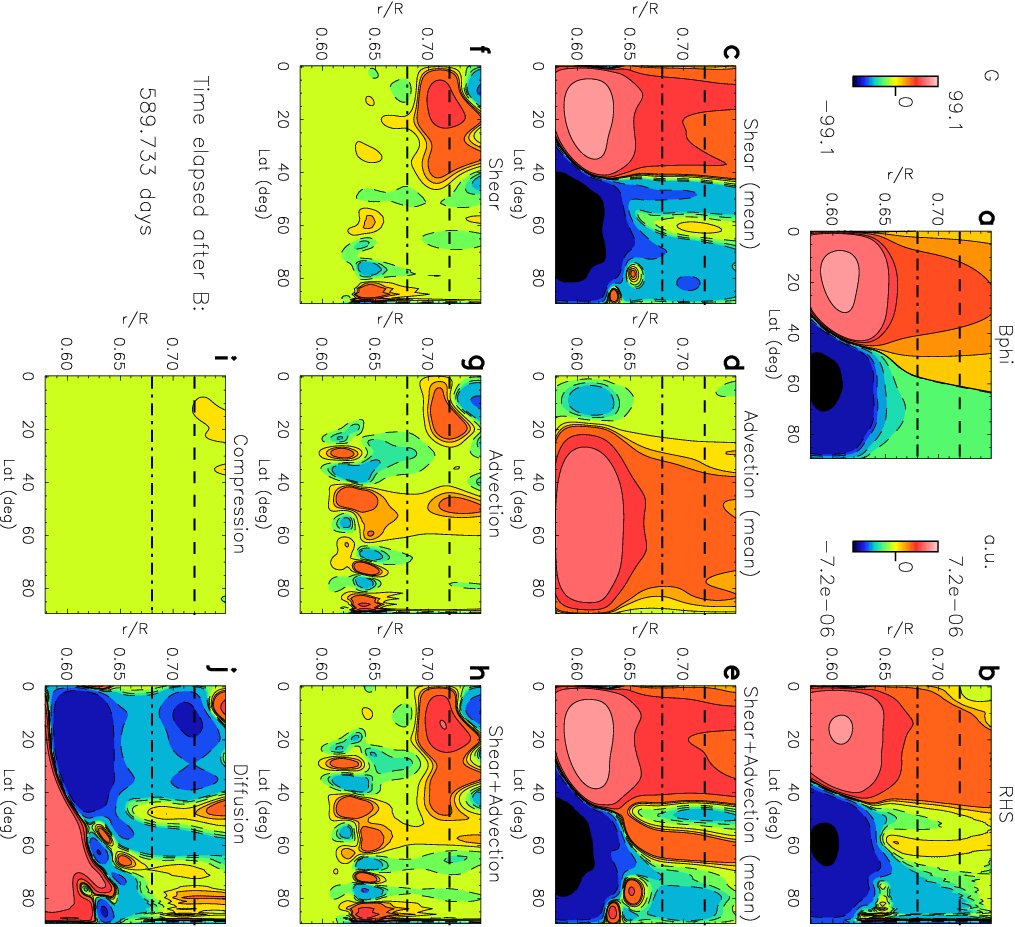}    
  \caption{Major terms of production of $\langle B_\varphi \rangle$ in
    the tachocline. The various contributions are taken at the same
    time as Fig. \ref{fig:btheta_evol_init}, but using expressions
    corresponding to $\langle B_\varphi \rangle$.}
\label{fig:Bphi_creation_begin}
\end{figure}

 We may now also examine two different
processes in the evolution of the azimuthal magnetic field: the
creation and sustainment of the magnetic layer, and the expansion of the magnetic
field into the convection zone. $\, \langle B_\varphi \rangle$ is firstly created underneath the
convection zone through the interaction between the dipolar
magnetic field and the differential rotation \mybold{around $r=0.62\, R_\odot$}.
Diffusion slightly
counterbalances the $\Omega$ effect, and advection does not act much in
this layer. In order to illustrate these processes, we plot in
Fig. \ref{fig:Bphi_creation_begin} $\langle B_\varphi
\rangle$ and the right hand side terms from
Eq. \eqref{eq:prod_of_magfield} \mybold{(as in
  Fig. \ref{fig:btheta_evol_init}, but taken in
the $\varphi$ direction)}
during the early evolution of our model.
\mybold{Diffusion (panel j) opposes the creation of magnetic field by
  mean shear (panel c) in the
magnetic layer (around $0.6\, R_\odot$).}
Although the magnetic layer seems to remain radially localized, the
magnetic field also exists in
the convective zone. It has a much smaller amplitude than in the
magnetic layer, but it is not negligible. 
We observe that $\langle B_\varphi \rangle$ is
preferentially created through the two terms of shear at latitudes
below $40^\circ$. 
Notice that the non-axisymmetric
velocity and magnetic field \mybold{(panels f, g, and h)} contribute equally \mybold{with} the axisymmetric terms
to produce some magnetic field \mybold{in the penetration layer
  (between the two horizontal lines)}. Diffusion generally
tends to oppose the mean shear \mybold{contribution} $P_{MS}$ at all
latitudes at the base of the convection zone, \mybold{but does not act
much in comparison with the other contributions in panels e and h}. The mean advection opposes
the creation of negative $\langle B_\varphi \rangle$ above latitude
$40^\circ$, which explains why the significant creation of coherent mean azimuthal
magnetic field is localized at lower latitudes. \\


To summarize, we showed that the magnetic field evolution through the
tachocline is caused by a complex
interplay between shear and advection mechanisms. The diffusive
process does not dictate the evolution of the magnetic field
\mybold{at the base of the convection zone, neither}
at the equator nor in the polar regions. Magnetic
pumping is present, but inefficient. As a result,
the magnetic field penetrates into the convection zone \mybold{and
  does not remain confined. We stress here that the nonlinear term
  involving Maxwell and Reynolds stresses is a key and fundamental
  feature of our 3D simulation.}

\section{Balances in the tachocline}
\label{sec:analysis}

In order to understand the global rotation profile \mybold{realized in
 our simulation}
(\textit{see} Sect. \ref{sec:general-evolution}), we study the MHD
meridional force balance and the angular momentum transport equation in the
following subsections. Even if the magnetic field does not modify the
meridional force balance (Sect. \ref{sec:thermal-wind-balance}),
it is proven \mybold{to be} responsible for
the \mybold{outward} transport of angular momentum \mybold{in the mid-
  to high-latitude regions} (Sect. \ref{sec:angul-moment-balance}).  

\subsection{MHD meridional force balance}
\label{sec:thermal-wind-balance}

Starting from the vorticity equation (Eq. \eqref{eq:vorticity}), one
can derive the full meridional force balance in the general magnetic case
(\textit{e.g.}, \citealt{Fearn:1998p1955,Brun:2005p1234}).

\begin{eqnarray}
  \label{eq:vorticity}
  \partial_t \boldsymbol{\omega} &=&
  -\boldsymbol{\nabla}\times\left(\boldsymbol{\omega}_a\times\mathbf{v}\right)
  +
  \frac{1}{\bar{\rho}^2}\boldsymbol{\nabla}\bar{\rho}\times\boldsymbol{\nabla}P
  - \boldsymbol{\nabla}\times\left( \frac{\rho g}{\bar{\rho}}
    \mathbf{e}_r\right) \nonumber \\ &+&
  \frac{1}{4\pi}\boldsymbol{\nabla}\times\left(\frac{1}{\bar{\rho}}\left(\boldsymbol{\nabla}
      \times \mathbf{B}\right)\times\mathbf{B}\right) -
\boldsymbol{\nabla}\times\left(\frac{1}{\bar{\rho}}\boldsymbol{\nabla}\cdot\mathcal{D}\right), 
\end{eqnarray}
with $\boldsymbol{\omega}_a=\boldsymbol{\nabla}\times\mathbf{v} +
2\boldsymbol{\Omega}_0$ the absolute vorticity and
$\boldsymbol{\omega}=\boldsymbol{\nabla}\times\mathbf{v}$ the
vorticity in the rotating frame. In a statistically stationary state,
equation \eqref{eq:vorticity} can be rewritten by taking the temporal and azimuthal
averages \mybold{of the longitudinal component}:
\begin{eqnarray}
2\Omega_0\frac{\partial \langle v_{\varphi} \rangle}{\partial z}&=&
\underbrace{-\left\langle(\mbox{\boldmath $\omega$}\cdot\mbox{\boldmath $\nabla$})v_{\varphi} + \frac{\omega_{\varphi}v_r}{r} +
  \frac{\omega_{\varphi}v_{\theta}\cot\theta}{r}
  \right\rangle}_{\mbox{\scriptsize{stretching}}}
\nonumber \\ 
&+&
\underbrace{\left\langle({\bf v}\cdot\mbox{\boldmath $\nabla$})\omega_{\varphi}
  +\frac{v_{\varphi}\omega_r}{r} +
  \frac{v_{\varphi}\omega_{\theta}\cot\theta}{r}
  \right\rangle}_{\mbox{\scriptsize{advection}}} \nonumber \\
&+& \underbrace{- \langle\omega_{\varphi}v_r\rangle\frac{d\ln \bar{\rho}}{d
    r}}_{\mbox{\scriptsize{compressibility}}} 
+
    \underbrace{\frac{g}{r c_p}\frac{\partial \langle S \rangle}{\partial
    \theta} + \frac{1}{r \bar{\rho}
  c_p}\frac{d\bar{S}}{dr}\frac{\partial \langle P
  \rangle}{\partial\theta} }_{\substack{\mbox{\scriptsize{baroclinicity due to
    latitudinal}} \\ \mbox{\scriptsize{entropy and pressure gradients}}}}
\nonumber \\
&+& \underbrace{\frac{1}{r}\left[\frac{\partial}{\partial r}(r \langle{\cal
      A}_{\theta}\rangle) - \frac{\partial}{\partial \theta}\langle{\cal
      A}_r\rangle \right]}_{\mbox{\scriptsize{viscous stresses}}} \nonumber \\
&+& \underbrace{\left\langle
  -\frac{1}{c\bar{\rho}}\left(\bf{B}\cdot\boldsymbol{\nabla}\right)
    j_\varphi  - \frac{ j_r B_\varphi
      }{c\bar{\rho}r} 
    - \frac{ j_\theta B_\varphi \cot\theta}{c\bar{\rho}r} \right\rangle
}_{\mbox{\scriptsize{magnetic torque 1}}} \nonumber \\
&+&\underbrace{\left\langle
  \frac{1}{c\bar{\rho}}\left(\bf{j}\cdot\boldsymbol{\nabla}\right)
    B_\varphi 
      + \frac{B_r j_\varphi
      }{c\bar{\rho}r} 
    + \frac{ B_\theta j_\varphi \cot\theta}{c\bar{\rho}r} \right\rangle
}_{\mbox{\scriptsize{magnetic torque 2}}}
\nonumber \\
&+& \underbrace{
 \frac{\left\langle B_r j_\varphi \right\rangle}{c\bar{\rho}}\frac{d}{dr}\ln{\bar{
        \rho}} -
 \frac{\left\langle B_\varphi j_r \right\rangle}{c\bar{\rho}}\frac{d}{dr}\ln{\bar{
        \rho}}
}_{\mbox{\scriptsize{compressibility 2}}}
\label{eq:TW_balance}
\end{eqnarray}
where $\displaystyle \frac{\partial}{\partial z} = \cos\theta \frac{\partial}{\partial r} -
\sin\theta\frac{\partial}{\partial \theta}$ and 
\begin{eqnarray}
{\cal A}_r&=& \frac{1}{\bar{\rho}}\left[\frac{1}{r^2}\frac{\partial(r^2{\cal D}_{rr})}{\partial r}+\frac{1}{r\sin\theta}\frac{\partial(\sin\theta{\cal D}_{\theta r})}{\partial\theta} - \frac{ {\cal D}_{\theta \theta} + {\cal D}_{\phi \phi}}{r} \right] \nonumber \\
{\cal A}_{\theta}&=& \frac{1}{\bar{\rho}}\left[\frac{1}{r^2}\frac{\partial(r^2{\cal
      D}_{r\theta})}{\partial
    r}+\frac{1}{r\sin\theta}\frac{\partial(\sin\theta{\cal D}_{\theta
      \theta})}{\partial\theta} \right] \nonumber \\
&& + \frac{1}{\bar{\rho}}\left[\frac{ {\cal D}_{\theta r} - cot \theta {\cal D}_{\phi \phi}}{r} \right]. \nonumber
\end{eqnarray}
In Eq. \eqref{eq:TW_balance} we underlined different terms:
\begin{itemize}
\item \textit{stretching} represents the stretching of the vorticity
  by velocity gradients;
\item \textit{advection} represents the advection of the vorticity by
  the flow;
\item \textit{compressibility} represents the flow compressibility
  reaction on the vorticity;
\item $\frac{g}{r c_p}\frac{\partial \langle S \rangle}{\partial
    \theta} $ is the baroclinic term, characteristic of non-aligned
  density and pressure gradients;
\item $\frac{1}{r \bar{\rho}
  c_p}\frac{d\bar{S}}{dr}\frac{\partial \langle P
  \rangle}{\partial\theta}$ is part of the baroclinic term but
arises from departures from adiabatic stratification;
\item \textit{viscous stresses} represent the diffusion of vorticity
  caused by viscous effects;
\item \textit{magnetic torques 1 and 2} represent the 'shear' and
  'transport' of the magnetic field by the current;
\item \textit{compressibility 2} represents the component of
  the curl of the magnetic torque owing to compressibility of the flow.
\end{itemize}
The meridional force balance equation \eqref{eq:TW_balance} helps understanding
which physical effects are responsible for breaking the
Taylor-Proudman constraint of cylindrical differential rotation in the
convection zone
(achieved for example with baroclinic flows, \textit{see}
\citealt{Miesch:2006p1957,Balbus:2009p1972,Brun:2010p1959}). 
Indeed, considering a small Rossby number and neglecting stratification, Reynolds,
Maxwell, and viscous stresses, Eq. \eqref{eq:TW_balance} reduces to
\begin{equation}
  \label{eq:TW_bal_baroclin}
  \frac{\partial\langle v_\varphi \rangle}{\partial z} =
  \frac{g}{2\Omega_0 r c_p} \frac{\partial\langle S\rangle}{\partial\theta},
\end{equation}
which is the original thermal wind equation (\textit{e.g.}
\citealt{Pedlosky:1987p2519,Durney:1999p1871}).
The baroclinic term induces motions that break the cylindrical rotation profile.
In the model of \citet{Wood:2011p1259}, this
balance is considered in the polar regions. Compositional gradients and
magnetic effects are added to equation
\eqref{eq:TW_bal_baroclin}. This balance is of primary importance in
their magnetic confinement layer scaling. We therefore present below the
full meridional force balance of our model to evaluate the role
played by all mechanisms in the tachocline, except compositional effects.

\begin{figure}[!htbp]
  \centering
  \includegraphics[width=\figwidth]{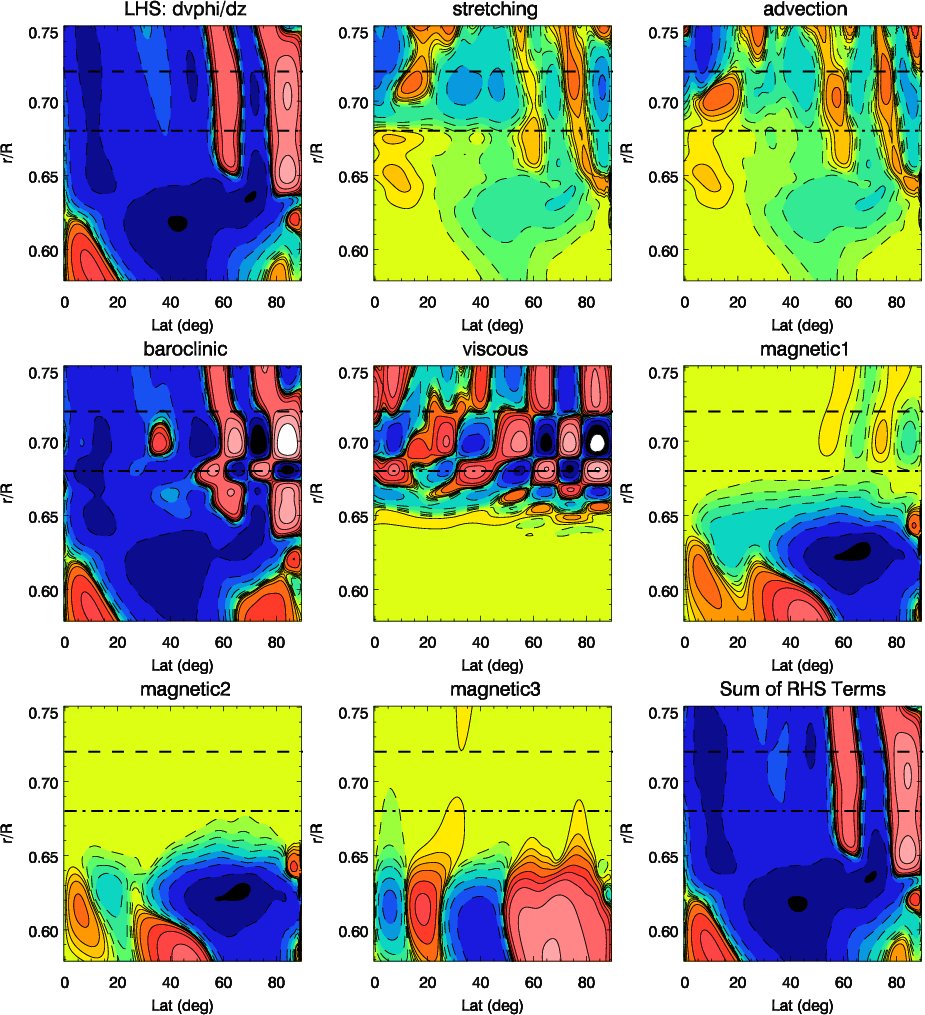}      
  \caption{Zoom around the tachocline of the different
    terms discussed in Eq. \eqref{eq:TW_balance}, averaged over
    longitude and two solar rotation periods \mybold{during the late
      evolution of the model}. Magnetic terms were
    \mybold{increased} by a factor $500$ to make their
    structure more evident.}
  \label{fig:TW_bal}
\end{figure}

Fig. \ref{fig:TW_bal} displays the main contributing terms from
Eq. \eqref{eq:TW_balance} in our simulation. The magnetic terms were
 increased (by a factor of $500$) to \mybold{make them more visible}.
Surprisingly, they are too low to contribute to the meridional force
balance, as such there is no 'magnetic wind' (\textit{see} \citealt{Fearn:1998p1955}).
As already noticed in \citet{Brun:2010p1234}, the baroclinic term
dominates the balance, especially under $0.65\, R_\odot$,
\textit{i.e.}, below the penetration. At the base of the convection
zone ($0.72\, R_\odot$), all terms play a significant
role. Stretching, advection, and viscous stresses \mybold{compete
against each other} at low
latitude, letting the baroclinic term dominate again. Near the pole,
the balance is more complex, and viscous stresses (helped a little by
stretching and advection) \mybold{counteract} the positive/negative pattern of
the baroclinic term. The strict thermal wind balance expressed in
Eq. \eqref{eq:TW_bal_baroclin} is thus not observed in the
polar region. The magnetic terms are located in the magnetic layer
where $B_\varphi$ is generated (\textit{see}
Sect. \ref{sec:general-evolution}), and are negligible.
\mybold{Indeed, our magnetic field is far from being force-free, and
  the Lorentz force is non-zero in the three components of the momentum
  equation. The poloidal velocity is mainly driven by advection
  of convective motions and by baroclinic terms in the radiation
  zone. As a result, the poloidal magnetic field does not act much on
  the poloidal velocity. Note that this does not imply the
  opposite. We indeed observed (\textit{see} Sect. \ref{sec:tach-regi-evol}) that
  the poloidal magnetic field evolution is dominated by advection and
  shear from convective motions.}}
We did not
recover the balance from \citet{Wood:2011p1259}, although we already
have a sufficiently strong magnetic field to transport angular momentum, as
seen in Sect. \ref{sec:angul-moment-balance}. Gyroscopic pumping is
not operating efficiently near the polar region. There is a small
downwelling, but it is not strong enough to modify the local dynamics
(\textit{see} \citealt{Brun:2010p1234} for more details).

\subsection{Angular momentum balance}
\label{sec:angul-moment-balance}

Identifying processes that redistribute angular momentum in
the bulk of the tachocline gives us clues to understand how this layer
evolves.
Because we choose stress-free
and torque free
magnetic boundary conditions, no external torque is applied to the
system. As a consequence, the angular momentum is conserved.
Following previous studies (\textit{e.g.}, \citealt{Brun:2004p1}), 
we examine the contribution of different terms in the balance of angular
momentum. Averaging the $\varphi$ component of the momentum equation 
(\ref{eq:mom_conserv}) over $\varphi$ and multiplying it by
$r\sin\theta$, we obtain the following equation for the specific angular
momentum $\mathcal{L}\equiv
r\sin\theta\left(\Omega_0 r\sin\theta + \langle v_\varphi \rangle \right)$:
\begin{equation}
  \label{eq:ang_mom_bal}
  \partial_t (\bar{\rho}\mathcal{L}) = - \mbox{\boldmath
    $\nabla$}\cdot( \underbrace{ \mathbf{F}^{MC} + \mathbf{F}^{RS} + \mathbf{F}^{VD} }_{\mbox{hydro}} + \underbrace{\mathbf{F}^{MT} + \mathbf{F}^{MS}}_{\mbox{MHD}} ),
\end{equation}
 where the different terms correspond to contributions
 from meridional circulation, reynolds stress, viscous diffusion,
 maxwell torque and  maxwell
 stress. They are defined by
 \begin{eqnarray}
   \label{eq:MC}
   \mathbf{F}^{MC} &\equiv&
   \bar{\rho}\langle\mathbf{v}_M\rangle\mathcal{L} \\
   \label{eq:RS}
   \mathbf{F}^{RS} &\equiv&
   r\sin\theta\bar{\rho}\left(\langle v_r'v_\varphi'\rangle\mathbf{e}_r +
     \langle v_\theta'v_\varphi'\rangle\mathbf{e}_\theta \right) \\
  \mathbf{F}^{VD} &\equiv&
   - \nu\bar{\rho}r^2\sin\theta\left\{ \partial_r\left(\frac{\langle
         v_\varphi \rangle}{r}\right)\mathbf{e}_r \right. \nonumber \\
   \label{eq:VD}
   &+& \left.
     \frac{\sin\theta}{r^2}\partial_\theta\left(\frac{\langle v_\varphi
       \rangle}{\sin\theta}\right)\mathbf{e}_\theta\right\} \\
   \label{eq:MT}
   \mathbf{F}^{MT} &\equiv&
   -\frac{r\sin\theta}{4\pi}\langle B_\varphi\rangle\langle\mathbf{B}_M\rangle \\
   \label{eq:MS}
   \mathbf{F}^{MS} &\equiv&
   -\frac{r\sin\theta}{4\pi}\left(\langle B_r'B_\varphi'\rangle\mathbf{e}_r +
     \langle B_\theta'B_\varphi'\rangle\mathbf{e}_\theta \right),
\end{eqnarray}
where the subscript $._M$ designates the meridional component of
$\mathbf{v}$ and $\mathbf{B}$.  In the previous equations, we
decomposed the velocity and the magnetic field into an azimuthally
averaged part $\langle .\rangle$ and $\varphi$-dependent part (with
a prime). The different contributions can be separated between radial
($\mathcal{F}_r$ along $\mathbf{e}_r$) and latitudinal
($\mathcal{F}_\theta$ along
 $\mathbf{e}_\theta$) contributions. We then compute
a radial flux of angular momentum defined by
\begin{eqnarray}
  \label{eq:radial_flux_ang_mom}
  \mathcal{I}_r(r) &=& \int_{\theta_1}^{\theta_2}
  \mathcal{F}_r(r,\theta)r^2\sin\theta \hspace{0.2cm}\mbox{d}{\theta},
\end{eqnarray}
where $(\theta_1,\theta_2)$ maybe chosen to study a particular
region of our simulation.

\begin{figure*}[!htpb]
  \centering
  \subfigure[]{
\includegraphics[width=.57\figwidth,angle=90]{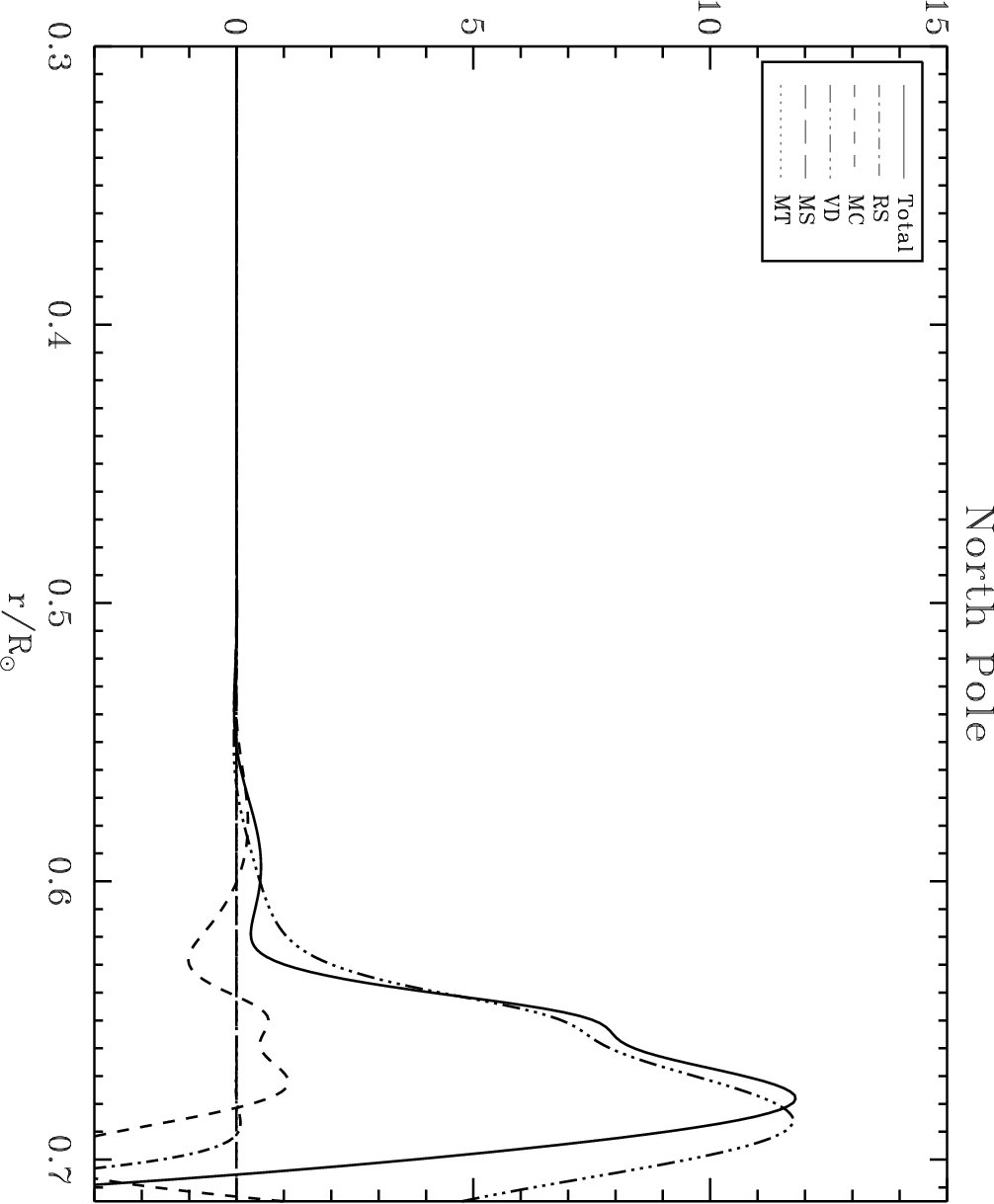} 
\includegraphics[width=.57\figwidth,angle=90]{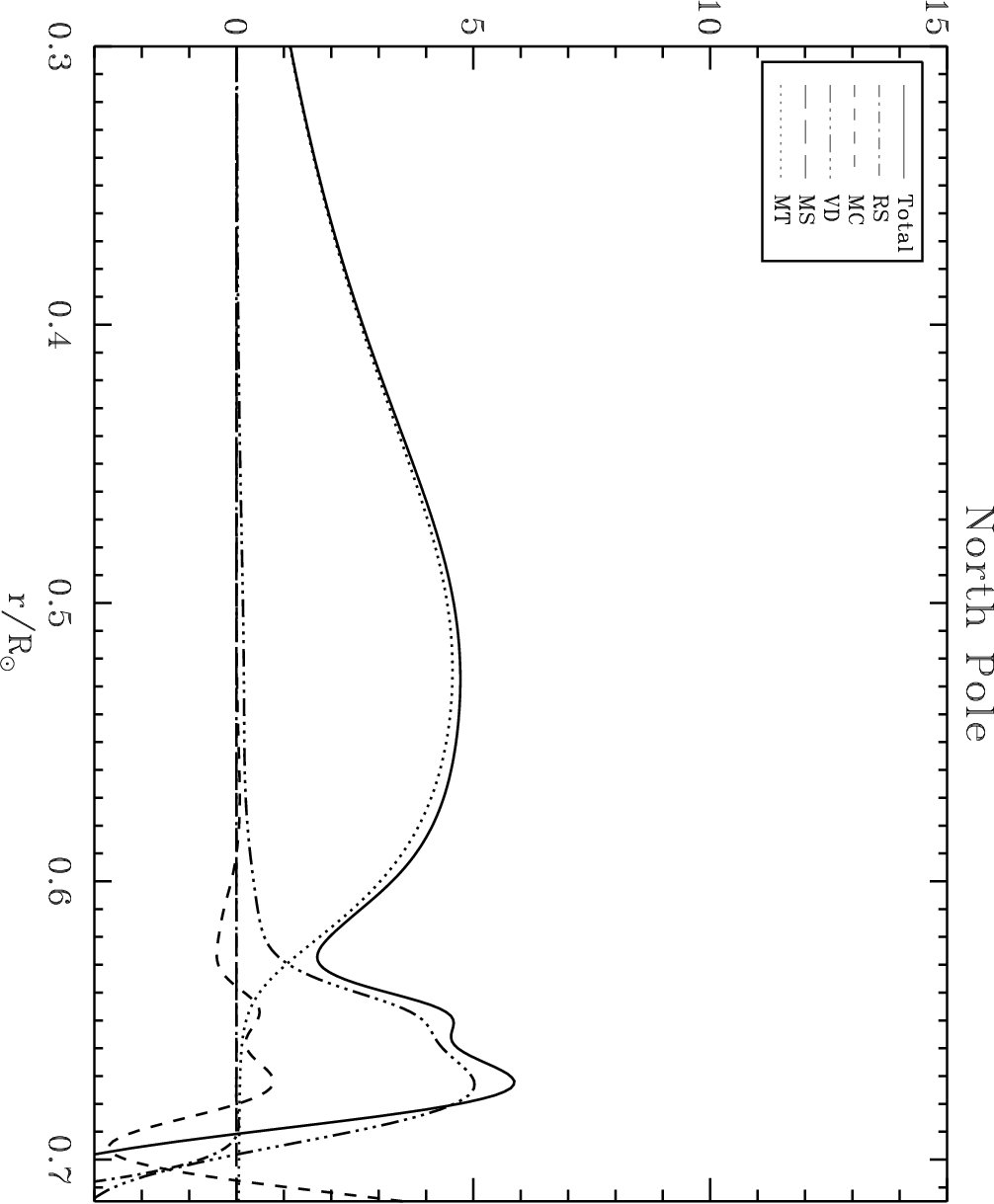} 
\includegraphics[width=.57\figwidth,angle=90]{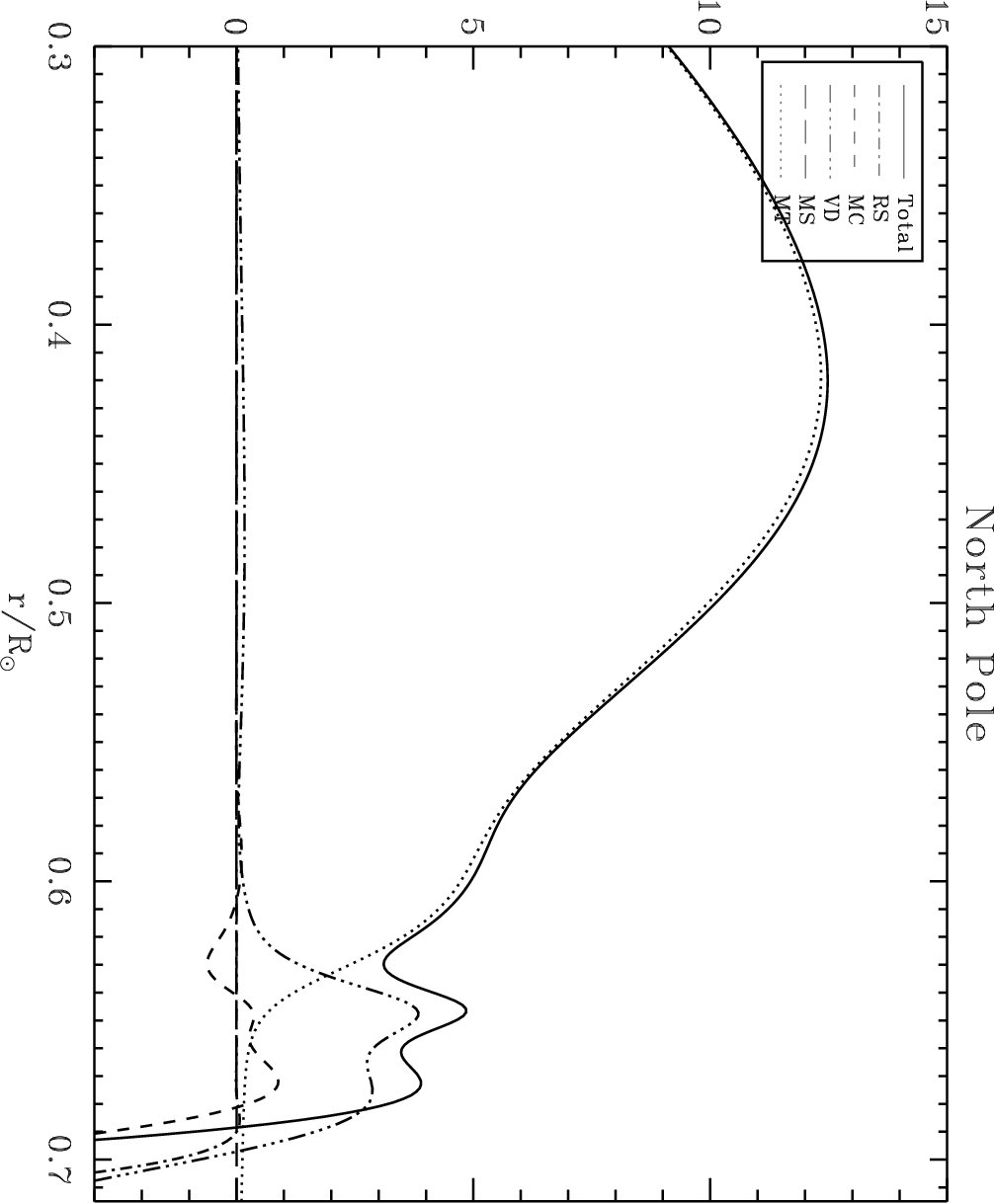} 
  \label{fig:northpole_amom}
}
  \subfigure[]{
\includegraphics[width=.57\figwidth,angle=90]{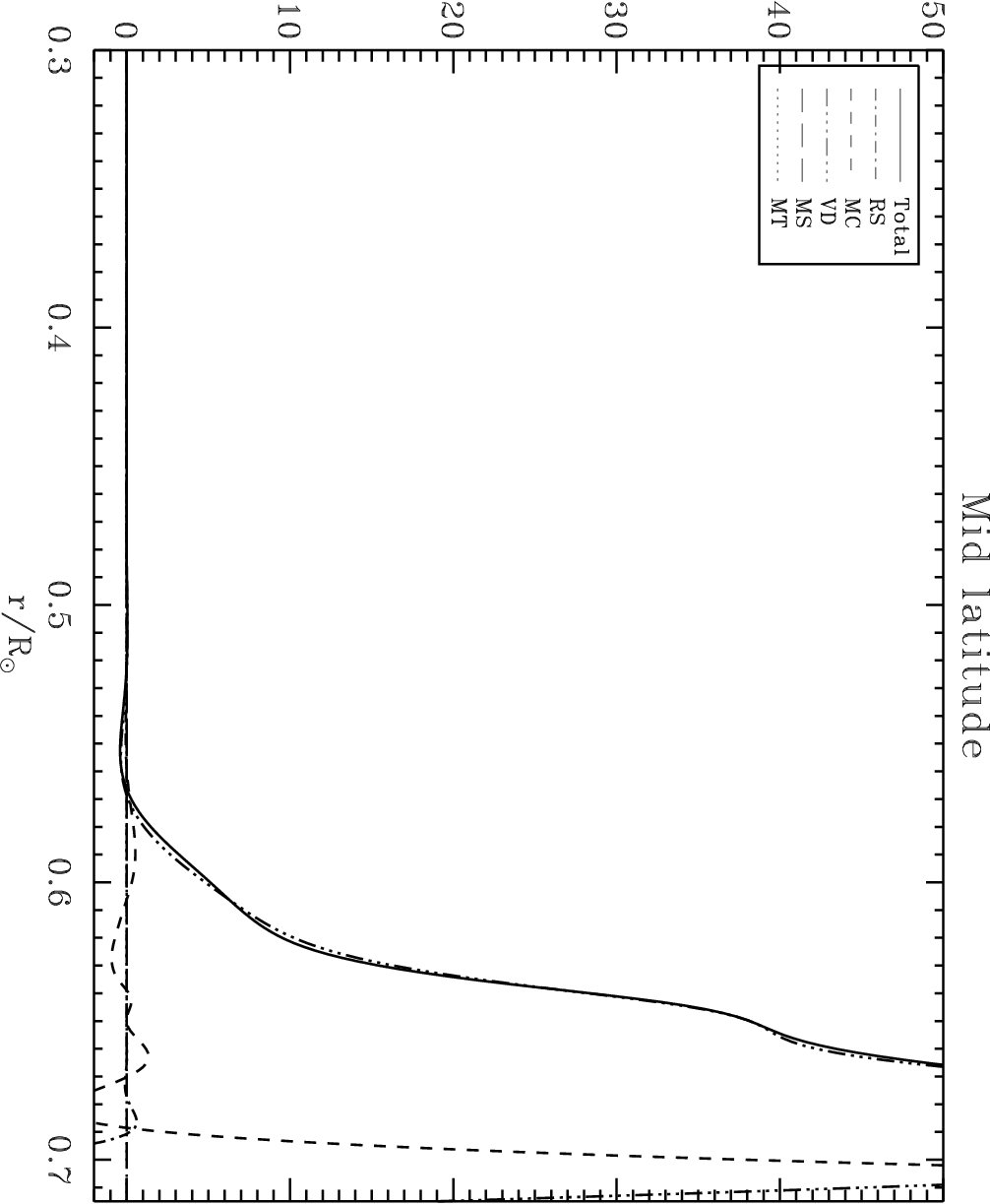} 
\includegraphics[width=.57\figwidth,angle=90]{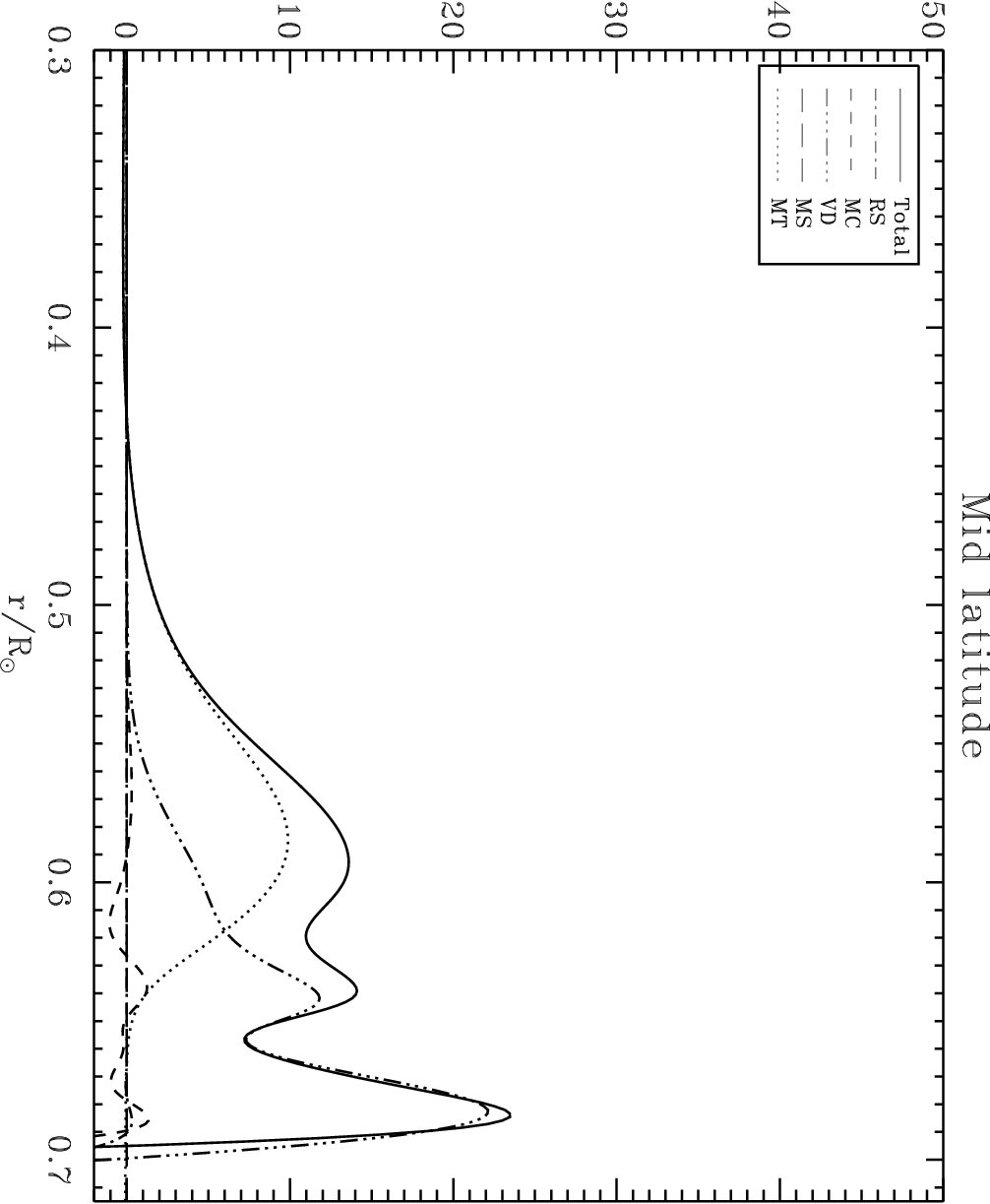} 
\includegraphics[width=.57\figwidth,angle=90]{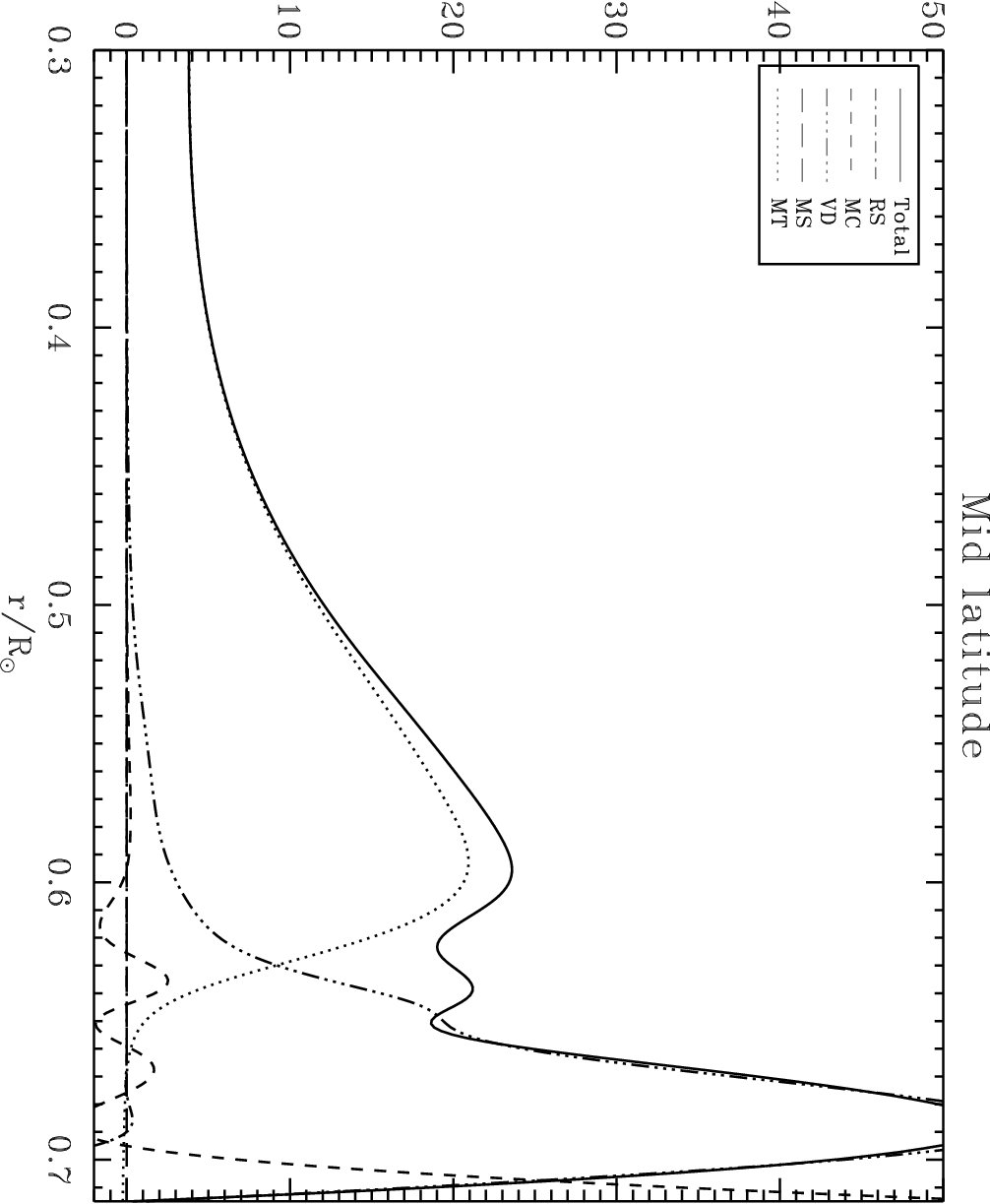} 
  \label{fig:midlat_amom}
}
 \caption{\mybold{Angular momentum redistribution in the radiative zone and
    the tachocline region. Panels (a) are summed over 
    $\theta_1=60$ and $\theta_2=90$ (north pole), and panels (b) are
    summed over $\theta_1=30$ and $\theta_2=60$ (mid latitudes).
   The different fluxes are averaged
    approximately over six rotation periods. From left to right, the
    three plots correspond to the instants used to make
    Figs. \ref{fig:Bphi_p_MC1}, \ref{fig:Bphi_p_MC3} and \ref{fig:Bphi_p_MC4}.
    Our magnetic field
    does not initially change the angular momentum balance.
}}
  \label{fig:angmombal_polar_region}
\end{figure*}
 
As emphasized in Sect. \ref{sec:glob-evol-magn}, magnetic field
lines seem to be advected and diffused \mybold{first outward and then poleward}. In Fig.
\ref{fig:angmombal_polar_region} we plot the different components of
the radial angular momentum flux in the radiative zone and the
tachocline at different times,  summed over the polar region
\mybold{and at mid latitudes}. Initially, the magnetic field
is axisymmetric so Maxwell stresses do not contribute to the angular momentum
balance. Because we chose a purely meridional magnetic field, there is
no large-scale Maxwell
torque. The radial angular momentum flux is then essentially
carried by the meridional circulation in the radiative zone and is very \mybold{weak}
(few percents of its value in the convective zone). \mybold{Note,
  however, that internal waves are present in our simulation and may
  contribute to the transport of angular momentum (\mybold{see}
  \citealt{Brun:2010p1234} for more details).}
In the tachocline,
meridional circulation, Reynolds stress and viscous diffusion transport the angular
momentum. 

While the magnetic field \mybold{more and more pervades the domain}, a magnetic torque \mybold{develops} in the
magnetic layer (along with $B_\varphi$ generation, see
Figs. \ref{fig:vphi_plus_Bstrl}-\ref{fig:Bphi_p_MC}) at the top of the
radiative zone both at high and mid latitudes. Angular momentum is
then extracted \mybold{by the Maxwell torque} from the radiative
zone (that locally rotates faster) into the \mybold{tachocline} (that
locally rotates slower), slowing down the radiative interior.
\mybold{This torque tends to align the $\Omega$ contours with the
poloidal magnetic field lines, thus bringing the system closer and
closer to a state known as
Ferraro's law of isorotation. We remark
in addition that the Maxwell torque operates initially in a wider part of the
radiative zone in the polar region than at mid latitudes. Because the
magnetic field primarily connects at the poles, it is natural that the
transport of angular momentum by the magnetic field starts earlier in this
region. One may also observe that this transport is more intense at
mid-latitude, acknowledging that the magnetic field lines involve
a magnetic field with much higher strength there. We stress again that
the angular momentum is conserved in our simulation, meaning that the
angular momentum lost in the radiative zone is redistributed in the
convective zone.} 

\mybold{We also observe that the magnetic torque applies essentially
  in the radiation zone and becomes negligible at the base of the
  convection zone. Because we are not yet in iso-rotation (\textit{see}
  Fig. \ref{fig:vphi_plus_Bstrl4}), this may only be interpreted by the
fact that $B_\varphi$ is weak in this region.}

Because the first magnetic field lines that transport angular
  momentum are localized at \mybold{mid- and high}-latitudes, we observe here only a slow-down of the radiative
  interior. If we had continued this simulation, magnetic field lines
  would have connected the radiative zone to the convection zone
  \mybold{near the equator}. This would imply a speed-up of the
  radiative interior, consistent with the results from
  \citet{Brun:2006p24}. However, we do not show this later evolution
  because it does not add any useful information.

Viscosity seems to act substantially in the bulk of
the tachocline, augmenting the angular momentum transport
locally. \mybold{This viscous transport is caused by
the radial shear and the relatively high viscosity value used to meet numerical requirements.
As a result, diffusive processes are over-evaluated in our simulation.
Angular momentum transport like this would not
contribute in the Sun, given the low viscosity of the solar plasma.
In our case, both thermal and viscous diffusion processes are acting in
our simulation thanks to our choice
of Prandtl number ($10^{-3}$) in the radiative zone. 
However, BZ06 showed that the simulated viscous
spreading of the tachocline does not make much qualitative difference with the
thermal spreading introduced by \citet{Spiegel:1992p5}.}

The typical viscous diffusive
time scale in our simulation is given by $\tau_\nu = {R^2}/{\nu}
\sim 2.2\, 10^{10} \, s$ at the top of the radiative zone. As the magnetic
field grows, the Alfv\'en time $\tau_A =
R{\sqrt{4\pi\bar{\rho}}}/{B}$ diminishes alike and magnetic effects
become more important than diffusion. The angular momentum balance
analysis emphasizes two phases in our simulation: a
diffusive transport of angular momentum and the action of large-scale Maxwell torque.
This confirms
that our magnetic field does not prevent the spread of the tachocline,
but on the contrary transports angular momentum outwards and speeds
the process up. 

\section{Discussion and conclusion}
\label{sec:disc-concl}

The existence of a magnetic field in the solar radiation zone has been
the subject of a longstanding debate. Such a field, for example of fossil origin,
could easily account for the uniform rotation of the radiative interior.
It could also prevent the inward spread of the tachocline, through thermal diffusion,
as was proposed by \citet{Gough:1998p34}.
However, it has been argued that such a field would diffuse into the convection zone and imprint
the latitude-dependent rotation of that region on the radiation zone
below (which is not observed by helioseismology inversions).

This objection has indeed been confirmed by the calculations of \citet{Garaud:2002p27}  
and \citet{Brun:2006p24}. However, their simulations treated the top of the radiation zone
as a non-penetrative boundary, which does not allow for a meridional flow that would originate in the convection zone
and could oppose the diffusion of the magnetic field.

This restriction was lifted in the $2$D simulations performed by
\citet{Sule:2005p31,Rudiger:2007p40,Garaud:2008p36}, who showed that
this meridional circulation could indeed prevent the magnetic field
from diffusing into the convection zone. \mybold{Very
recent work in 2D \citep{Rogers:2011p1234} has studied the role of
a fossil magnetic field in a simulation coupling a radiative and a
convective zone. It confirms our result that the field is not confined at the
base of the convection zone. However, the author does not find any
magnetic transport of angular momentum, possibly because of a
force-free state for the magnetic field.} These $2$D simulations assumed large-scale axisymmetric flows,
unlike the three-dimensional motions of the real convection zone. For this reason we decided
to treat the problem using a 3D code, with the best possible
representation of the turbulent convective motions.

We showed by \mybold{considering time-dependent 3D motions} that the interior magnetic
field does not stay confined in the radiative interior,
but expands into the convection zone. Non-axisymmetric convective motions interact
much differently with a dipolar-type magnetic field, as would a monolithic, single cell
large-scale meridional circulation.

\mybold{That our magnetic field is able to expand from the
  radiative interior into the convection zone results in an efficient outward
transport of angular momentum
through the tachocline, imposing Ferraro's law of iso-rotation. It
is likely a direct consequence of the non force-free configuration of our
magnetic field that yields this efficient transport through the large-scale
magnetic torque.} 

We are aware that our simulations are far from representing the real
Sun, because of the enhanced diffusivities we had to use to meet numerical
requirements. \mybold{The overshooting convective motions are
  overestimated owing to our enhanced thermal diffusivity $\kappa$.
The high solar Peclet number (lower $\kappa$) regime would make the flow amplitude drop
even faster at the base of the convection zone, which would result in an
even weaker meridional circulation velocity. This would not help
the confinement of the magnetic field. \\
Although our enhanced magnetic diffusivity is quite large, we
demonstrated (\textit{see} Sect. \ref{sec:general-evolution}) that our
choice of parameters ensured that we are in the correct regime for
magnetic field confinement, with
advection dominating diffusion.} Nevertheless, more
efforts will be made  in
future work to increase the spatial resolution and to improve the
subgrid-scales representation.\\
 \mybold{The initial magnetic topology
  may also be questioned. The gradient of our primordial dipolar field possesses a
  maximum at the equator. Because the field is initialized in the radiation
  zone where its evolution is diffusive, the magnetic field initially evolves
  faster at its maximum gradient location, \textit{i.e.}, at the equator. As a result, it first encounters
  the complex motions coming from the convection zone at the equator. Other
  simulations conducted with a modified topology to change the
  location of the maximum gradient (and thus the latitude where the
  magnetic field primarily enters the tachocline) led to similar
  results as those reported here.}

\mybold{Given the extension of our tachocline}, the differential rotation is already initially present
below the base of the convection zone before we \mybold{introduce} the magnetic
field.
A confined magnetic field would naturally erode this angular velocity gradient,
though we do not know much about the real initial conditions. In our
simulation the magnetic field opens into the convection zone, consequently there
is no chance to obtain a solid rotator in the radiative zone because
there is no confinement. In fact, our magnetic field lines primarily open at the
equator. We point out that this is a major \mybold{difficulty with} the GM98 scenario:
even if magnetic pumping can be a way to slow down the penetration of the 
magnetic field into the convective zone, it will not prevent it from
connecting the two zones, and/or prevent the magnetic field lines from
opening there.  Furthermore, the
meridional circulation is not a laminar \mybold{uni-cellular} flow in our 3D simulation. It has a
complex instantaneous multicellular pattern, which affects the
efficiency of a polar confinement. 

We therefore conclude that a dipolar axisymmetric fossil magnetic field cannot
prevent the spread of the solar tachocline. We intend to explore other
mechanisms and different magnetic field topologies. Our 3D model will
allow us to test non-axisymmetric configurations, \textit{e.g.}, an
oblique and confined dipole.

\begin{acknowledgements}
\mybold{The authors thank Nick Brummell, Gary Glatzmaier, Michael McIntyre and
Toby Wood for helpful discussions. We are also grateful to Sean Matt
for a careful reading of our manuscript.}
The authors acknowledge funding
by the European Research Council through ERC grant STARS2 207430
(www.stars2.eu).
3D renderings in Figs. \ref{fig:3D} and \ref{fig:3D_up} were made with SDvision (\textit{see}
\citealt{Pomarede:2010p4299}).
\end{acknowledgements}

\bibliographystyle{aa}
\bibliography{bib_astro.new}

\begin{thebibliography}{51}
\expandafter\ifx\csname natexlab\endcsname\relax\def\natexlab#1{#1}\fi

\bibitem[{Balbus {et~al.}(2009)Balbus, Bonart, Latter, \&
  Weiss}]{Balbus:2009p1972}
Balbus, S.~A., Bonart, J., Latter, H.~N., \& Weiss, N.~O. 2009, MNRAS, 400, 176

\bibitem[{Barnes {et~al.}(1999)Barnes, Charbonneau, \&
  MacGregor}]{Barnes:1999p2302}
Barnes, G., Charbonneau, P., \& MacGregor, K.~B. 1999, \apj, 511, 466

\bibitem[{Braithwaite \& Spruit(2004)}]{Braithwaite:2004p1944}
Braithwaite, J. \& Spruit, H.~C. 2004, \nat, 431, 819

\bibitem[{Brown {et~al.}(2010)Brown, Browning, Brun, Miesch, \&
  Toomre}]{Brown:2010p371}
Brown, B.~P., Browning, M.~K., Brun, A.~S., Miesch, M.~S., \& Toomre, J. 2010,
  \apjs, 711, 424

\bibitem[{Brown {et~al.}(1989)Brown, Christensen-Dalsgaard, Dziembowski, Goode,
  Gough, \& Morrow}]{Brown:1989p7}
Brown, T.~M., Christensen-Dalsgaard, J., Dziembowski, W.~A., {et~al.} 1989,
  \apjs, 343, 526

\bibitem[{Brun(2005)}]{Brun:2005p1234}
Brun, A.~S. 2005, Habilitation in Physics (University of Denis-Diderot - Paris
  VII)

\bibitem[{Brun(2007)}]{Brun:2007p2626}
Brun, A.~S. 2007, Astr. Nach., 328, 1137

\bibitem[{Brun {et~al.}(2010)Brun, Antia, \& Chitre}]{Brun:2010p1959}
Brun, A.~S., Antia, H.~M., \& Chitre, S.~M. 2010, \aap, 510, 33

\bibitem[{Brun {et~al.}(2002)Brun, Antia, Chitre, \& Zahn}]{Brun:2002p831}
Brun, A.~S., Antia, H.~M., Chitre, S.~M., \& Zahn, J.-P. 2002, \aap, 391, 725

\bibitem[{Brun {et~al.}(2011)Brun, Miesch, \& Toomre}]{Brun:2010p1234}
Brun, A.~S., Miesch, M., \& Toomre, J. 2011, accepted in ApJ

\bibitem[{Brun {et~al.}(2004)Brun, Miesch, \& Toomre}]{Brun:2004p1}
Brun, A.~S., Miesch, M.~S., \& Toomre, J. 2004, \apjs, 614, 1073

\bibitem[{Brun \& Toomre(2002)}]{Brun:2002p996}
Brun, A.~S. \& Toomre, J. 2002, \apj, 570, 865

\bibitem[{Brun \& Zahn(2006)}]{Brun:2006p24}
Brun, A.~S. \& Zahn, J.-P. 2006, A{\&}A, 457, 665

\bibitem[{Cattaneo {et~al.}(2003)Cattaneo, Emonet, \&
  Weiss}]{Cattaneo:2003p2833}
Cattaneo, F., Emonet, T., \& Weiss, N. 2003, \apj, 588, 1183

\bibitem[{Charbonneau {et~al.}(1999)Charbonneau, Christensen-Dalsgaard,
  Henning, Larsen, Schou, Thompson, \& Tomczyk}]{Charbonneau:1999p19}
Charbonneau, P., Christensen-Dalsgaard, J., Henning, R., {et~al.} 1999, \apjs,
  527, 445

\bibitem[{Clune {et~al.}(1999)Clune, Elliott, Miesch, Toomre, \&
  Glatzmaier}]{Clune:1999p42}
Clune, T.~L., Elliott, J.~R., Miesch, M.~S., Toomre, J., \& Glatzmaier, G.~A.
  1999, Parrallel Computing, 25, 361

\bibitem[{Dritschel \& McIntyre(2008)}]{Dritschel:2008p4287}
Dritschel, D.~G. \& McIntyre, M.~E. 2008, Journal of the Atmospheric Sciences,
  65, 855

\bibitem[{Duez \& Mathis(2010)}]{Duez:2010p1947}
Duez, V. \& Mathis, S. 2010, \aap, 517, 58

\bibitem[{Durney(1999)}]{Durney:1999p1871}
Durney, B.~R. 1999, \apj, 511, 945

\bibitem[{Elliott(1997)}]{Elliott:1997p822}
Elliott, J.~R. 1997, \aap, 327, 1222

\bibitem[{Fearn(1998)}]{Fearn:1998p1955}
Fearn, D.~R. 1998, Reports on Progress in Physics, 61, 175

\bibitem[{Ferraro(1937)}]{Ferraro:1937p830}
Ferraro, V. C.~A. 1937, MNRAS, 97, 458

\bibitem[{Garaud(2002)}]{Garaud:2002p27}
Garaud, P. 2002, MNRAS, 329, 1

\bibitem[{Garaud \& Garaud(2008)}]{Garaud:2008p36}
Garaud, P. \& Garaud, J.-D. 2008, MNRAS, 391, 1239

\bibitem[{Gough \& McIntyre(1998)}]{Gough:1998p34}
Gough, D.~O. \& McIntyre, M.~E. 1998, \nat, 394, 755

\bibitem[{Kim(2005)}]{Kim:2005p623}
Kim, E.-J. 2005, \aap, 441, 763

\bibitem[{Kim \& Leprovost(2007)}]{Kim:2007p555}
Kim, E.-J. \& Leprovost, N. 2007, \aap, 468, 1025

\bibitem[{Leprovost \& Kim(2006)}]{Leprovost:2006p549}
Leprovost, N. \& Kim, E.-J. 2006, \aap, 456, 617

\bibitem[{Miesch(2003)}]{Miesch:2003p523}
Miesch, M.~S. 2003, \apj, 586, 663

\bibitem[{Miesch {et~al.}(2006)Miesch, Brun, \& Toomre}]{Miesch:2006p1957}
Miesch, M.~S., Brun, A.~S., \& Toomre, J. 2006, \apj, 641, 618

\bibitem[{Miesch {et~al.}(2000)Miesch, Elliott, Toomre, Clune, Glatzmaier, \&
  Gilman}]{Miesch:2000p29}
Miesch, M.~S., Elliott, J.~R., Toomre, J., {et~al.} 2000, \apjs, 532, 593

\bibitem[{Moffatt(1978)}]{Moffatt:1978p1953}
Moffatt, H.~K. 1978, Cambridge University Press

\bibitem[{Morel(1997)}]{Morel:1997p1788}
Morel, P. 1997, A {\&} A Supplement series, 124, 597

\bibitem[{Pedlosky(1987)}]{Pedlosky:1987p2519}
Pedlosky, J. 1987, Journal of Phys. Oc., 17, 1978

\bibitem[{Pomarede \& Brun(2010)}]{Pomarede:2010p4299}
Pomarede, D. \& Brun, A. 2010, Astronomical Data Analysis Software and Systems
  XIX. Proceedings of a conference held October 4-8, 434, 378

\bibitem[{Rogers(2011)}]{Rogers:2011p1234}
Rogers, T.~M. 2011, accepted in ApJ

\bibitem[{Rudiger \& Kitchatinov(1997)}]{Rudiger:1997p17}
Rudiger, G. \& Kitchatinov, L.~L. 1997, Astro. Nach., 318, 273

\bibitem[{Rudiger \& Kitchatinov(2007)}]{Rudiger:2007p40}
Rudiger, G. \& Kitchatinov, L.~L. 2007, NJP, 9, 302

\bibitem[{Schou {et~al.}(1998)Schou, Antia, Basu, Bogart, Bush, Chitre,
  Christensen-Dalsgaard, di~Mauro, Dziembowski, Eff-Darwich, Gough, Haber,
  Hoeksema, Howe, Korzennik, Kosovichev, Larsen, Pijpers, Scherrer, Sekii,
  Tarbell, Title, Thompson, \& Toomre}]{Schou:1998p3791}
Schou, J., Antia, H.~M., Basu, S., {et~al.} 1998, \apj, 505, 390

\bibitem[{Spiegel \& Zahn(1992)}]{Spiegel:1992p5}
Spiegel, E.~A. \& Zahn, J.-P. 1992, A{\&}A, 265, 106

\bibitem[{Starr(1968)}]{starr1968physics}
Starr, V. 1968, {Physics of negative viscosity phenomena} (McGraw-Hill)

\bibitem[{Sule {et~al.}(2005)Sule, Rudiger, \& Arlt}]{Sule:2005p31}
Sule, A., Rudiger, G., \& Arlt, R. 2005, A{\&}A, 437, 1061

\bibitem[{Tayler(1973)}]{Tayler:1973p1727}
Tayler, R.~J. 1973, MNRAS, 161, 365

\bibitem[{Thompson {et~al.}(2003)Thompson, Christensen-Dalsgaard, Miesch, \&
  Toomre}]{Thompson:2003p2511}
Thompson, M.~J., Christensen-Dalsgaard, J., Miesch, M.~S., \& Toomre, J. 2003,
  ARA{\&}A, 41, 599

\bibitem[{Tobias {et~al.}(2001)Tobias, Brummell, Clune, \&
  Toomre}]{Tobias:2001p472}
Tobias, S.~M., Brummell, N.~H., Clune, T.~L., \& Toomre, J. 2001, \apj, 549,
  1183

\bibitem[{Weiss {et~al.}(2004)Weiss, Thomas, Brummell, \&
  Tobias}]{Weiss:2004p464}
Weiss, N.~O., Thomas, J.~H., Brummell, N.~H., \& Tobias, S.~M. 2004, \apj, 600,
  1073

\bibitem[{Wood \& McIntyre(2007)}]{Wood:2007p4288}
Wood, T.~S. \& McIntyre, M.~E. 2007, UNSOLVED PROBLEMS IN STELLAR PHYSICS: A
  Conference in Honor of Douglas Gough. AIP Conference Proceedings, 948, 303

\bibitem[{Wood \& McIntyre(2011)}]{Wood:2011p1259}
Wood, T.~S. \& McIntyre, M.~E. 2011, accepted in JFM

\bibitem[{Zahn(1991)}]{Zahn:1991p2459}
Zahn, J.-P. 1991, \aap, 252, 179

\bibitem[{Zahn {et~al.}(2007)Zahn, Brun, \& Mathis}]{Zahn:2007p1749}
Zahn, J.-P., Brun, A.~S., \& Mathis, S. 2007, \aap, 474, 145

\bibitem[{Zanni \& Ferreira(2009)}]{Zanni:2009p2526}
Zanni, C. \& Ferreira, J. 2009, \aap, 508, 1117

\end{thebibliography}

\end{document}